\documentclass[a4paper, 11pt]{article}
\pdfoutput=1
\usepackage{jheppub}

\usepackage{cancel}
\usepackage{verbatim} 
\usepackage{multirow}
\newcommand{\al}{\ensuremath{\alpha} }
\newcommand{\be}{\ensuremath{\beta} }
\newcommand{\ga}{\ensuremath{\gamma} }
\newcommand{\Ga}{\ensuremath{\Gamma} }
\newcommand{\de}{\ensuremath{\delta} }
\newcommand{\De}{\ensuremath{\Delta} }
\newcommand{\si}{\ensuremath{\sigma} }
\newcommand{\cO}{\ensuremath{\mathcal O} }
\newcommand{\cN}{\ensuremath{\mathcal N} }
\newcommand{\gc}{\ensuremath{g_c^2} }
\newcommand{\gceq}[1]{\ensuremath{g_{c = #1}^2} }
\newcommand{\gGF}{\ensuremath{g_{\text{GF}}^2} }
\newcommand{\gtc}{\ensuremath{\widetilde g_c^2} }
\newcommand{\gtceq}[1]{\ensuremath{\widetilde g_{c = #1}^2} }
\newcommand{\gtGF}{\ensuremath{\widetilde g_{\text{GF}}^2} }
\newcommand{\gstar}{\ensuremath{g_{\star}^2} }
\newcommand{\topt}{\ensuremath{\tau_{\text{opt}}} }
\newcommand{\gsim}{\ensuremath{\gtrsim} }
\newcommand{\lsim}{\ensuremath{\lesssim} }
\newcommand{\Lra}{\ensuremath{\Longrightarrow} }
\newcommand{\Sb}{\ensuremath{\cancel{S^4}} }
\newcommand{\MSbar}{\ensuremath{\overline{\textrm{MS}} } }
\newcommand{\chidof}{\ensuremath{\mbox{$\chi^2/\text{d.o.f.}$}}}
\newcommand{\vev}[1]{\ensuremath{\left\langle #1 \right\rangle} }
\newcommand{\nn}{\nonumber}
\newcommand{\eq}[1]{eq.~\ref{#1}}
\newcommand{\fig}[1]{figure~\ref{#1}}
\newcommand{\Fig}[1]{Figure~\ref{#1}}
\newcommand{\tab}[1]{table~\ref{#1}}
\newcommand{\appref}[1]{appendix~\ref{#1}}
\newcommand{\secref}[1]{section~\ref{#1}}
\newcommand{\refcite}[1]{ref.~\cite{#1}}
\newcommand{\Refcite}[1]{Ref.~\cite{#1}}

\title{Nonperturbative \be function \\ of twelve-flavor SU(3) gauge theory}

\author{Anna~Hasenfratz$^1$}
\author{and David~Schaich$^{2,3}$}
\affiliation{$^1$Department of Physics, University of Colorado, Boulder, Colorado 80309, United States}
\affiliation{$^2$Department of Physics, Syracuse University, Syracuse, New York 13244, United States}
\affiliation{$^3$AEC Institute for Theoretical Physics, University of Bern, 3012 Bern, Switzerland}
\emailAdd{anna.hasenfratz@colorado.edu}
\emailAdd{schaich@itp.unibe.ch}

\abstract{ 
  We study the discrete \be function of SU(3) gauge theory with $N_f = 12$ massless fermions in the fundamental representation.
  Using an nHYP-smeared staggered lattice action and an improved gradient flow running coupling $\gtc(L)$ we determine the continuum-extrapolated discrete \be function up to $\gc \approx 8.2$.
  We observe an IR fixed point at $\gstar = 7.3\left(_{-2}^{+8}\right)$ in the $c = \sqrt{8t} / L = 0.25$ scheme, and $\gstar = 7.3\left(_{-3}^{+6}\right)$ with $c = 0.3$, combining statistical and systematic uncertainties in quadrature.
  \mbox{The systematic effects we} \mbox{investigate include the stability} of the $(a / L) \to 0$ extrapolations, the interpolation of $\gtc(L)$ as a function of the bare coupling, the improvement of the gradient flow running coupling, and the discretization of the energy density.
  In an appendix we observe that the resulting systematic errors increase dramatically upon combining \mbox{smaller $c \lsim 0.2$ with smaller} ${L \leq 12}$, leading to an IR fixed point at $\gstar = 5.9(1.9)$ in the $c = 0.2$ scheme, \mbox{which resolves to} $\gstar = 6.9\left(_{-1}^{+6}\right)$ upon considering only $L \geq 16$.
  At the IR fixed point we measure the leading irrelevant critical exponent to be $\ga_g^{\star} = 0.26(2)$, comparable to perturbative estimates.
}
\arxivnumber{1610.10004}

\begin{document}
\maketitle
\flushbottom

\section{\label{sec:intro}Introduction} 
SU(3) gauge theory with $N_f = 12$ flavors of massless fermions in the fundamental representation has been considered by many independent lattice studies in recent years.
This effort is motivated by the expectation that the 12-flavor system exhibits conformal or near-conformal dynamics qualitatively different than QCD.
That is, $N_f = 12$ is likely either within or close to the lower boundary of the SU(3) conformal window $N_f^{(c)} \leq N_f < 16.5$, where the theory flows to a chirally symmetric conformal fixed point in the infrared (IRFP)~\cite{Caswell:1974gg, Banks:1981nn}.
Should the system undergo spontaneous chiral symmetry breaking (i.e., $12 < N_f^{(c)}$), then it provides an example of a strongly coupled theory in which lattice calculations have observed a light $0^{++}$ scalar~\cite{Aoki:2013zsa, Fodor:2014pqa}.
In this case investigations of $N_f = 12$ are relevant to explore possible strongly coupled new physics beyond the standard model (BSM), in which such a light composite scalar could be consistent with the observed SM-like Higgs boson~\cite{Chatrchyan:2013lba, Aad:2013wqa}.
Alternatively, if the 12-flavor system is within the conformal window, as our results indicate, it provides a useful testbed in which to develop and apply non-perturbative methods to investigate IR-conformal systems.
Even in this case there can be connections to BSM phenomenology, in models where the mass of some of the fermions is lifted to guarantee spontaneous chiral symmetry breaking.
Lattice investigations of this situation have shown that this system follows hyperscaling, a highly non-QCD-like behavior, exhibiting natural large scale separation and UV dynamics dominated by the 12-flavor IRFP~\cite{Brower:2015owo, Hasenfratz:2016gut}.

Initial indications that $N_f = 12$ would be interesting came from continuum field theory analyses.
For example, two-, three-, and four-loop perturbative computations of the \be function all predict an IRFP for the system~\cite{Caswell:1974gg, Banks:1981nn, Ryttov:2010iz, Pica:2010xq}.\footnote{A recent five-loop \be function computation~\cite{Baikov:2016tgj, Herzog:2017ohr} appears to change this trend, although the subsequent refs.~\cite{Stevenson:2016mnv, Ryttov:2016ner, Ryttov:2016asb, Ryttov:2016hal} argue that all systems with $9 \leq N_f \leq 16$ exhibit IRFPs at the five-loop level.  We address this development in \secref{sec:conclusion}.}
Analyses that combine perturbation theory with Schwinger--Dyson equations~\cite{Appelquist:1996dq, Appelquist:1998rb} produce estimates for the location of the lower boundary of the SU(3) conformal window that range from $N_f^{(c)} \approx 8$ in \refcite{Bashir:2013zha} to $N_f^{(c)} \approx 12$ in refs.~\cite{Appelquist:1996dq, Appelquist:1998rb, Dietrich:2006cm}. 
Similarly, functional renormalization group (RG) methods suggest $N_f^{(c)} \approx 10$--13~\cite{Braun:2006jd, Braun:2010qs} while a conjectured thermal inequality predicts the bound $N_f^{(c)} \lsim 12$~\cite{Appelquist:1999hr}.

Numerical studies of the 12-flavor system have employed a wide variety of methods, including investigation of the running coupling and its discrete \be function~\cite{Appelquist:2007hu, Appelquist:2009ty, Bilgici:2009nm, Itou:2010we, Ogawa:2011ki, Lin:2012iw, Itou:2012qn, Itou:2013faa, Cheng:2014jba, Hasenfratz:2015xpa, Lin:2015zpa, Fodor:2016zil}; exploration of the phase diagram through calculations at zero and finite temperature~\cite{Deuzeman:2009mh, Jin:2009mc, Deuzeman:2010fn, Cheng:2011ic, Deuzeman:2011pa, Jin:2012dw, Fodor:2012uu, Schaich:2012fr, Deuzeman:2012ee, daSilva:2012wg, Hasenfratz:2013uha, Ishikawa:2013tua, Ishikawa:2015iwa}; analysis of hadron masses and decay constants~\cite{Deuzeman:2009mh, Fodor:2009wk, Jin:2009mc, Fodor:2011tu, Appelquist:2011dp, DeGrand:2011cu, Deuzeman:2012pv, Jin:2012dw, Fodor:2012uu, Aoki:2012eq, Fodor:2012et, Deuzeman:2013kma, Aoki:2013zsa, Cheng:2013xha, Lombardo:2014cqa, Fodor:2014pqa, Lombardo:2014pda, Lombardo:2014mda}; study of the eigenmodes of the Dirac operator~\cite{Fodor:2009wk, Hasenfratz:2012fp, Cheng:2013eu, Cheng:2013bca, Itou:2014ota}; and more~\cite{Hasenfratz:2010fi, Hasenfratz:2011xn, Hasenfratz:2011da, Meurice:2012sj, Fodor:2012uw, Petropoulos:2012mg, Itou:2013kaa, Itou:2013ofa, Aoki:2016yrm}.
See also the recent reviews~\cite{DeGrand:2015zxa, Giedt:2015alr, Nogradi:2016qek}.
Except for refs.~\cite{Ishikawa:2013tua, Ishikawa:2015iwa}, all of these studies use staggered fermions (with or without various forms of improvement), which conveniently represent $N_f = 12$ continuum flavors as three (unrooted) lattice fields.\footnote{At the perturbative $g^2 = 0$ fixed point staggered lattice fermions are equivalent to continuum fermions.  At a non-trivial IRFP this is not necessarily the case; instead, the different chiral symmetry properties of different lattice fermion formulations could correspond to different fixed points.  Such behavior has been studied in three-dimensional spin systems~\cite{Calabrese:2002bm, Hasenfratz:2015ssa}.}
The different approaches considered have complementary strengths, and the most reliable information about the IR dynamics of the system is obtained by attempting to integrate the available results.

For example, step-scaling studies of the discrete \be function directly search for an IRFP within a particular range of renormalized couplings.
The exactly massless fermions typically employed by such studies make it more difficult for them to explore spontaneous chiral symmetry breaking, which finite-temperature or spectral techniques \mbox{are better} suited to investigate.
If no IRFP is observed by step-scaling studies (as in recent work on \mbox{$N_f = 8$~\cite{Hasenfratz:2014rna, Fodor:2015baa}}), then additional computations with $am > 0$ are needed to investigate chiral symmetry breaking in the considered range of couplings.
Without \mbox{identifying spontaneous} chiral symmetry breaking in the $am = 0$ limit it remains possible for there to be an IRFP at some stronger coupling beyond the range in which the discrete \be function was explored.
As spontaneous chiral symmetry breaking is an inherently non-perturbative phenomenon we wish to probe it using lattice calculations rather than relying on imprecise estimates of the critical coupling strength $g_{\MSbar}^2 \sim 10$~\cite{Appelquist:1996dq, Appelquist:1998rb}.

In the case of $N_f = 12$, the pioneering step-scaling study of refs.~\cite{Appelquist:2007hu, Appelquist:2009ty} identified an IRFP at $g_{SF}^2 \approx 5$ in the Schr\"odinger functional scheme (with purely statistical uncertainties $\gsim 10\%$).
Subsequent investigations~\cite{Bilgici:2009nm, Itou:2010we, Ogawa:2011ki, Lin:2012iw, Itou:2012qn, Itou:2013faa, Cheng:2014jba, Hasenfratz:2015xpa, Lin:2015zpa, Fodor:2016zil} have attempted to improve upon this result by considering larger lattice volumes, different schemes for the running coupling, and improved lattice actions with smaller discretization artifacts.
Two recent large-scale projects are of particular note.
\Refcite{Lin:2015zpa} explores the discrete \be function up to $\gc \lsim 6$ in the ${c = \sqrt{8t} / L = 0.45}$ and 0.5 gradient flow schemes with color-twisted boundary conditions (BCs) and an unimproved lattice action.
Although the resulting step-scaling function approaches zero it does not vanish in the accessible range of couplings, and a bulk transition into a lattice phase obstructs progress to larger $\gc$.
\Refcite{Fodor:2016zil} employs very large lattice volumes and an improved action to explore the very narrow region $6 \lsim \gc \lsim 6.4$ in the $c = 0.2$ gradient flow scheme, also obtaining a non-zero discrete \be function.\footnote{This particular range of \gc was chosen based on some results in our earlier publication~\cite{Cheng:2014jba}, which identified a 12-flavor IRFP at $\gstar = 6.2(2)$.  In \appref{app:c02} we compare that previous work with the full results presented here.}
As we show in \fig{fig:compare}, both of these investigations are consistent with our full $c = 0.25$ and 0.3 results that predict an IRFP at $\gstar = 7.3\left(_{-3}^{+8}\right)$ (despite the slightly different renormalization schemes considered).

In addition to the step-scaling studies summarized above, most other $N_f = 12$ investigations offer further evidence supporting the existence of a conformal, chirally symmetric IR fixed point.
Investigations of the phase diagram both at zero and finite temperature have observed a first-order bulk phase transition that extends from the $am = 0$ chiral limit to non-zero mass~\cite{Deuzeman:2009mh, Cheng:2011ic, Fodor:2012uu, Schaich:2012fr, Deuzeman:2012ee, daSilva:2012wg, Hasenfratz:2013uha, Ishikawa:2013tua}.
At finite temperature $T = 1 / (aN_t)$, where $a$ is the lattice spacing and $N_t$ is the temporal extent of the lattice, the chiral transition lines run into the bulk phase at non-zero mass~\cite{Schaich:2012fr, Hasenfratz:2013uha}.
This is a necessary condition for IR-conformality, where the finite-temperature transitions in the chiral limit must accumulate at a finite coupling as $N_t \to \infty$, and remain separated from the weak-coupling conformal phase.
No lattice investigations of the 12-flavor phase diagram have been able to identify spontaneous chiral symmetry breaking in the form of chiral transitions that remain in the weakly coupled phase upon extrapolation to the chiral limit.

Spectral studies offer another means to explore the IR dynamics, by confronting nonzero-mass lattice data with expectations based on either chiral perturbation theory or conformal finite-size scaling.
While refs.~\cite{Aoki:2012eq, Cheng:2013xha, Lombardo:2014pda} observe consistency with conformal hyperscaling for $N_f = 12$, \refcite{Fodor:2011tu} reported a very low level of confidence in conformality.
However, subsequent re-analyses of the data published by \refcite{Fodor:2011tu} suggest that this conclusion is sensitive to the details of the analyses~\cite{Appelquist:2011dp, DeGrand:2011cu, Cheng:2013xha}.
In particular, by taking into account corrections to scaling arising from the nearly marginal (i.e., slowly running) nature of the gauge coupling, in \refcite{Cheng:2013xha} we were able to carry out consistent finite-size scaling analyses that included both our own spectrum data as well as those published by refs.~\cite{Fodor:2011tu, Aoki:2012eq}.

Our finite-size scaling study predicted the scheme-independent mass anomalous dimension $\ga_m^{\star} = 0.235(15)$ at the 12-flavor IR fixed point.
A similar result $\ga_m^{\star} = 0.235(46)$ was reported by \refcite{Lombardo:2014pda}.\footnote{Finite-size scaling analyses without corrections to scaling typically obtained larger values that often varied non-universally depending on the observables analyzed: $0.2 \lsim \ga_m \lsim 0.4$~\cite{Fodor:2012et}, $\ga_m^{\star} = 0.403(13)$~\cite{Appelquist:2011dp}, $\ga_m^{\star} \simeq 0.35$~\cite{DeGrand:2011cu} and $\ga_m^{\star} = 0.4$--0.5~\cite{Aoki:2012eq}.  A recent study of the mass dependence of the topological susceptibility obtained a similar $\ga_m^{\star} = 0.3$--0.5 by fitting $\chi_t \propto (am)^{4 / (1 + \ga_m^{\star})}$~\cite{Aoki:2016yrm}.  These results are all consistent with an upper bound $\ga_m^{\star} \leq 1.29$ from the conformal bootstrap program~\cite{Iha:2016ppj}, though not with the perturbative $\ga_m^{\star} \approx 1.3$--1.5 reported by \refcite{Doff:2016jzk}.}
In addition, our studies of the massless Dirac operator eigenmodes independently predict $\ga_m^{\star} \approx 0.25$~\cite{Cheng:2013eu, Cheng:2013bca}.
These results are quite close to the four-loop perturbative prediction $\ga_m^{\star} = 0.253$ in the $\overline{\textrm{MS}}$ scheme~\cite{Ryttov:2010iz}, and the new five-loop result $\ga_m^{\star} = 0.255$~\cite{Ryttov:2016ner}, though a recent scheme-independent series expansion~\cite{Ryttov:2016hdp} obtains a larger $\ga_m^{\star} = 0.400(5)$~\cite{Ryttov:2016asb, Ryttov:2016hal}.
This small, potentially perturbative mass anomalous dimension, in combination with the assumption that $\ga_m^{\star} \simeq 1$ around the lower edge of the conformal window, may suggest that $N_f = 12$ is quite deep within the conformal regime.

Despite the many high-quality, large-scale investigations of the 12-flavor system summarized above, there is still progress to be made in resolving its IR properties.
In this work we report our final results on the step-scaling calculation of the discrete \be function for $N_f = 12$.
These results supersede the partial analysis included in \refcite{Cheng:2014jba}, and predict a conformal IR fixed point at $\gstar = 7.3\left(_{-2}^{+8}\right)$ in the gradient flow scheme with $c = 0.25$.
We also investigate the slope of the step-scaling function at the IRFP, both directly and via finite-size scaling as in refs.~\cite{Appelquist:2009ty, Lin:2015zpa}.
This slope is related to the leading irrelevant critical exponent $\ga_g^{\star}$, for which we find $\ga_g^{\star} = 0.26(2)$, consistent with the four-loop perturbative prediction $\ga_g^{\star} = 0.282$.

Compared to \refcite{Cheng:2014jba} we have accumulated significantly more data, in particular generating several new lattice ensembles at relatively strong couplings $\be_F \lsim 4$ on each lattice volume up to $36^4$.
This allows us to explore the discrete \be function up to $\gc \lsim 8.2$, extending past the IRFP that we observe (though it would be nice to push further into the regime of backward flow in future work).
We now compare multiple discretizations of the energy density $E(t)$ in the gradient flow renormalized coupling, obtaining consistent results.
Finally, we add two new lattice volumes, $20^4$ and $30^4$, that allow us to omit the $12^4$ volume used in \refcite{Cheng:2014jba}.
As we show in \appref{app:c02}, analyses that include $12^4$ volumes in the $c = 0.2$ gradient flow scheme suffer from particularly large systematic uncertainties that were not comprehensively considered in \refcite{Cheng:2014jba}.

Although our 12-flavor results are qualitatively different than those we previously obtained for the 8-flavor discrete \be function~\cite{Hasenfratz:2014rna}, much of our analysis follows the same procedure as that work, and the next three sections are organized in the same way.
We begin by reviewing gradient flow step scaling in the next section, including the improvement of the gradient flow running coupling.
In \secref{sec:setup} we describe our numerical setup and lattice ensembles.
We use an nHYP-smeared staggered fermion lattice action~\cite{Hasenfratz:2001hp, Hasenfratz:2007rf}, with both fundamental and adjoint plaquette terms in the gauge action~\cite{Hasenfratz:2011xn, Cheng:2011ic, Schaich:2012fr, Hasenfratz:2013uha}.
We employ this same action in our 12-flavor finite-temperature~\cite{Cheng:2011ic, Schaich:2012fr, Hasenfratz:2013uha}, spectral~\cite{Cheng:2013xha} and eigenmode~\cite{Cheng:2013eu, Cheng:2013bca} studies summarized above, which can therefore be consistently compared.
On each of eight $L^4$ volumes with $12 \leq L \leq 36$ we generate between 14--35 ensembles at different bare couplings in the range $3 \leq \be_F \leq 9$.

Our step-scaling analyses and results are presented in \secref{sec:results}, including discussion of systematic uncertainties from the stability of the $(a / L) \to 0$ extrapolations, the interpolation of \gtc as a function of the bare coupling, and the improvement of the gradient flow running coupling.
We compare the clover and plaquette discretizations of $E(t)$ as another consistency check, obtaining agreement in all cases we consider.
Finally, we also confirm the consistency of our results with those recently reported by refs.~\cite{Lin:2015zpa, Fodor:2016zil}.
In \secref{sec:slope} we investigate the leading irrelevant critical exponent from the slope of the step-scaling function at the IRFP, observing $\ga_g^{\star} = 0.26(2)$, comparable to perturbative estimates.
We check this result by carrying out a finite-size scaling analysis.
We conclude in \secref{sec:conclusion} with some brief discussion of how our new results affect the broader context of 12-flavor lattice investigations summarized above, and highlight a few directions that merit further study in the future.

We include three appendices collecting some supplemental checks of our results.
In \appref{app:s2} we briefly consider the discrete \be functions resulting from two scale changes $s = 2$ and $s = 4 / 3$ different from the $s = 3 / 2$ considered in the body of the paper.
In contrast to $s = 3 / 2$, for both of $s = 2$ and $4 / 3$ we are forced to include small-volume $12^4$ lattice ensembles in our analyses.
We obtain consistent results from all three scale changes, as summarized in \tab{tab:gstar}.
However, as we show in \appref{app:c02}, systematic uncertainties increase dramatically when combining smaller $c \lsim 0.2$ with smaller $L \leq 12$.
These systematic uncertainties were not comprehensively considered in the partial analysis we included in \refcite{Cheng:2014jba}, which reported $\gstar = 6.2(2)$ with $c = 0.2$ and $L \geq 12$, compared to the $\gstar = 5.9(1.9)$ we now obtain with this choice of $c$ and $L_{\text{min}}$ (\tab{tab:gstar}).
Finally, \appref{app:data} provides a subset of our data.

\section{\label{sec:gradflow}Gradient flow step scaling and its improvement} 
We investigate a renormalized coupling defined through the gradient flow, which is a continuous transformation that smooths lattice gauge fields to systematically remove short-distance lattice cutoff effects~\cite{Narayanan:2006rf}.
The demonstration that the gradient flow is mathematically well defined and invertible~\cite{Luscher:2009eq} inspired its use in a wide variety of applications (recently reviewed by \refcite{Luscher:2013vga}).
Here we consider the coupling~\cite{Luscher:2010iy}
\begin{equation}
  \gGF(\mu) = \frac{1}{\cN} \vev{t^2 E(t)} = \frac{128\pi^2}{3(N^2 - 1)} \vev{t^2 E(t)},
\end{equation}
where the energy density $E(t)$ is evaluated after `flow time' $t$, corresponding to the energy scale $\mu = 1 / \sqrt{8t}$.
We will compare two lattice operators that can be used to define the energy density, first $E(t) = -\frac{1}{2}\mbox{ReTr}\left[G_{\mu\nu}(t) G^{\mu\nu}(t)\right]$ with the symmetric clover-leaf definition of $G_{\mu\nu}$, and second $E(t) = 12 (3 - \Box(t))$ where $\Box$ is the plaquette normalized to 3.
The overall normalization \cN is set by matching $\gGF(\mu)$ with the continuum \MSbar coupling at tree level.
To carry out step-scaling analyses we tie the energy scale to the lattice volume $L^4$ by fixing the ratio $c = \sqrt{8t} / L$, as proposed by refs.~\cite{Fodor:2012td, Fodor:2012qh, Fritzsch:2013je}.
Each choice of $c$ defines a different renormalization scheme, producing different results for the renormalized coupling $\gc(L)$ and for the discrete \be function in the continuum limit.
When periodic BCs are used for the gauge fields, these \be functions are only one-loop (and not two-loop) universal~\cite{Fodor:2012td}.

Extrapolating $(a / L) \to 0$ is required to remove cutoff effects in the gradient flow renormalized couplings $\gc$.\footnote{We refer to these as `continuum extrapolations' in some places, but this is strictly true only for couplings weaker than the \gstar of the IR fixed point.}
These cutoff effects depend on the lattice action used to generate the configurations, on the gauge action used in the gradient flow transformation, and on the lattice operator used to define the energy density $E(t)$.
It is possible to systematically remove lattice artifacts by improving all three quantities simultaneously~\cite{Ramos:2014kka, Ramos:2015baa}.
Here we take a simpler approach, using the Wilson plaquette action in the gradient flow transformation (i.e., the ``Wilson flow'') and combining two improvements that suffice to greatly reduce---and often essentially remove---cutoff effects.
First, following \refcite{Fodor:2014cpa}, we modify the definition of the renormalized coupling to perturbatively correct for cutoff effects,
\begin{equation}
  \label{eq:pert_g2}
  \gc(L) = \frac{128\pi^2}{3(N^2 - 1)} \frac{1}{C(L, c)} \vev{t^2 E(t)}.
\end{equation}
In this expression $C(L, c)$ is a four-dimensional finite-volume sum in lattice perturbation theory, which depends on the action, flow and operator.
We use the tree-level computation of $C(L, c)$ from \refcite{Fodor:2014cpa}, including a term that accounts for the zero-mode contributions allowed by the periodic BCs for the gauge fields.

As we will see in \fig{fig:extrap}, even this perturbatively improved gradient flow coupling can exhibit significant cutoff effects.
While larger values of $c \gsim 0.3$ reduce these artifacts to some extent, this is accomplished only at the price of increased statistical uncertainties~\cite{Fritzsch:2013je}.
A better option, introduced in \refcite{Cheng:2014jba}, is to slightly shift the flow time at which the energy density is computed:
\begin{equation}
  \label{eq:t-shift}
  \gtGF(\mu; a) = \gGF(\mu; a) \frac{\vev{E(t + \tau_0 a^2)}}{\vev{E(t)}}
\end{equation}
with $|\tau_0| \ll t / a^2$.
This $t$-shift $\tau_0$ can be either positive or negative.
Its effects vanish in the continuum limit where $\tau_0 a^2 \to 0$ so that $\gtGF(\mu) = \gGF(\mu)$.
For $\cO(a)$-improved actions like those we use, choosing an optimal $\tau_0$ value \topt allows the removal of all $\cO(a^2)$ corrections of the coupling $\gtGF(\mu; a)$ defined in \eq{eq:t-shift}.
Although this optimal \topt changes as a function of $\gtGF(\mu)$, in this work we observe that \topt depends only weakly on $\gtGF(\mu)$, as in our previous studies of 4-, 8- and 12-flavor SU(3) systems~\cite{Cheng:2014jba, Hasenfratz:2014rna}.
Therefore we simply use a constant value of \topt for all $\gtGF(\mu)$, which suffices to remove most observable lattice artifacts throughout the ranges of couplings we explore.

Since we optimize $\tau_0$ after applying the tree-level perturbative corrections discussed above, these two improvements do not interfere with each other.
Nor do either of them require any additional computation, since the numerical integration through which we evaluate the gradient flow already provides all the data needed to shift $t \to t + \tau_0 a^2$.
Using the resulting \gtc gradient flow running coupling, we will investigate the 12-flavor discrete \be function corresponding to scale change $s$,
\begin{equation}
  \label{eq:beta}
  \be_s(\gtc; L) = \frac{\gtc(sL; a) - \gtc(L; a)}{\log(s^2)}.
\end{equation}
We will also refer to this quantity as the step-scaling function $\si_s(u, L)$ with $u \equiv \gtc(L; a)$.
To obtain our final results for the continuum discrete \be function $\be_s(\gc) = \lim_{(a / L) \to 0} \be_s(\gtc, L)$ we extrapolate $(a / L) \to 0$.

We emphasize that different values of $\tau_0$ should all produce the same $\be_s(\gc)$ in the continuum limit~\cite{Cheng:2014jba}.
In \appref{app:c02} we will show that this requirement is not satisfied for the lattice volumes we can access when $c \lsim 0.2$. 
In this case continuum extrapolations with different $t$-shifts disagree by statistically significant amounts, which likely contributes to the discrepancy between refs.~\cite{Cheng:2014jba} and \cite{Fodor:2016zil}.
In this work, when such sensitivity to the $t$-shift is present we will account for it as a source of systematic uncertainty, which was not done in \refcite{Cheng:2014jba}.

The different discretizations of $E(t)$ should also produce the same $\be_s(\gc)$ in the continuum limit.
We will separately analyze the plaquette and clover definitions of $E(t)$, and find that they produce consistent results within uncertainties when $c \geq 0.25$ and $L \geq 16$.
In \appref{app:s2} we note that reducing $L \geq 12$ requires increasing $c \geq 0.3$ in order to maintain the good agreement between these two sets of results.
When identifying the location of the IR fixed point, we will include the predictions of both discretizations in our determination of the total uncertainties on $\gstar$.

\section{\label{sec:setup}Numerical setup and lattice ensembles} 
Our numerical calculations use nHYP-smeared staggered fermions~\cite{Hasenfratz:2001hp, Hasenfratz:2007rf} with smearing parameters $\al = (0.5, 0.5, 0.4)$, and a gauge action including fundamental and adjoint plaquette terms with couplings related by $\be_A / \be_F = -0.25$~\cite{Hasenfratz:2011xn, Cheng:2011ic, Schaich:2012fr, Hasenfratz:2013uha}.
The fermions are exactly massless ($am = 0$), which freezes the topological charge at $Q = 0$.
We impose anti-periodic BCs for the fermions in all four directions, while the gauge fields are periodic.
Previous studies of this lattice action observed an ``$\Sb$'' lattice phase in which the single-site shift symmetry ($S^4$) of the staggered action is spontaneously broken~\cite{Cheng:2011ic, Schaich:2012fr, Hasenfratz:2013uha}.
At $am = 0$ a first-order transition into the \Sb phase occurs at $\be_F^{(c)} \approx 2.75$. 
In this work we only consider weaker couplings safely distant from the \Sb lattice phase.

\begin{table}[tbp]
  \centering
  \renewcommand\arraystretch{1.2}  
  \addtolength{\tabcolsep}{3 pt}   
  \begin{tabular}{ccc}
    \hline
     $s = 2$          &  $s = 3 / 2$  &  $s = 4 / 3$  \\
    \hline
    ~$L = 12 \to 24$~ & ~$12 \to 18$~ & ~$12 \to 16$~ \\
     $L = 16 \to 32$  &  $16 \to 24$  &  $18 \to 24$  \\
     $L = 18 \to 36$  &  $20 \to 30$  &  $24 \to 32$  \\
                      &  $24 \to 36$  &               \\
    \hline
  \end{tabular}
  \caption{\label{tab:pairs}Pairs of lattice volumes available for the three scale changes $s = 2$, $3 / 2$ and $4 / 3$.}
\end{table}

We generate ensembles of gauge configurations with eight different $L^4$ lattice volumes with $L = 12$, 16, 18, 20, 24, 30, 32 and 36.
Depending on $L$ we study 14--35 values of the bare coupling in the range $3 \leq \be_F \leq 9$.
The 158 resulting ensembles are summarized in tables~\ref{tab:ensembles12}--\ref{tab:ensembles36} in \appref{app:data}.
These volumes allow us to consider three scale changes $s = 2$, $3 / 2$ and $4 / 3$, each with at least three pairs of volumes for continuum extrapolations as listed in \tab{tab:pairs}.
In the body of the paper we focus on $s = 3 / 2$ where we can retain three points with $L \geq 16$; we will see in the next section that the $L = 12$ ensembles exhibit potentially significant cutoff effects.
Even so, we obtain comparable results for $s = 2$ and $4 / 3$ analyses including $L = 12$ data, which are collected in \appref{app:s2}.

We use the hybrid Monte Carlo (HMC) algorithm to generate configurations.
Even at the strongest bare couplings we investigate we retain good HMC acceptance and reversibility in the $am = 0$ chiral limit with unit-length molecular dynamics trajectories and step sizes $\de\tau \approx 0.1$ at the outer level of our standard multi-timescale Omelyan integrator.
While the performance of the HMC algorithm is not a robust means to monitor the phase structure of the system, this behavior indicates that none of our ensembles exhibit chiral symmetry breaking.
This conclusion is supported by our observation of a gap in the Dirac operator eigenvalue spectrum on many of these ensembles, including the strongest couplings $\be_F \geq 3$ that we consider~\cite{Cheng:2013eu, Cheng:2013bca}.

\begin{figure}[btp]
  \includegraphics[width=0.45\textwidth]{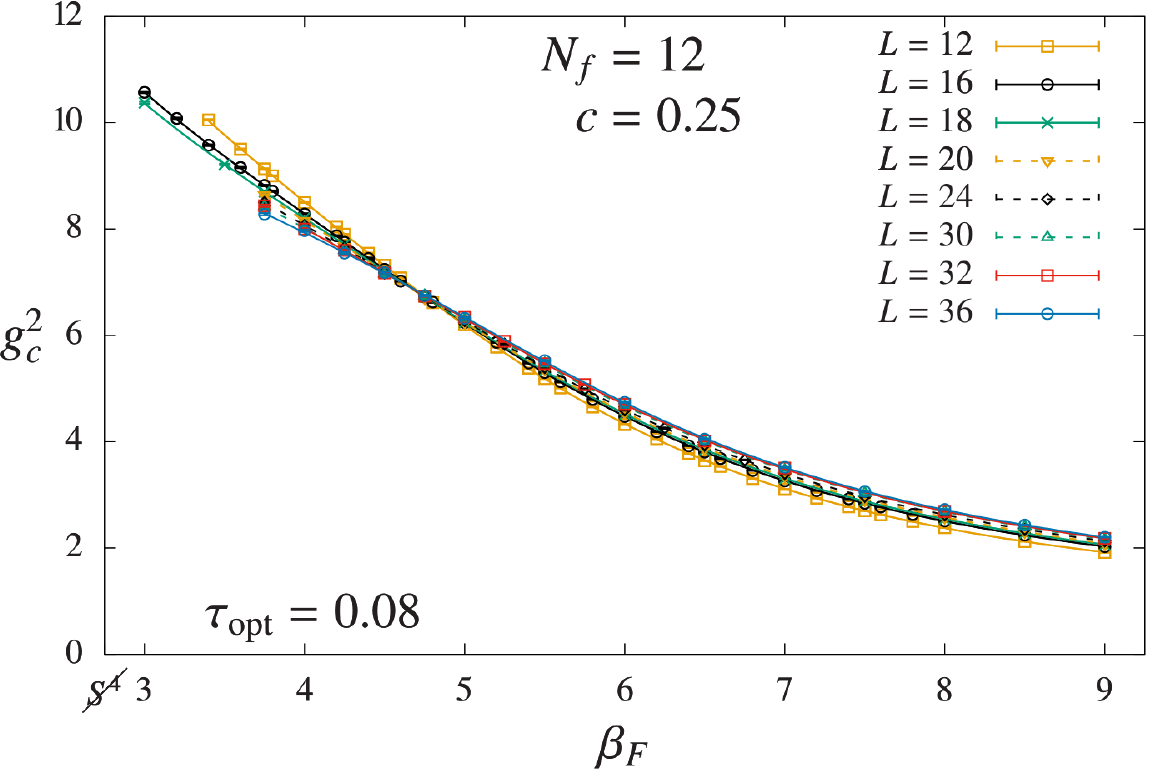}\hfill \includegraphics[width=0.45\textwidth]{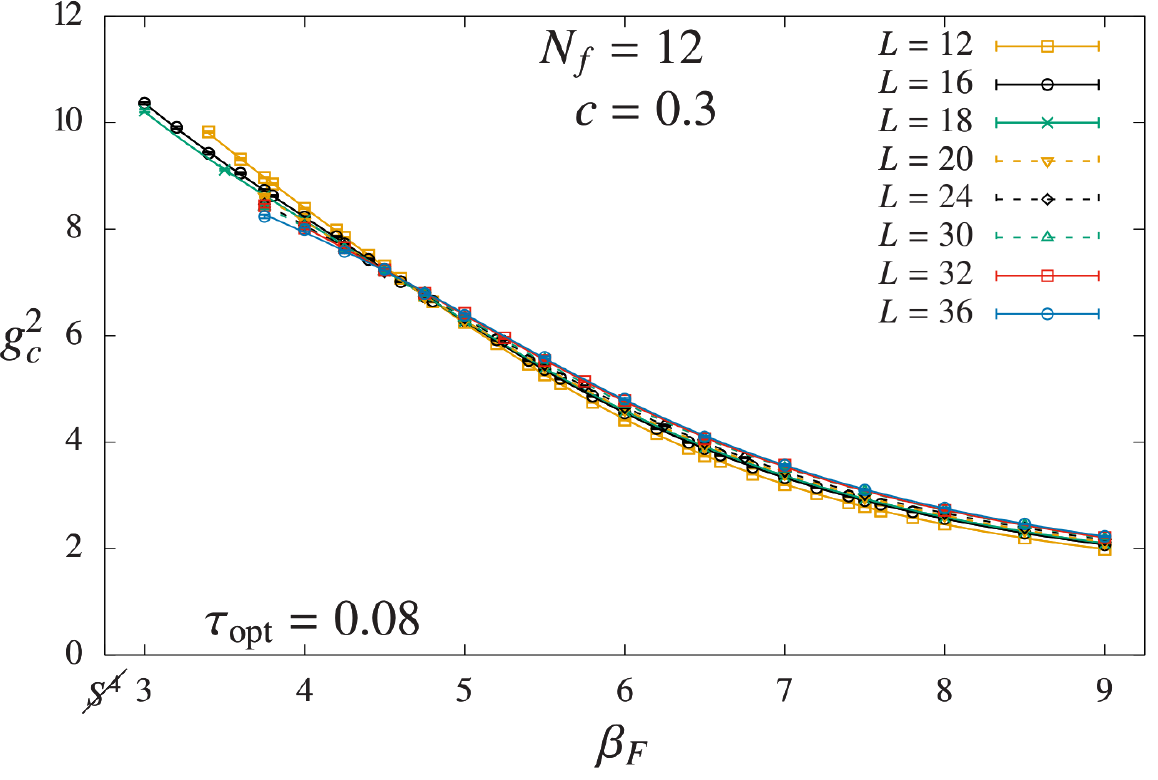} \\[12 pt]
  \includegraphics[width=0.45\textwidth]{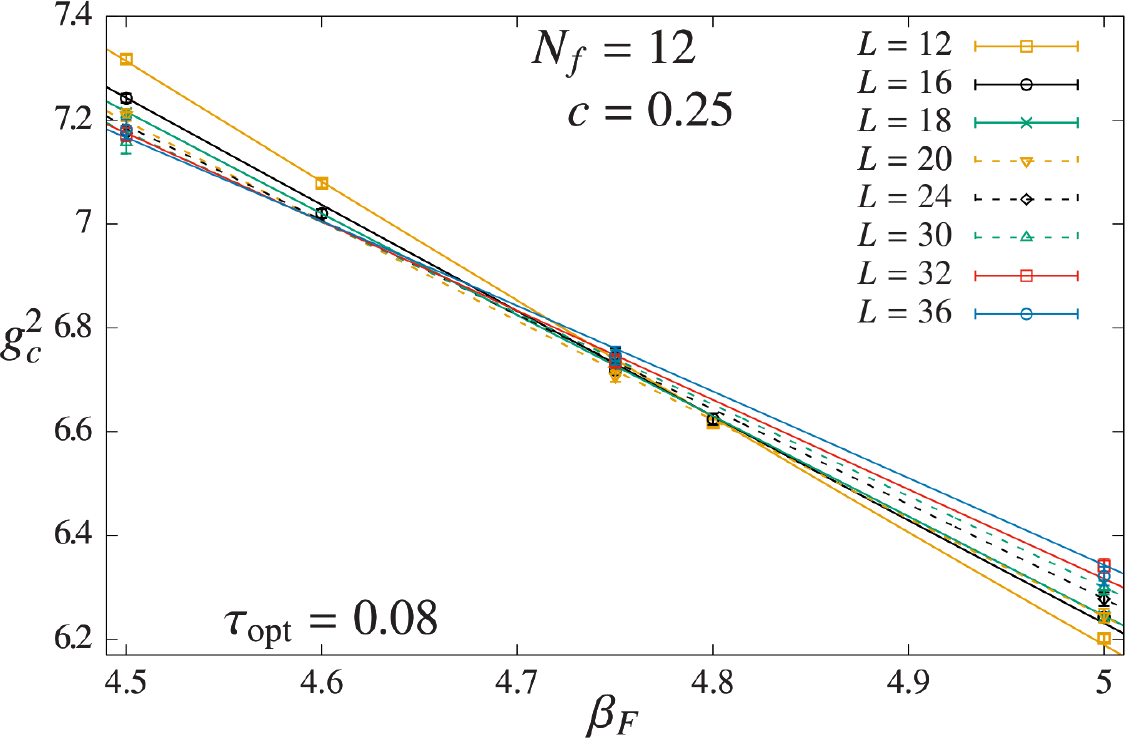}\hfill \includegraphics[width=0.45\textwidth]{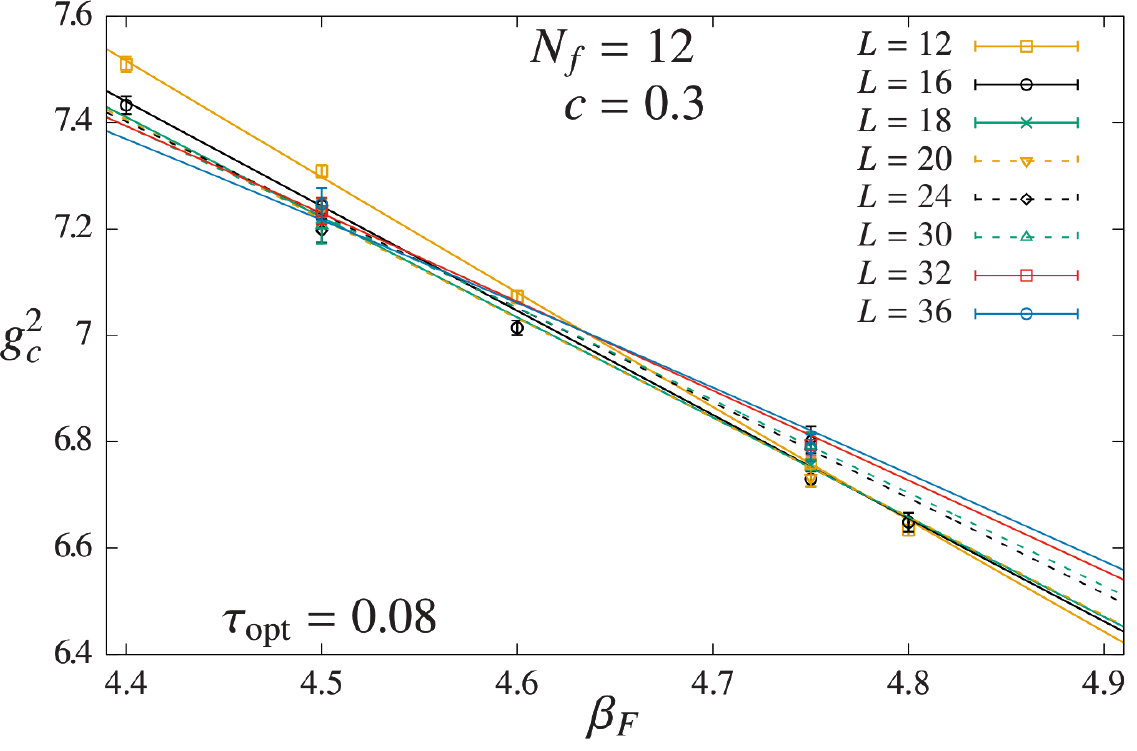}
  \caption{\label{fig:gcSq}Gradient flow renormalized coupling $\gtc(L)$ vs.\ $\be_F$ for $c = 0.25$ (left) and $c = 0.3$ (right), both with optimal $\topt = 0.08$.  The lines are interpolations using the rational function form in \protect\eq{eq:pade}.  The left edge of the plots in the top row indicates the boundary of the \Sb phase, $\be_F^{(c)} \approx 2.75$.  The plots in the bottom row zoom in on the narrow regions around $\be_F \lsim 5$ where the results from different lattice volumes all cross each other.  For clarity we omit the uncertainties on the interpolations.  Within uncertainties the crossings for $L \geq 16$ and $c = 0.3$ are all consistent.}
\end{figure}

In \fig{fig:gcSq} we show the gradient flow renormalized coupling $\gtc(L)$ measured on each ensemble for $c = 0.25$ and 0.3 (using the clover discretization of the energy density).
These data use the optimal $t$-shift value $\topt = 0.08$ that we discuss in the next section, and also include the tree-level perturbative correction factor $C(L, c)$ in \eq{eq:pert_g2}.
The perturbative corrections are fairly mild for our lattice action, Wilson flow, and clover or plaquette discretization of the energy density.
The largest is $C(12, 0.25) \approx 1.12$ for the plaquette discretization, with all others smaller than 6.2\% effects.
From these plots we can already see that the 12-flavor coupling runs very slowly, with little change in $\gtc(L)$ as the volume increases by a factor of three, especially for $\be_F > 4.0$.
This feature of the system was mentioned in \secref{sec:intro}, as the reason that finite-size scaling analyses need to account for the corresponding corrections to scaling.

\clearpage
The lines in \fig{fig:gcSq} are interpolations using the rational function form in \eq{eq:pade}.
The plots in the bottom row zoom in on narrow regions of width $\De\be_F = 0.5$ where the interpolations from different lattice volumes all cross each other.
At the weak-coupling edge of these plots, $\be_F = 5$ (4.9) for $c = 0.25$ (0.3), the interpolated \gtc monotonically increase with $L$ from 12 to 36.
At the strong-coupling edge, $\be_F = 4.5$ (4.4), the order has completely reversed and the interpolated \gtc monotonically decrease as the lattice volume increases.
Of course there are statistical uncertainties in the data that make the full analysis more complicated: To reduce clutter in these figures we don't display the uncertainties on the interpolations, within which most of the interpolations remain consistent with each other throughout much or all of this range.

The finite-volume crossings visible in these plots could be extrapolated to the infinite-volume limit to predict a 12-flavor IRFP, as in the $c = 0.2$ analysis of \refcite{Cheng:2014jba}.\footnote{The finite-volume crossings in figure~3 of \refcite{Cheng:2014jba} are at weaker couplings $\be_F \approx 6$ due to the absence of $t$-shift improvement as well as the smaller value of $c = 0.2$.  Crossing analyses for $c = 0.25$ and 0.3, using the same data sets and procedures as \refcite{Cheng:2014jba}, previously predicted $\gstar = 6.8(3)$ and 7.1(5), respectively~\cite{Hasenfratz:2014SCGT}.}
With $c = 0.25$ the crossings occur at  $\gstar(L) \lsim 7$ but extrapolate to a slightly larger value $\gstar \approx 7.3$ in the continuum limit.
With $c = 0.3$ the crossings all cluster around $\gstar(L) \lsim 7.3$, with a nearly constant continuum extrapolation.
Instead of taking this approach, however, in this work we construct the full continuum-extrapolated discrete \be function across a broad range of couplings, the topic to which we now turn.

\section{\label{sec:results}Step-scaling analyses and results} 
Following the standard procedure for lattice step-scaling analyses, for each $L$ we first fit the renormalized couplings $\gtc(L)$ to some interpolating function in the bare coupling ${\be_F \equiv 12 / g_0^2}$, then use those interpolations to determine the finite-volume discrete \be functions $\be_s(\gtc, L)$ from \eq{eq:beta}, which we extrapolate to the $(a / L) \to 0$ limit.
We will refer to the last step as the `continuum extrapolation', although this is strictly true only for couplings weaker than the \gstar of the IR fixed point.
While the choice of interpolating function is essentially arbitrary, typically some functional form motivated by lattice perturbation theory is used.
For example, refs.~\cite{Appelquist:2009ty, Fodor:2012td, Fodor:2015baa} fit $\frac{1}{g^2} - \frac{1}{g_0^2}$ to polynomials in $g_0^2$.
Following refs.~\cite{Karavirta:2011zg, Hasenfratz:2014rna} we instead use the rational function
\begin{equation}
  \label{eq:pade}
  \gtc(L) = \left(\frac{12}{\be_F}\right) \frac{1 + a_1 \be_F + a_2 \be_F^2}{b_0 + b_1 \be_F + b_2 \be_F^2},
\end{equation}
which also produces the expected $\gtc \propto g_0^2$ at weak coupling.
These interpolations are shown in \fig{fig:gcSq}.
Most of the fits shown are of good quality, although there are some outliers with $\chidof \gsim 1$ corresponding to confidence levels $\mbox{CL} \lsim 0.1$.
For reference we collect all this information in tables~\ref{tab:interp02}--\ref{tab:interp03} in \appref{app:data}.
Notably, the worst-quality interpolations are for the $L = 12$ data that we omit from our $s = 3 / 2$ step-scaling analyses.

To investigate potential systematic effects from our choice of interpolating function we also carry out analyses using~\cite{Fodor:2015baa}
\begin{equation}
  \label{eq:poly}
  \frac{1}{\gtc(L)} = \frac{\be_F}{12}\sum_{i = 0}^4 c_i \left(\frac{12}{\be_F}\right)^i,
\end{equation}
where we include five terms to produce the same number of fit parameters as \eq{eq:pade}.
Although these interpolations appear satisfactory upon visual inspection, they generally produce much larger \chidof\ than the rational function in \eq{eq:pade} (tables~\ref{tab:interp02}--\ref{tab:interp03}).
Therefore we will use the rational function for our final results, and treat any statistically significant differences between these two analyses as another source of systematic uncertainty.

\begin{figure}[btp]
  \includegraphics[width=0.45\textwidth]{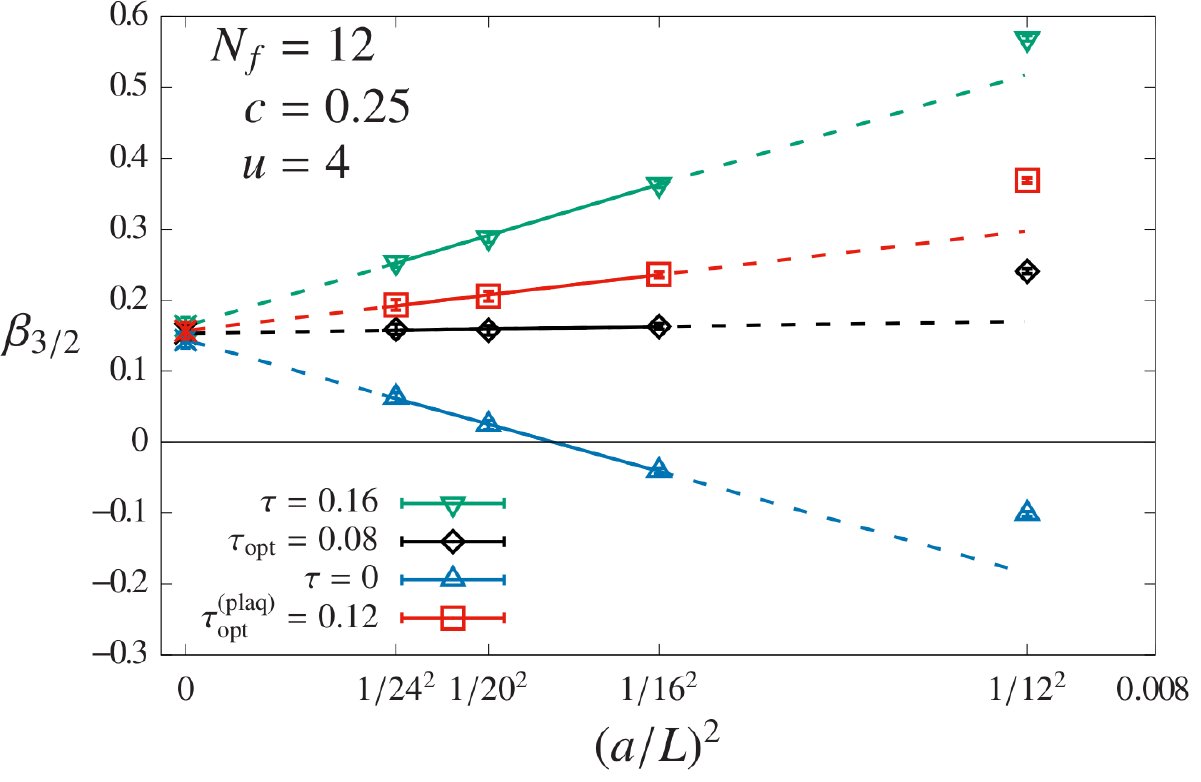}\hfill \includegraphics[width=0.45\textwidth]{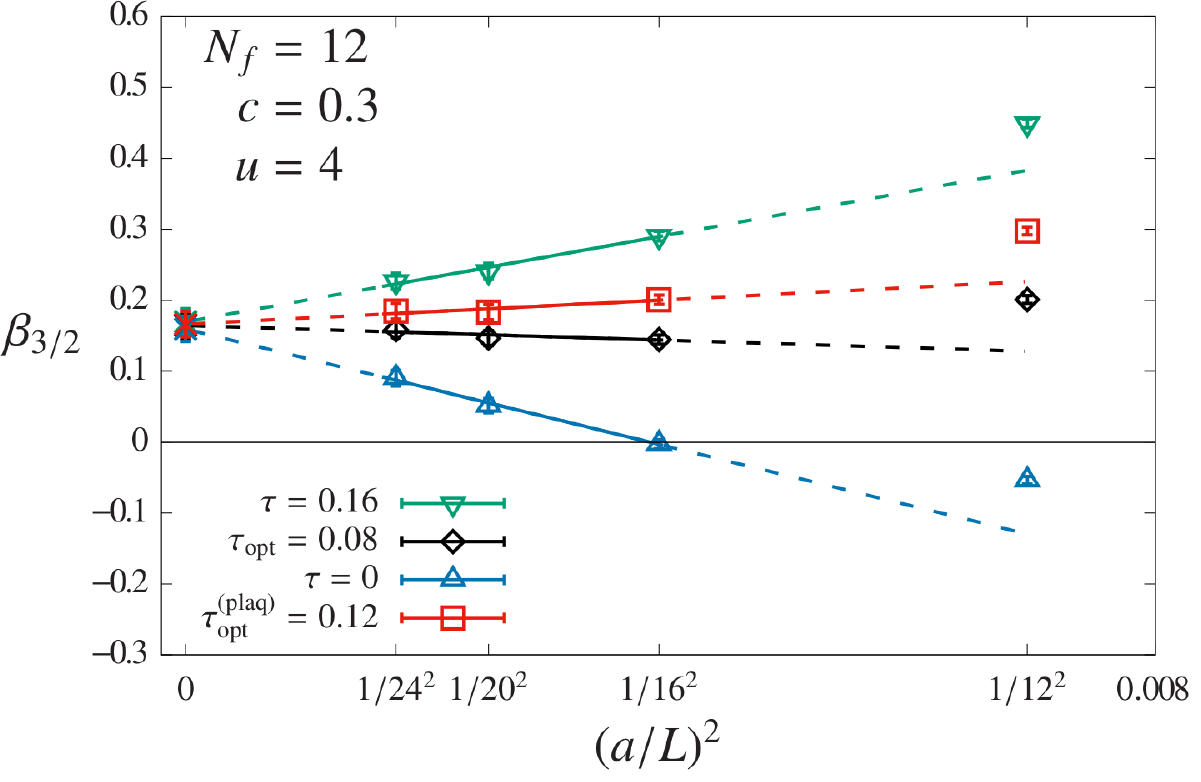} \\[12 pt]
  \includegraphics[width=0.45\textwidth]{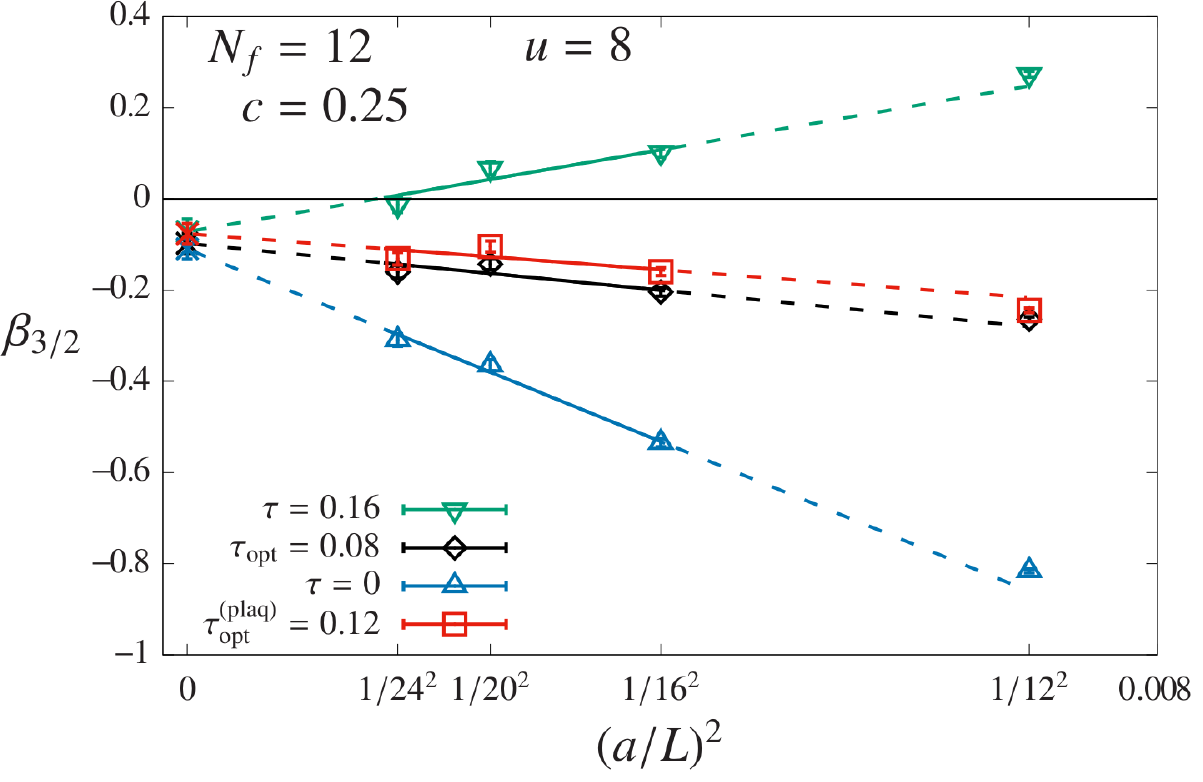}\hfill \includegraphics[width=0.45\textwidth]{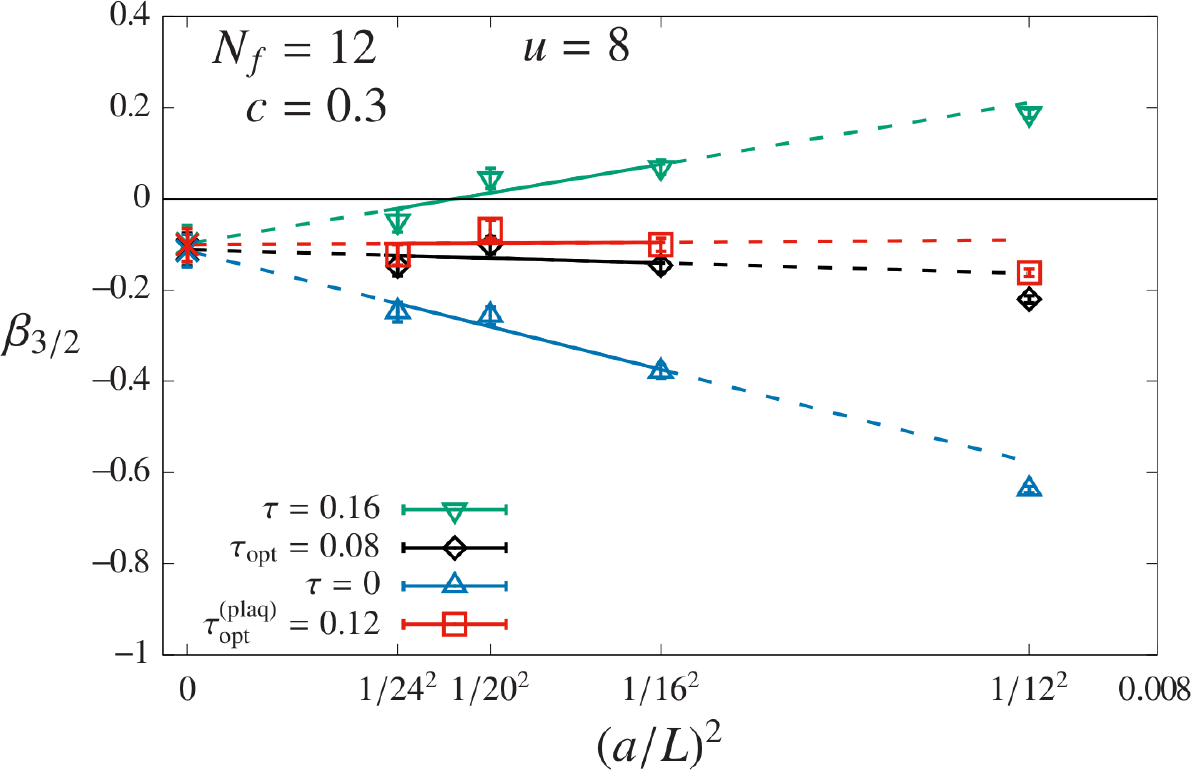}
  \caption{\label{fig:extrap}Linear $(a / L)^2 \to 0$ extrapolations of the $s = 3 / 2$ discrete \be function for $c = 0.25$ (left) and 0.3 (right), at two values of $u = 4$ (top) and 8 (bottom) on either side of the IR fixed point.  In each plot we compare $\tau_0 = 0$ and 0.16 to the optimal $\topt = 0.08$, and also include results from the plaquette discretization of the energy density $E(t)$ in \protect\eq{eq:t-shift} at the corresponding optimal $\topt^{\text{(plaq)}} = 0.12$.  As required, all different $\tau_0$ produce extrapolations to consistent values in the $(a / L)^2 \to 0$ limit.  Only $L \geq 16$ are included in the fits, though $L = 12 \to 18$ points are shown for comparison.  The bottom row of plots shows that restricting $L \geq 20$ at $u = 8$ would produce $(a / L)^2 \to 0$ extrapolations farther below zero, reinforcing the existence of the IR fixed point.}
\end{figure}

\begin{figure}[btp]
  \includegraphics[width=0.45\textwidth]{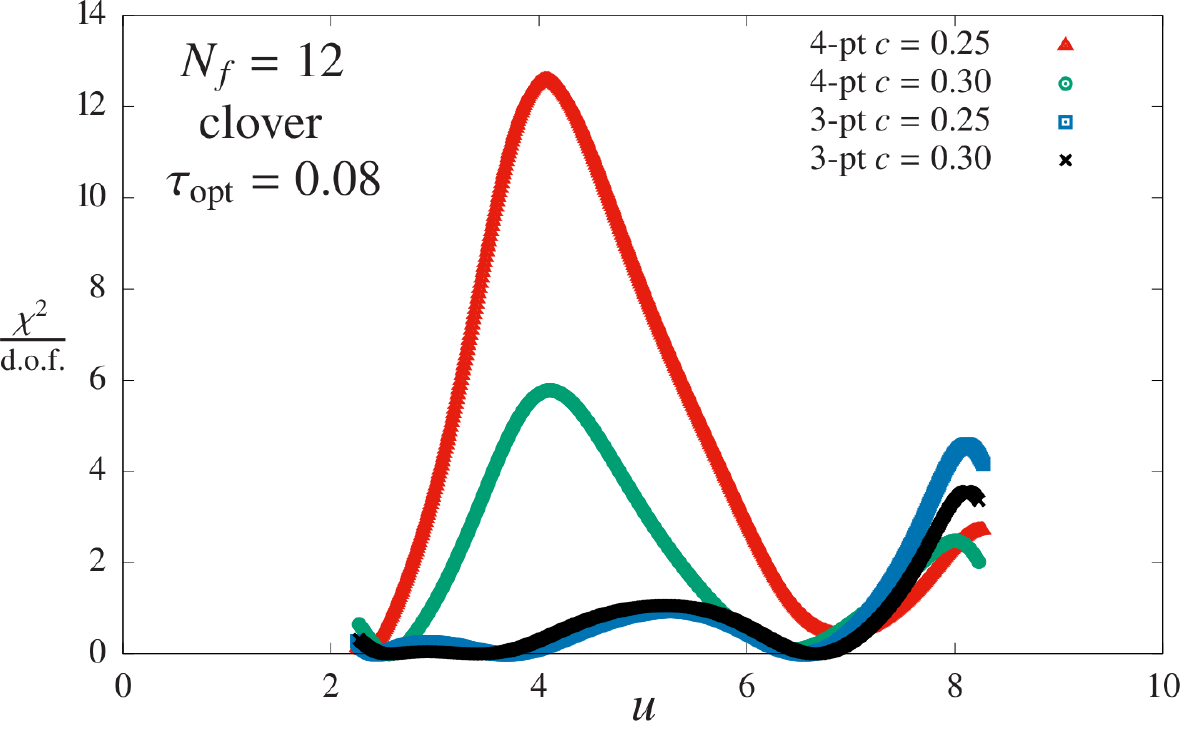}\hfill \includegraphics[width=0.45\textwidth]{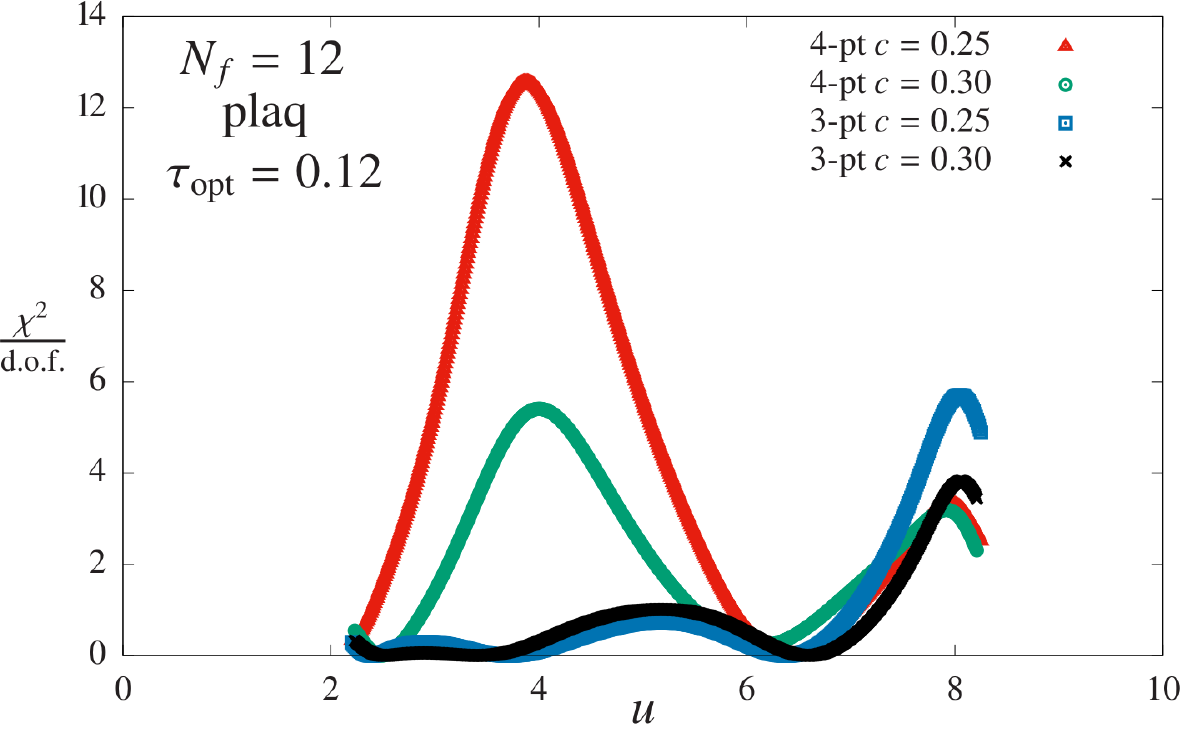}
  \caption{\label{fig:extrap_compare}\chidof\ from linear $(a / L)^2 \to 0$ extrapolations of the $s = 3 / 2$ discrete \be function vs.\ the renormalized coupling $u$.  For each of the clover (left, with $\topt = 0.08$) and plaquette (right, with $\topt^{\text{(plaq)}} = 0.12$) discretizations of the energy density $E(t)$ in \protect\eq{eq:t-shift} we compare three-point extrapolations using $L \geq 16$ against four-point extrapolations using $L \geq 12$, for both $c = 0.25$ and 0.3.  While the larger $c$ improves the quality of the extrapolations as expected~\protect\cite{Fritzsch:2013je}, for most $u$ the dominant contribution to the $\chi^2$ comes from the $L = 12 \to 18$ point.  Dropping $L = 12$ significantly improves the $(a / L)^2 \to 0$ extrapolations, except for $u \gsim 7$.  In this regime \protect\fig{fig:extrap} shows that additionally dropping $L = 16$ would produce results for $\be_s$ farther below zero, reinforcing the existence of the IR fixed point.}
\end{figure}

Turning to the $(a / L) \to 0$ extrapolations, we show several representative extrapolations in \fig{fig:extrap}, for $c = 0.25$ and 0.3 at two values of the renormalized coupling $u = 4$ and 8 on either side of the IR fixed point.
Since staggered fermions are $\cO(a)$ improved, we extrapolate linearly in $(a / L)^2$.
In each figure we compare results from the clover discretization of the energy density $E(t)$ in \protect\eq{eq:t-shift} for several values of the $t$-shift improvement parameter $\tau_0$, including $\tau_0 = 0$ and the optimal $\topt = 0.08$.
We also include one set of results from the plaquette discretization of $E(t)$, at the corresponding optimal $\topt^{\text{(plaq)}} = 0.12$.
We use the same vertical scale for both $c = 0.25$ and 0.3, to illustrate how the larger value of $c$ reduces the size of cutoff effects for fixed $\tau_0$, as expected~\cite{Fritzsch:2013je}.

The unshifted ($\tau_0 = 0$) results in \fig{fig:extrap} all show significant dependence on $(a / L)^2$, despite the tree-level perturbative correction discussed in \secref{sec:gradflow}.
We optimize $\tau_0$ by finding the value \topt for which these cutoff effects are minimized.
Since we use constant \topt for all couplings, at most values of $u$ the $\cO(a^2)$ effects are only reduced and not entirely removed.
For both $c = 0.25$ and 0.3 we find that $\topt = 0.08$~(0.12) for the clover~(plaquette) discretization of $E(t)$ is satisfactory for the full range of couplings we consider.
\Fig{fig:extrap} demonstrates the resulting reduction of cutoff effects on both sides of the IR fixed point.

At $u = 4$ the expected linear dependence on $(a / L)^2$ provides a good description of the data for $L \geq 16$, with average confidence levels of 0.70 for $c = 0.25$ and 0.58 for $c = 0.3$.
However, the $L = 12 \to 18$ points clearly deviate from this linear scaling, which is our motivation for omitting these data from our main analyses.
\Fig{fig:extrap_compare} illustrates the effects of the $L = 12 \to 18$ data on the quality of the $(a / L)^2 \to 0$ extrapolations, by plotting the resulting \chidof\ for the full range of $u$ that we access.
While the larger $c = 0.3$ improves the quality of the extrapolations as expected~\protect\cite{Fritzsch:2013je}, for most $u$ the dominant contribution to the $\chi^2$ comes from the $L = 12 \to 18$ point.
The exception is the region at stronger couplings $u \gsim 7$, where \fig{fig:extrap} suggests that the $L = 16 \to 24$ points start to deviate from the larger-volume results.
To account for this effect we repeat all continuum extrapolations with only the two points involving $L \geq 20$, and include any differences between these results and the full $L \geq 16$ prediction as another systematic uncertainty.
From \fig{fig:extrap} we note that dropping $L = 16$ at strong coupling will produce $(a / L)^2 \to 0$ extrapolations farther below zero, reinforcing the existence of the IR fixed point.

\Refcite{Lin:2015zpa} comments that `Symanzik-type' continuum extrapolations of the form shown in \fig{fig:extrap}---employing polynomials in $(a / L)^2$---are guaranteed to be valid only in the basin of attraction of the gaussian UV fixed point, and not necessarily in the vicinity of the non-trivial IR fixed point.
Our improvement of the gradient flow running coupling, discussed in \secref{sec:gradflow}, addresses this issue.
First, for any $u$ we can find a value of the $t$-shift $\tau_0$ for which the extrapolation is independent of $L$ and therefore insensitive to the power of $(a / L)$ in the extrapolation.
Then, by demanding that all $\tau_0$ produce the same result upon extrapolating $(a / L)^2 \to 0$ we can check the validity of these extrapolations, and include any deviations as a systematic uncertainty.
In this context, it is interesting to note that the resulting systematic uncertainties often increase significantly at couplings comparable to and stronger than $\gstar$ (cf.~\fig{fig:errorBudgets} in \appref{app:c02}), which may be related to this underlying issue.

\begin{figure}[btp]
  \includegraphics[width=0.45\textwidth]{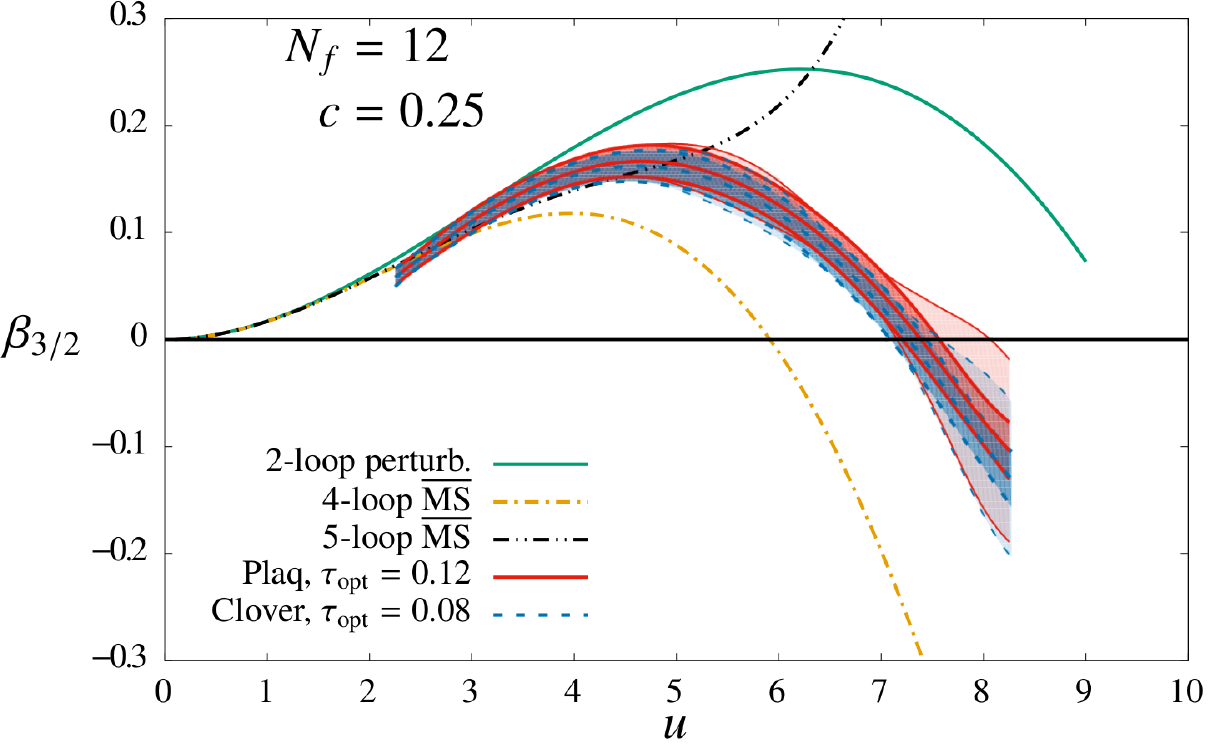}\hfill
  \includegraphics[width=0.45\textwidth]{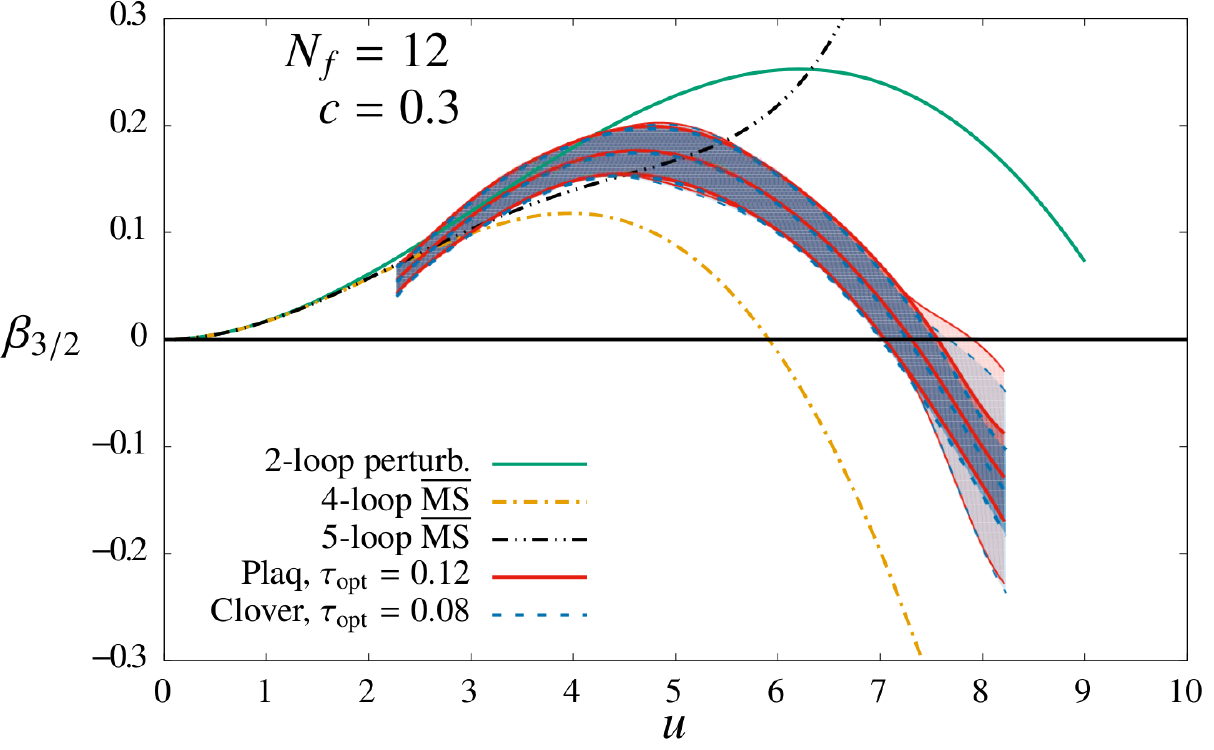}
  \caption{\label{fig:beta}Continuum-extrapolated discrete \be function for scale change $s = 3 / 2$ with $c = 0.25$ (left) and 0.3 (right).  In each plot we include both the plaquette (solid red) and clover (dashed blue) discretizations of $E(t)$ using the optimal $\topt^{\text{(plaq)}} = 0.12$ and $\topt = 0.08$, respectively, as well as two-loop perturbation theory (solid line) and the four- and five-loop perturbative predictions in the \MSbar scheme (dash-dotted and dash-double-dotted lines, respectively).  Both gradient flow renormalization schemes indicate an IR fixed point at $\gstar = 7.26$.  The darker error bands indicate statistical uncertainties, while the lighter error bands show the total uncertainties, with statistical and systematic errors added in quadrature.  Although the systematic errors are symmetrized, \protect\fig{fig:extrap} shows that at strong coupling the larger lattice volumes would produce results for $\be_s$ farther below zero, reinforcing the existence of the IR fixed point.}
\end{figure}

So far we have discussed three potential sources of systematic error that we account for in our analyses.
For convenience we briefly summarize them here:
\begin{description}
  \setlength{\itemsep}{1 pt}
  \setlength{\parskip}{0 pt}
  \setlength{\parsep}{0 pt}
  \vspace{-6 pt}
  \item[Interpolation:] We interpolate $\gtc(L)$ as functions of $\be_F$ on each lattice volume, fitting the data to both a rational function (\eq{eq:pade}) and a polynomial (\eq{eq:poly}).
    We take our final results from the rational function interpolations, and include any discrepancies between the two approaches as a systematic error.
    For $c = 0.25$ and intermediate $u \approx 5$--6 this is the source of the largest systematic uncertainty, which is comparable to the statistical uncertainty.
    For $c = 0.3$ the different interpolations are much more consistent.

  \item[Extrapolation:] To assess the stability of the linear $(a / L)^2 \to 0$ extrapolations we repeat all analyses without including the smallest-volume $L = 16 \to 24$ data, considering only $20 \to 30$ and $24 \to 36$ points.
    We take our final results from the three-point extrapolations, with another systematic uncertainty defined by any disagreement between the two- and three-point analyses.
    This systematic uncertainty is largest at our stronger couplings $u \gsim 7$, where it can be approximately 2.5~times the statistical uncertainty, for both $c = 0.25$ and 0.3.
    As we emphasized in \fig{fig:extrap}, the larger volumes produce extrapolated results for $\be_s$ farther below zero, reinforcing the existence of the IR fixed point.

  \item[Optimization:] Finally, we account for any sensitivity to the $t$-shift improvement parameter $\tau_0$.
    Recall from \secref{sec:gradflow} that different values of $\tau_0$ should all produce the same $\be_s(\gtc)$ in the continuum limit.
    Whenever our final results using the optimal \topt differ from the results we would have obtained from unshifted ($\tau_0 = 0$) analyses, we include the difference as a third systematic error.
    This is a conservative prescription, because we introduced the $t$-shift improvement to \emph{remove} these cutoff artifacts, by enabling more reliable $(a / L) \to 0$ extrapolations.
    Even so, this systematic uncertainty vanishes for all the $s = 3 / 2$ analyses considered in the body of this paper, which involve $c \geq 0.25$ and $L \geq 16$.
    In appendices~\ref{app:s2} and \ref{app:c02} we report that this is not the case for some supplemental checks that include $L = 12$ data.
    Including $L = 12$, this systematic uncertainty vanishes only for $c \geq 0.3$, and can even be the largest source of uncertainty if we consider the small $c = 0.2$ analyzed by refs.~\cite{Cheng:2014jba, Fodor:2016zil}.
\end{description}
\vspace{-6 pt}
In all three cases, to ensure that statistical fluctuations are not double-counted as both statistical and systematic errors we take the latter to correspond to the amount by which the results being compared differ beyond their 1$\si$ statistical uncertainties.
That is, the systematic error estimates vanish when the results being compared agree within 1$\si$ statistical uncertainties, ensuring that no spurious systematic errors are assigned as a consequence of statistical fluctuations.
Different schemes to estimate systematic uncertainties could be explored in future works, or by re-analysis of the raw data we provide in \appref{app:data}.
We carry out separate error analyses for each of the clover and plaquette discretizations of the energy density $E(t)$ in \eq{eq:t-shift}.
Additional systematic effects from the choice of $E(t)$ discretization can be assessed by comparing the two sets of numerical results that we include in \fig{fig:beta}.

We now present our final results for the 12-flavor system in \fig{fig:beta}, which shows the continuum-extrapolated $s = 3 / 2$ discrete \be function for two different renormalization schemes, $c = 0.25$ and 0.3.
In each panel we include our non-perturbative results for both the clover and plaquette discretizations of the energy density $E(t)$ in \eq{eq:t-shift}.
Statistical uncertainties are shown by the darker error bands, while the lighter error bands indicate the total uncertainties, with statistical and systematic errors added in quadrature.

Along with our numerical results, \fig{fig:beta} also shows the two-, four- and five-loop perturbative predictions for the $s = 3 / 2$ discrete \be function.
These perturbative predictions are based on
\begin{align}
  \be(g^2) \equiv L \frac{dg^2}{dL} & = \frac{2g^4}{16\pi^2} \sum_{i = 0} b_i \left(\frac{g^2}{16\pi^2}\right)^i \label{eq:continuum} \\
  b_0 & = \frac{11}{3}C_2(G) - \frac{4}{3}N_f T(R) \cr
  b_1 & = \frac{34}{3}\left[C_2(G)\right]^2 - N_f T(R) \left[\frac{20}{3}C_2(G) + 4C_2(R)\right] \nn
\end{align}
for $N_f$ fermions transforming in representation $R$ of the gauge group.
For the fundamental representation of SU(3) gauge theory we have
\begin{align}
  C_2(G) & = 3 &
  T(F) & = \frac{1}{2} &
  C_2(F) & = \frac{4}{3},
\end{align}
so that $N_f = 12$ gives $b_0 = 3$ and $b_1 = -50$.
Higher-order coefficients $b_i$ depend on the renormalization scheme.
In the \MSbar scheme, \refcite{Ryttov:2010iz} reports numerical values $b_2 \approx -1060$ and $b_3 \approx 6808$ for 12-flavor SU(3) gauge theory (see also \refcite{Pica:2010xq}).
For most \gc our results in \fig{fig:beta} lie in between the two- and four-loop perturbative curves, both of which predict an IR fixed point.
At the weakest couplings we explore our results agree with the four- and five-loop predictions, which remain slightly below the two-loop value.
Since the discrete \be function is scheme dependent these various results do not need to agree at non-zero $u$, and the five-loop curve suggests that perturbation theory does not converge for $g_{\MSbar}^2 \gsim 4$.
Our comparisons with perturbation theory are for illustration only.

\begin{figure}[btp]
  \centering
  \includegraphics[width=0.55\textwidth]{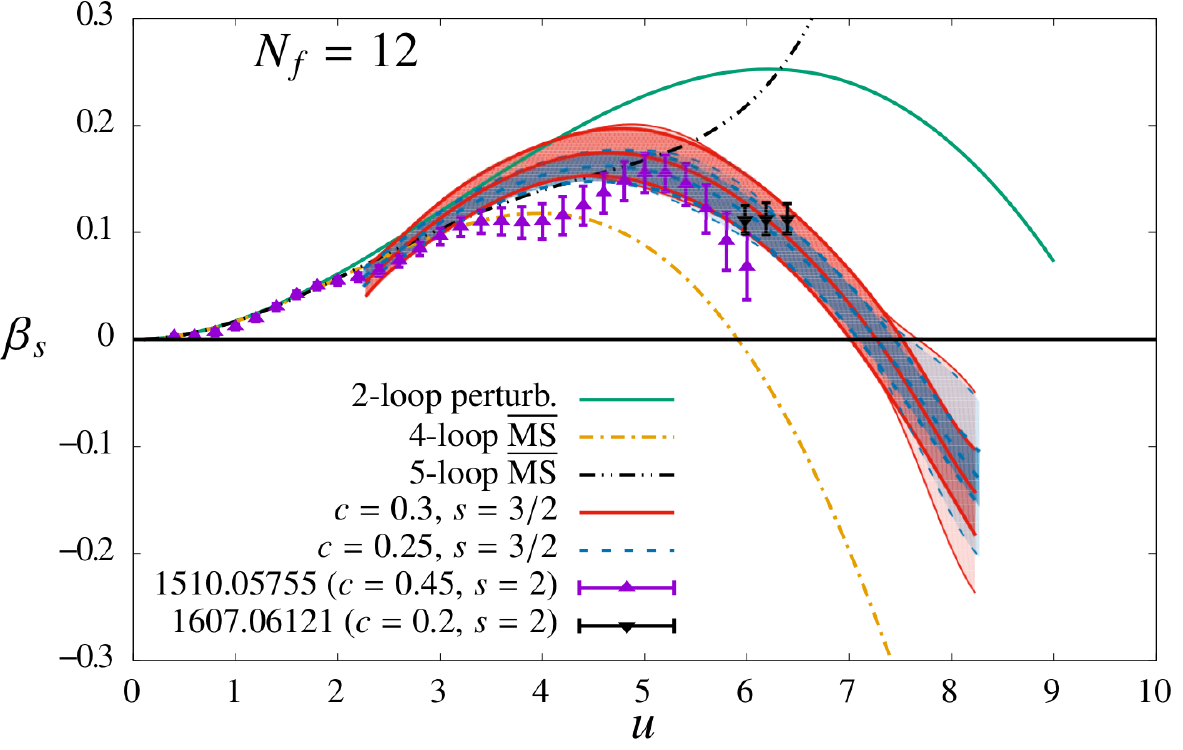}
  \caption{\label{fig:compare}Comparison of our discrete \be function results with data from refs.~\protect\cite{Lin:2015zpa} and \protect\cite{Fodor:2016zil} at couplings $\gc \lsim 6.4$.  All four data sets use the clover discretization of $E(t)$ but different gradient flow renormalization schemes: We overlay our $c = 0.25$ (dashed blue) and $c = 0.3$ (solid red) results (both with $\topt = 0.08$ and scale change $s = 3 / 2$) along with $c = 0.45$ results from \protect\refcite{Lin:2015zpa} and $c = 0.2$ results from \protect\refcite{Fodor:2016zil} (both with $s = 2$).  Given the different renormalization schemes and analysis details the results are all in good agreement.  (The perturbative curves continue to use $s = 3 / 2$ as in \protect\fig{fig:beta}.)}
\end{figure}

\newpage 
Finally, in \fig{fig:compare} we compare our new results with the two recent large-scale step-scaling projects discussed in \secref{sec:intro}~\cite{Lin:2015zpa, Fodor:2016zil}.
We overlay our $c = 0.25$ and 0.3 results from \fig{fig:beta}, adding $c = 0.45$ results from \refcite{Lin:2015zpa} and $c = 0.2$ results from \refcite{Fodor:2016zil}, all using the clover discretization of $E(t)$.
Both of the latter analyses employ scale change $s = 2$ rather than the $s = 3 / 2$ that we use.
Considering that all four sets of numerical results in \fig{fig:compare} use different renormalization schemes, they are in good agreement throughout their common range of couplings.
Had refs.~\cite{Lin:2015zpa, Fodor:2016zil} been able to explore the stronger couplings $u \lsim 8$ that we reach, we expect that they would have observed the same IR fixed point that we report.\footnote{\textbf{Note added:} While this paper was under review, the authors of \refcite{Fodor:2016zil} presented some preliminary results at stronger couplings $u \simeq 7$, which suggest potential tension with the IR fixed point that we observe~\cite{Kuti:2017BU, Fodor:2017Lat}.  While the authors of \refcite{Fodor:2016zil} emphasize the large lattice volumes they consider, we note that their $L = 16$, 18, 20, 24 and 28 are mostly the same as the $L = 16$, 20 and 24 that we use; their larger $sL = 32$, 36, 40, 48 and 56 mainly result from the larger scale change $s = 2$ they consider compared to our $s = 3 / 2$.  Therefore the continuum extrapolation appears unlikely to be an issue and instead, should the final results resemble these preliminary reports, we would be most interested in investigating the different forms of improvement used in the two studies, in particular comparing the ``Symanzik flow'' used by refs.~\cite{Fodor:2016zil, Kuti:2017BU, Fodor:2017Lat} with the Wilson flow we employ.}
In addition, because \refcite{Fodor:2016zil} considers larger $sL \leq 56$ than we do, the good agreement with our results provides evidence that our continuum extrapolations with $sL \leq 36$ are stable and our results would not change if we were to explore larger lattice volumes.
By coincidence, our IRFP is located at the same $\gstar = 7.26$ for both $c = 0.25$ and 0.3.
Combining statistical and systematic errors in quadrature produces the lighter error bands shown in figures~\ref{fig:beta} and \ref{fig:compare}, which cross the $\be_{3 / 2} = 0$ axis at $\gstar = 7.26\left(_{-17}^{+80}\right)$ for $c = 0.25$ and $\gstar = 7.26\left(_{-25}^{+64}\right)$ for $c = 0.3$.

\section{\label{sec:slope}The leading irrelevant critical exponent} 
Now that we have observed an IR fixed point at $\gstar = 7.26$, we will extract the universal critical exponent related to the slope of the discrete \be function at this IRFP.
Linearizing $\be(g^2) \approx \ga_g^{\star} \left(g^2 - \gstar\right)$ around the fixed point, \eq{eq:continuum} implies
\begin{equation}
  \log s = \int_L^{sL} d\log L = \int_{g^2}^{g^2 + \De} \frac{du}{\be(u)} \approx \frac{1}{\ga_g^{\star}} \log\left(1 + \frac{\be_s \log(s^2)}{g^2 - \gstar}\right),
\end{equation}
where $\De \equiv g^2(sL) - g^2(L) = \be_s \log(s^2)$ from \eq{eq:beta}.
Solving for the discrete \be function allows us to relate its slope at the IRFP to $\ga_g^{\star}$,
\begin{equation}
  \be_s(g^2) \approx \be_s'\left(g^2 - \gstar\right) \quad \mbox{with} \quad \be_s' = \frac{s^{\ga_g^{\star}} - 1}{\log(s^2)} \Lra \ga_g^{\star} = \frac{\log\left(1 + 2\be_s'\log s\right)}{\log s}.
\end{equation}
Our convention in \eq{eq:continuum} of considering the RG flow from the UV to the IR, $L \to sL$, produces both $\be_s' < 0$ and $\ga_g^{\star} < 0$.
We omit this negative sign to simplify comparisons with continuum predictions.
\Fig{fig:beta} already shows that we should obtain results comparable to four-loop perturbation theory in the \MSbar scheme, which predicts $\ga_g^{\star} = 0.282$ about 20\% smaller than the two-loop result $\ga_g^{\star} = 0.360$.
A recent scheme-independent estimate $\ga_g^{\star} = 0.228$ from \refcite{Ryttov:2016hal} is somewhat smaller still.

Directly fitting the data shown in \fig{fig:beta} to a linear form in the range $\gstar \pm 0.25$ produces \\[-36 pt] 
\begin{table}[!h]
  \centering
  \renewcommand\arraystretch{1.2}  
  \addtolength{\tabcolsep}{3 pt}   
  \begin{tabular}{c|cc}
    \hline
              & $c = 0.25$              & $c = 0.3$               \\
    \hline
    Clover    & $\ga_g^{\star} = 0.253$ & $\ga_g^{\star} = 0.280$ \\
    Plaquette & $\ga_g^{\star} = 0.249$ & $\ga_g^{\star} = 0.275$ \\
    \hline
  \end{tabular}
\end{table} \\[-36 pt] 
\clearpage
\noindent The high degree of correlation evident in \fig{fig:beta} makes it challenging to determine meaningful statistical uncertainties from these fits.
Since both observables as well as the $c = 0.25$ and 0.3 renormalization schemes should produce the same universal critical exponent, we can estimate a systematic uncertainty from the spread in the numbers above.
If we make the reasonable assumption that this systematic effect dominates over the statistical uncertainties and other systematics, then we end up with $\ga_g^{\star} = 0.26(2)$.

\begin{figure}[btp]
  \includegraphics[width=0.45\textwidth]{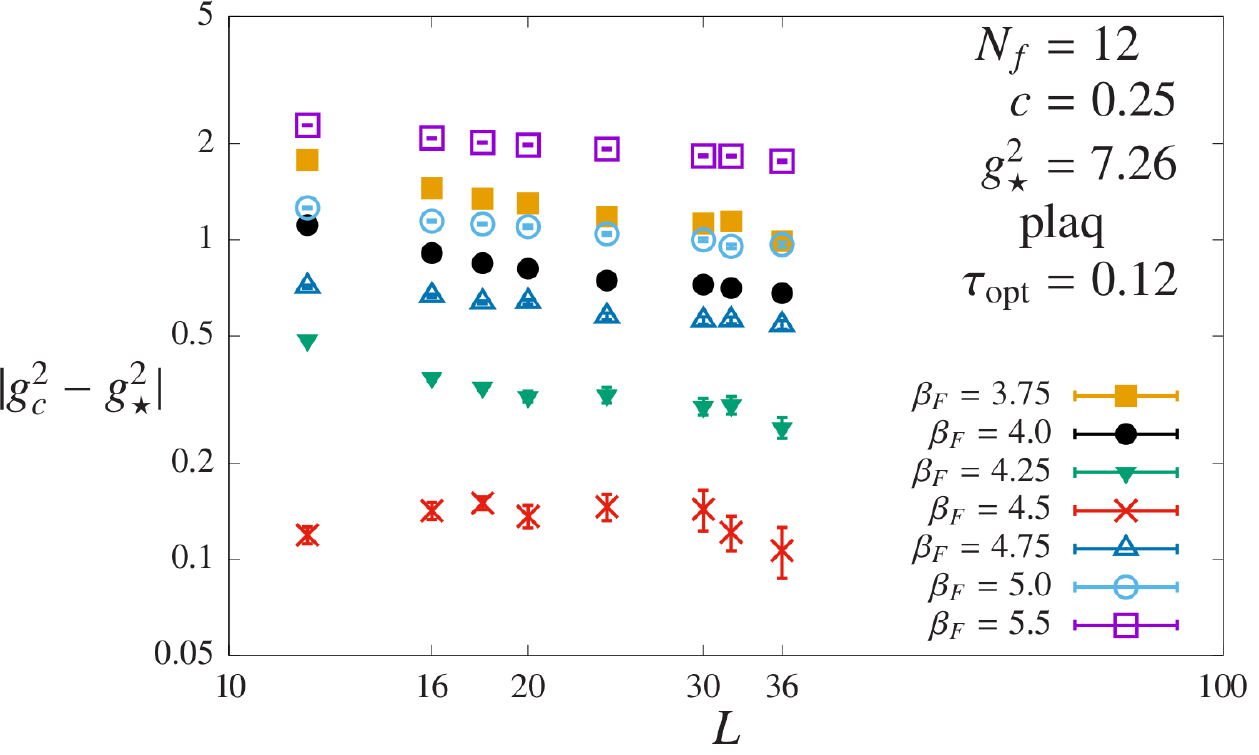}\hfill
  \includegraphics[width=0.45\textwidth]{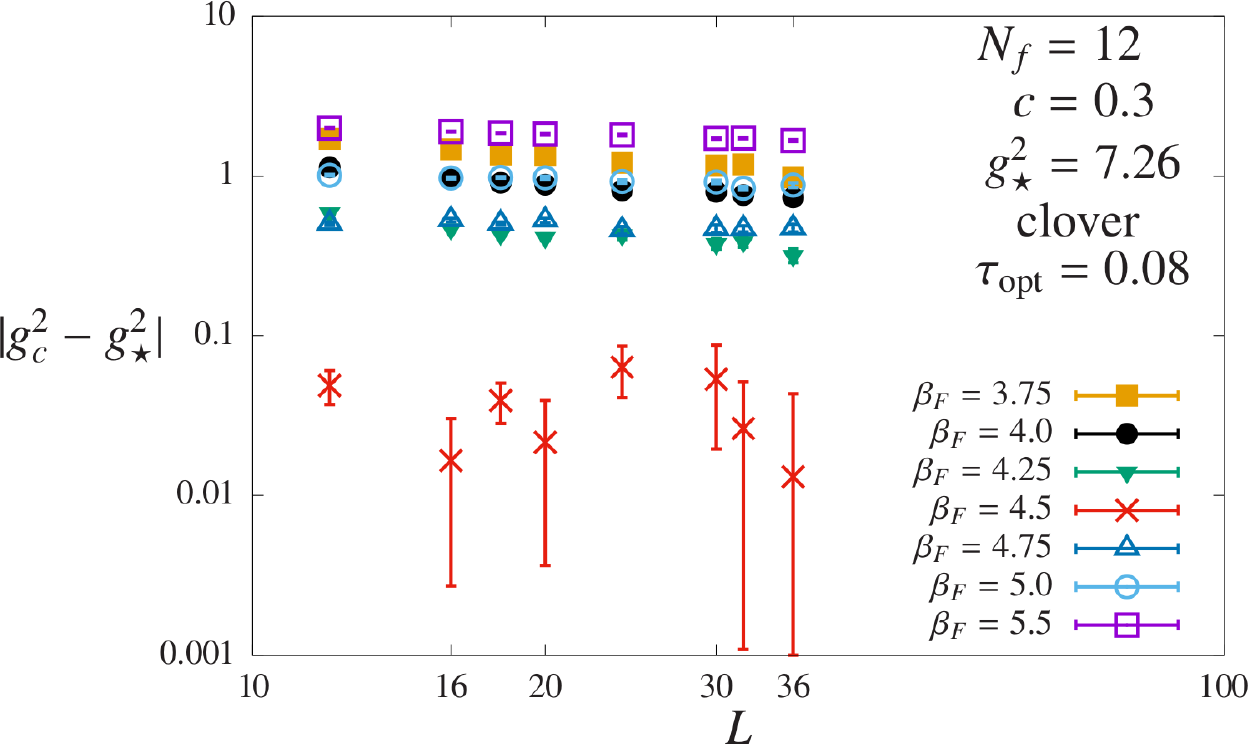}
  \caption{\label{fig:FSS} Raw data for finite-size scaling analyses of the critical exponent $\ga_g^{\star}$.  The \mbox{scaling relation} in \protect\eq{eq:scaling} corresponds to straight lines on these log--log plots of $|\gtc - \gstar|$ vs.\ $L$.  For both the plaquette discretization of $E(t)$ at $c = 0.25$ (left, with $\topt^{\text{(plaq)}} = 0.12$) and the clover discretization at $c = 0.3$ (right, with $\topt = 0.08$) we see \gtc increase towards $\gstar = 7.26$ as the bare coupling increases from $\be_F  = 5.5$ to 4.75 (empty symbols), then move to even stronger renormalized couplings for $4.25 \leq \be_F \leq 3.75$ (filled symbols).  Around $\be_F \approx 4.5$ the signal effectively vanishes since \gtc is so close to \gstar for all $L$.  The other combinations of $c$ and $E(t)$ discretizations produce similar figures.}
\end{figure}

Alternately, we can carry out a finite-size scaling analysis to determine $\ga_g^{\star}$, as in refs.~\cite{Appelquist:2009ty, Lin:2015zpa}.
The basic scaling relation is
\begin{equation}
  \label{eq:scaling}
  \gtc(\be_F, L) - \gstar \propto L^{\ga_g^{\star}}
\end{equation}
for fixed bare coupling $\be_F$.
In principle we could attempt to extract both \gstar and $\ga_g^{\star}$ from these fits, but to simplify the analysis we will use as input our determination of \gstar from \fig{fig:beta}.
In \fig{fig:FSS} we show some of the data available to be analyzed, plotting $|\gtc(\be_F, L) - \gstar|$ vs.\ $L$ on log--log axes for the $c = 0.25$ plaquette discretization and $c = 0.3$ clover discretization.
The other two data sets ($c = 0.25$ clover and $c = 0.3$ plaquette) are similar.
In all cases we can see $\gtc(\be_F, L)$ passing through the fixed-point $\gstar = 7.26$ around $\be_F \approx 4.5$, causing the signal to vanish.

\begin{figure}[btp]
  \includegraphics[width=0.45\textwidth]{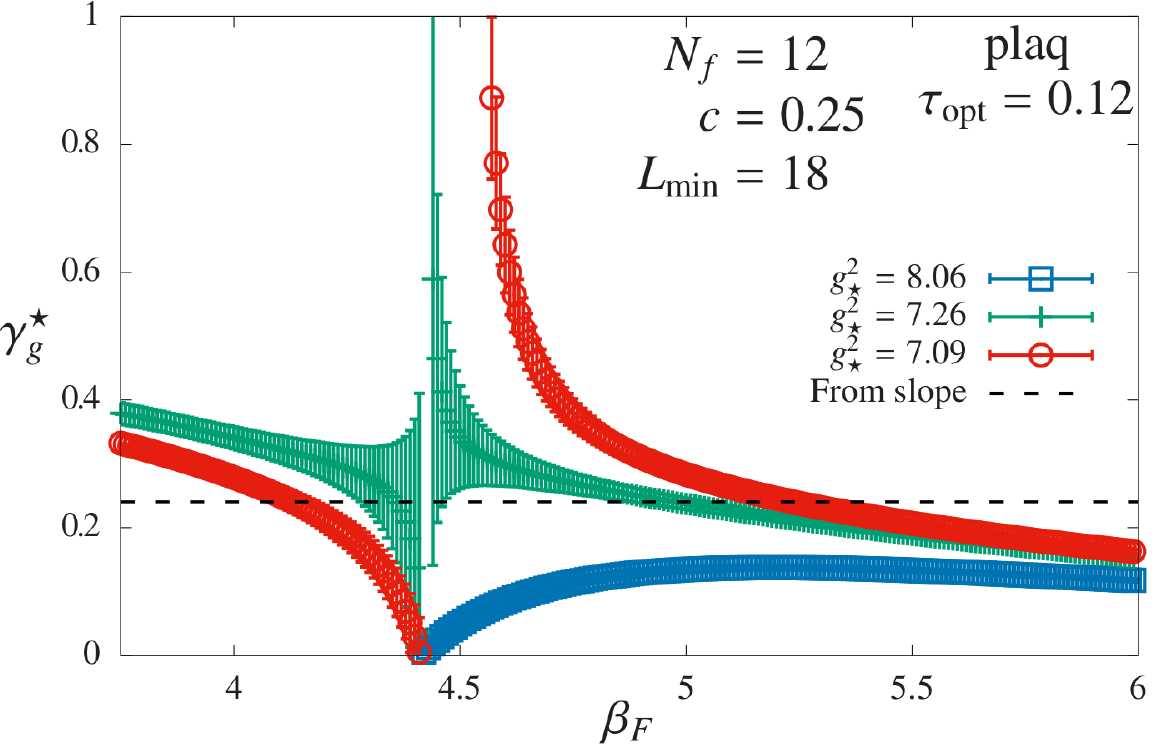}\hfill
  \includegraphics[width=0.45\textwidth]{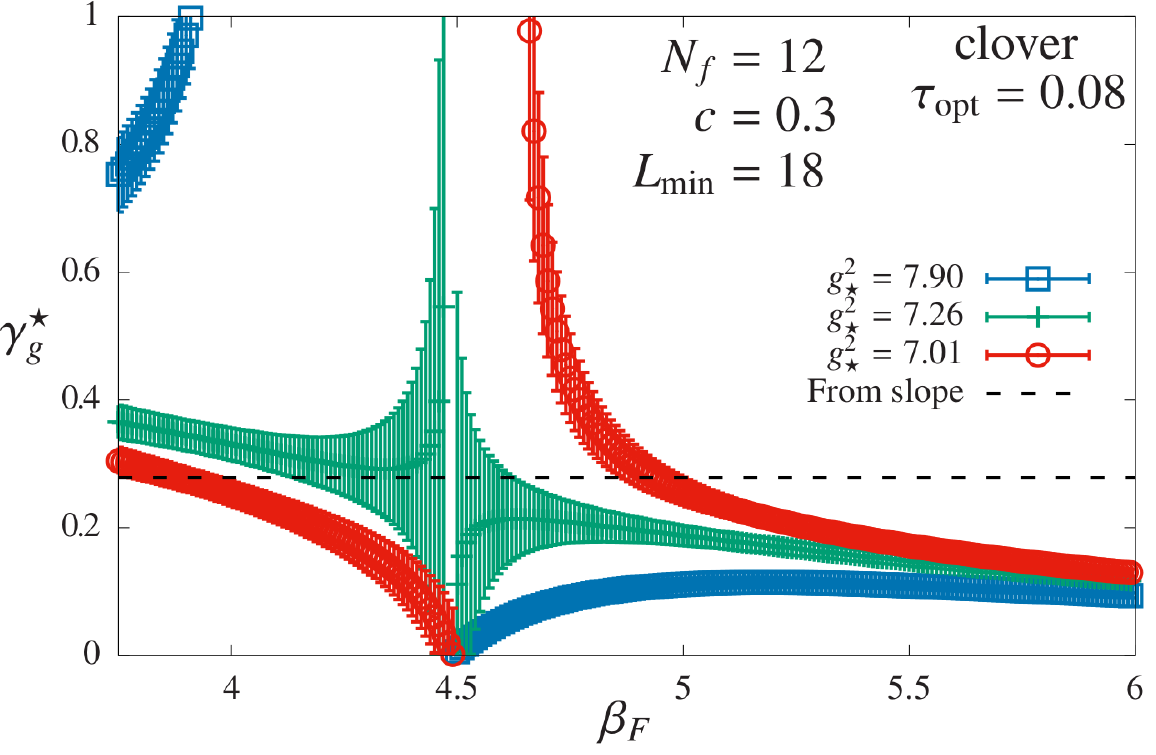}
  \caption{\label{fig:gamma} Finite-size scaling results for $\ga_g^{\star}$ using rational-function interpolations of the data in \protect\fig{fig:FSS}, for the same plaquette discretization of $E(t)$ at $c = 0.25$ (left, with $\topt^{\text{(plaq)}} = 0.12$) and clover discretization at $c = 0.3$ (right, with $\topt = 0.08$).  As expected, the signal vanishes around $\be_F \approx 4.5$ where \gtc is close to \gstar for all $L$, but the results are clearly consistent with the more precise predictions for $\ga_g^{\star}$ from the slopes of the discrete \be functions in \protect\fig{fig:beta} (dashed lines).  In each plot the three curves correspond to the central value of $\gstar = 7.26$ (green crosses) plus the minimum and maximum values of \gstar consistent with the combined statistical and systematic errors (red circles and blue squares, respectively).}
\end{figure}

The finite-size scaling analysis amounts to linear fits of these data, the slopes of which correspond to $\ga_g^{\star}$.
Several significant systematic effects are visible in \fig{fig:FSS}.
First we can see that the slopes of linear fits will change slightly for different bare couplings $\be_F$.
The scaling relation becomes more accurate closer to the IR fixed point, but the slow evolution of the coupling with $L$ (\fig{fig:gcSq}) means that near the IRFP the signal in $|\gtc - \gstar|$ effectively vanishes for all $L$.
Next, the slopes also depend on the range of $L$ included in the fits.
Empirically, we find that omitting the $L = 12$ data significantly increases the confidence levels of the fits.
Additionally omitting $L = 16$ also tends to improve fit quality, while there are no obvious trends upon omitting larger $L$.
Therefore we fit only $L \geq 18$, and should account for any sensitivity to the fit range as a systematic uncertainty.
We can also expect some systematic dependence on $c$ and the $E(t)$ discretization, as in the inline table above, which should be included in the final uncertainties as well.
Finally, and most significantly, we obtain \fig{fig:FSS} by fixing $\gstar = 7.26$.
Allowing \gstar to vary within the total uncertainties determined in the previous section leads to very wide variations in the resulting $\ga_g^{\star}$.

In combination, these systematic uncertainties only allow us to use the finite-size scaling analysis as a consistency check on the value $\ga_g^{\star} = 0.26(2)$ determined directly from the slopes of the discrete \be functions.
This is shown in \fig{fig:gamma}, where we plot finite-size scaling results for the critical exponent vs.\ the bare coupling $\be_F$, considering the same data sets shown in \fig{fig:FSS}.
In order to fill in more values of $\be_F$ we interpolate these data, using the rational function discussed in \secref{sec:results} (\eq{eq:pade}).
We see that the finite-size scaling results for fixed $\gstar = 7.26$ are clearly consistent with the $\ga_g^{\star}$ obtained from the corresponding $\be_s'$ (shown as dashed lines).
As expected, the fit uncertainties blow up around $\be_F \approx 4.5$ where the signal in $|\gtc - \gstar|$ effectively vanishes.
Accounting for the uncertainties on \gstar produces the other two curves in each plot.
Although the systematic spread of the results is enormous around the IRFP, the uncertainties are more manageable for $\be_F \gsim 5$, where they show a steady evolution towards the $\ga_g^{\star} = 0.26(2)$ determined above.

\section{\label{sec:conclusion}Discussion and conclusions} 
We have presented our final results for step-scaling calculations of the 12-flavor SU(3) discrete \be function, using nHYP-smeared staggered fermions and an improved gradient flow running coupling.
In the gradient flow scheme with $c = 0.25$ we observe an IR fixed point at $\gstar = 7.3\left(_{-2}^{+8}\right)$, which changes to $\gstar = 7.3\left(_{-3}^{+6}\right)$ when $c = 0.3$.
We are able to explore the discrete \be function up to $\gc \lsim 8.2$, extending past the IRFP, if not as far past as might be ideal.
We account for systematic effects from the stability of the $(a / L) \to 0$ extrapolations, the interpolation of $\gtc(L)$ as a function of the bare coupling, the improvement of the gradient flow running coupling, and the discretization of the energy density.
These results, including systematic uncertainties, are collected in \fig{fig:beta}.
At the IRFP we measure the leading irrelevant critical exponent to be $\ga_g^{\star} = 0.26(2)$, comparable to perturbative estimates.
This value for $\ga_g^{\star}$ comes from the slope of the discrete \be function and we checked that it is consistent with a finite-size scaling analysis, even though the very slow running of the 12-flavor coupling makes finite-size scaling challenging for $12 \leq L \leq 36$.

We have also shown (\fig{fig:compare}) that our results are consistent with the two recent large-scale step-scaling projects discussed in \secref{sec:intro}~\cite{Lin:2015zpa, Fodor:2016zil}, which were able to \mbox{investigate only} $\gc \lsim 6.4$.
\Refcite{Lin:2015zpa} emphasized the importance of comparing multiple discretizations \mbox{of the} energy density $E(t)$ in the definition of the gradient flow running coupling (\eq{eq:t-shift}), which motivated our investigation of both the plaquette- and clover-based observables.
Considering $L = 8 \to 16$, $10 \to 20$ and $12 \to 24$, \refcite{Lin:2015zpa} found that $c \geq 0.45$ was required \mbox{to avoid} systematic dependence on the choice of discretization.
By moving to larger volumes $L \geq 16$, we find good agreement between both discretizations for $c \geq 0.25$.
In \appref{app:s2} we report that investigations including $L = 12$ need $c \geq 0.3$ to obtain comparably good behavior.
In particular, $c = 0.2$ analyses that include $L = 12$ data suffer from severe systematic uncertainties, which were not comprehensively considered in \refcite{Cheng:2014jba} where we reported $\gstar = 6.2(2)$.
With $c = 0.2$ and $L \geq 12$ we now obtain $\gstar = 5.9(1.9)$, where the uncertainties are almost entirely systematic as we discuss in \appref{app:c02} (\tab{tab:gstar} and \fig{fig:beta_c02}).

Compared to perturbation theory, our results for the scheme-dependent \gstar lie in between the two-loop and four-loop \MSbar values.
At the weakest couplings we explore our $s = 3 / 2$ discrete \be function agrees with the four-loop scheme, which remains slightly below the two-loop case.
The scheme-independent critical exponent $\ga_g^{\star} = 0.26(2)$ that we obtain is consistent with the value 0.282 predicted by four-loop perturbation theory, which was also the case for the mass anomalous dimension $\ga_m^{\star} \approx 0.235$ found by refs.~\cite{Cheng:2013xha, Lombardo:2014pda}.
This close agreement with four-loop \MSbar perturbation theory may be partly coincidental.
Recent investigations of a scheme-independent series expansion~\cite{Ryttov:2016hdp} predict slightly different values $\ga_g^{\star} = 0.228$ and $\ga_m^{\star} = 0.400(5)$~\cite{Ryttov:2016asb, Ryttov:2016hal}, while an initial investigation of the five-loop \MSbar \be function~\cite{Baikov:2016tgj, Herzog:2017ohr} finds that the perturbative expansion breaks down at couplings weaker than $\gstar$, despite the apparently convergent behavior of the two-, three- and four-loop contributions.
Even so, subsequent investigations using the five-loop \be function as input argue that all systems with $9 \leq N_f \leq 16$ exhibit perturbative IRFPs~\cite{Stevenson:2016mnv, Ryttov:2016ner, Ryttov:2016asb, Ryttov:2016hal}.

The accumulating evidence for an IR fixed point in the discrete \be function~\cite{Appelquist:2007hu, Appelquist:2009ty, Bilgici:2009nm, Itou:2010we, Ogawa:2011ki, Lin:2012iw, Itou:2012qn, Itou:2013faa, Cheng:2014jba, Hasenfratz:2015xpa, Lin:2015zpa, Fodor:2016zil}, in addition to further supporting evidence (summarized in \secref{sec:intro}) from the phase diagram at zero and finite temperature~\cite{Deuzeman:2009mh, Cheng:2011ic, Fodor:2012uu, Schaich:2012fr, Deuzeman:2012ee, daSilva:2012wg, Hasenfratz:2013uha, Ishikawa:2013tua} as well as hyperscaling of the hadron masses and decay constants~\cite{Aoki:2012eq, Cheng:2013xha, Lombardo:2014pda} increases our confidence in the conclusion that the 12-flavor system is conformal in the IR.
The many existing investigations leave open a few directions that are particularly important to explore in the future.
First, the existence of a conformal IRFP makes $N_f = 12$ a useful basis for lattice studies of composite Higgs models in which the mass of some of the fermions is lifted to guarantee spontaneous chiral symmetry breaking~\cite{Brower:2015owo, Hasenfratz:2016gut}.
Although there is some motivation for moving to a smaller $N_f \simeq 10$ where the mass anomalous dimension may be larger, $\ga_m^{\star} \simeq 1$, it is still advantageous to test this approach for $N_f = 12$ where we have more information about the existence and characteristics of the IR fixed point.
(There are relatively few lattice studies of the 10-flavor system so far~\cite{Hayakawa:2010yn, Appelquist:2012nz, Chiu:2016uui}.)
Finally, the fact that almost all 12-flavor lattice studies have employed staggered fermions makes it important to investigate the universality (or lack thereof) of the observed IRFP.
As in three-dimensional spin systems~\cite{Calabrese:2002bm, Hasenfratz:2015ssa}, it is not guaranteed that different lattice fermion formulations with different chiral symmetry properties will produce identical predictions at a non-trivial fixed point.
This provides particular motivation for studies using Ginsparg--Wilson (overlap or domain wall) fermions that possess continuum-like chiral symmetries, despite their increased computational cost.

\section*{Acknowledgments} 
We thank Julius Kuti, David Lin and Kieran Holland for useful discussions of step scaling and many-flavor lattice investigations, as well as Robert Shrock for information about recent developments in perturbative analyses and scheme-independent series expansions.
We are also grateful to David Lin for providing numerical results from \refcite{Lin:2015zpa}, and to Stefan Meinel for sharing code to analyze correlations.
This work was supported by the U.S.~Department of Energy (DOE), Office of Science, Office of High Energy Physics, under Award Numbers DE-SC0010005 (AH), DE-SC0008669 and DE-SC0009998 (DS). 
Numerical calculations were carried out on the HEP-TH and Janus clusters at the University of Colorado, the latter supported by the U.S.~National Science Foundation through Grant No.~CNS-0821794, and on the DOE-funded USQCD facilities at Fermilab.

\appendix
\section{\label{app:s2}Results with different scale changes} 
\begin{figure}[tb]
  \includegraphics[width=0.45\textwidth]{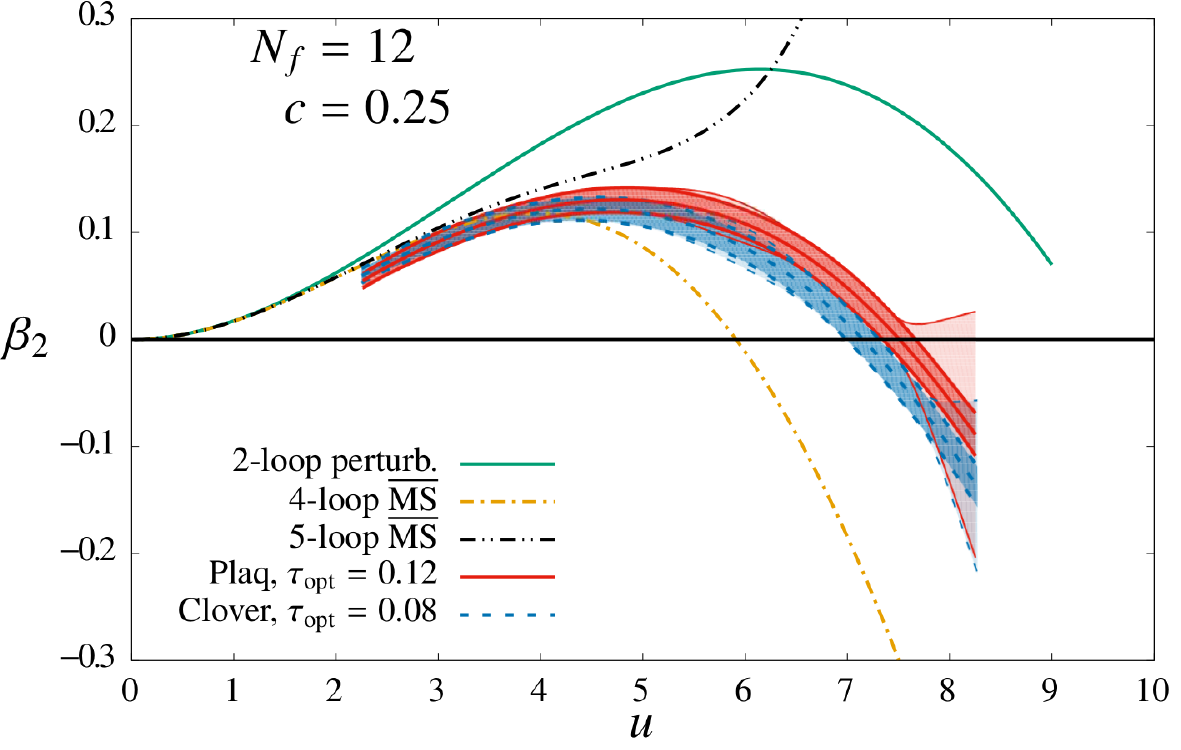}\hfill \includegraphics[width=0.45\textwidth]{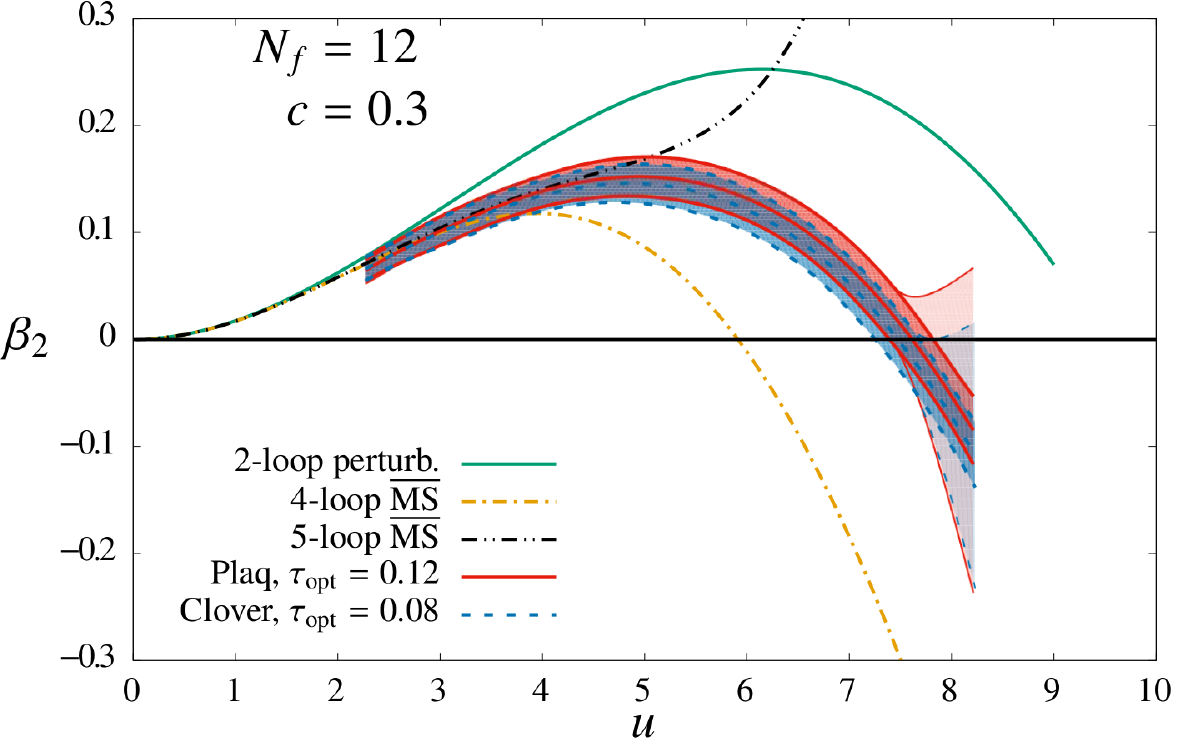} \\[12 pt]
  \includegraphics[width=0.45\textwidth]{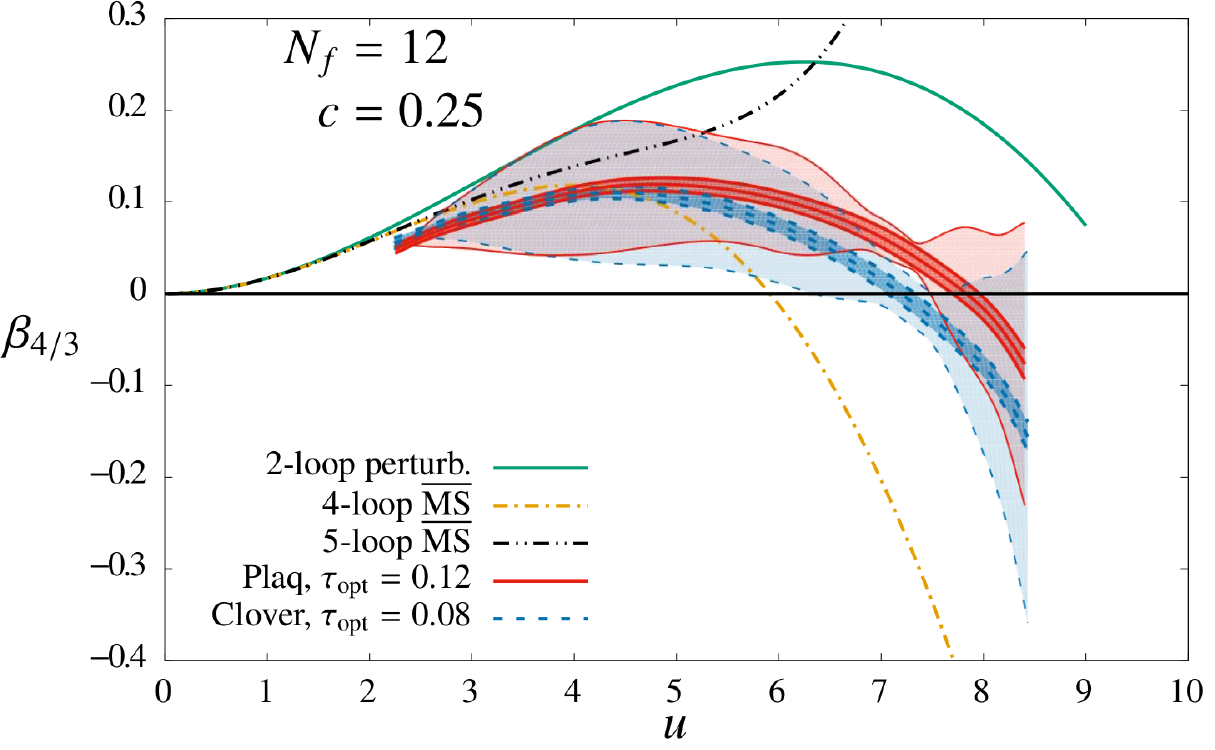}\hfill \includegraphics[width=0.45\textwidth]{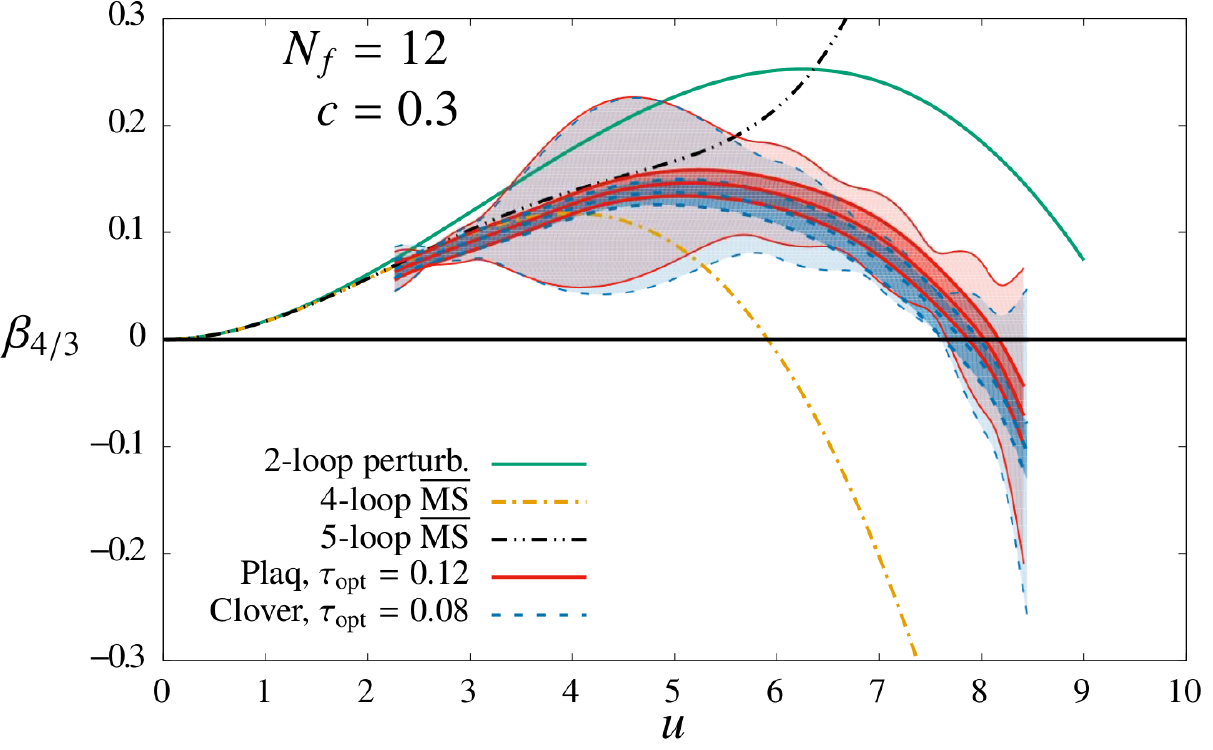}
  \caption{\label{fig:beta_s}Continuum-extrapolated discrete \be function for scale changes $s = 2$ (top) and $s = 4 / 3$ (bottom) with $c = 0.25$ (left) and 0.3 (right), plotted in the same style as \protect\fig{fig:beta} and also predicting an IR fixed point consistent with the $s = 3 / 2$ analyses considered in the body of the paper.  The inclusion of $L = 12$ data in the analyses leads to larger systematic uncertainties, especially for the smaller $s = 4 / 3$ where the slow flow of the coupling is more difficult to resolve.}
\end{figure}

As shown by \tab{tab:pairs} in \secref{sec:setup}, our data also allow us to carry out step-scaling analyses with scale changes $s = 2$ and $4 / 3$ in addition to the $s = 3 / 2$ considered in the body of the paper, if we are willing to include the smallest lattice volume $12^4$.
Following the same procedures described in \secref{sec:results} produces the continuum-extrapolated discrete \be function results shown in \fig{fig:beta_s} for $c = 0.25$ and 0.3.
While all of these analyses predict an IR fixed point consistent with that found for $s = 3 / 2$, the inclusion of the $L = 12$ data increases the systematic uncertainties, especially for the smaller $s = 4 / 3$ where the slow flow of the coupling is more difficult to resolve.

In particular, it is interesting to note that in the $s = 2$ case ($L \geq 12$) where the uncertainties are better controlled, we need $c \geq 0.3$ in order to obtain good agreement between results from the plaquette vs.\ clover discretizations of the energy density $E(t)$ in \eq{eq:t-shift}.
This contrasts with the good agreement we observe even for $c = 0.25$ in \fig{fig:beta} when considering only $L \geq 16$.
That is, larger lattice volumes improve the agreement between these two discretizations, which is consistent with expectations and with the results reported by \refcite{Lin:2015zpa}: considering $L \geq 8$, \refcite{Lin:2015zpa} found that $c \geq 0.45$ was needed to obtain comparable agreement.
One other notable change from the $L \geq 16$ results in the body of the paper is that the systematic uncertainty due to $t$-shift optimization discussed in \secref{sec:results} no longer vanishes for $c = 0.25$.
However, this systematic uncertainty continues to vanish for $c = 0.3$, suggesting that it---like the effect of $E(t)$ discretization---is also sensitive to the combination of $c$ and lattice volume.

From \fig{fig:beta_s} we can again estimate the leading irrelevant critical exponent $\ga_g^{\star}$ from the slopes of the discrete \be functions at the IRFP.
(The finite-size scaling consistency check discussed in \secref{sec:slope} already included all of the data going into the $s = 2$ and $4 / 3$ analyses.)
Following the same procedure described in \secref{sec:slope} (i.e., neglecting statistical uncertainties and setting systematic uncertainties by demanding agreement for $c = 0.25$ and 0.3 with both plaquette and clover discretizations) produces $\ga_g^{\star} = 0.24(3)$ from $s = 2$ and $\ga_g^{\star} = 0.22(6)$ from $s = 4 / 3$. 
Both of these values agree with our result $\ga_g^{\star} = 0.26(2)$ from $s = 3 / 2$ with $L \geq 16$, as well as the four-loop perturbative value 0.282 and the scheme-independent 0.228 from \refcite{Ryttov:2016hal}.
In summary, all scale changes $s$ that we can consider with our data set consistently predict a 12-flavor IR fixed point and a leading irrelevant critical exponent comparable to perturbative estimates.

\section{\label{app:c02}Results with smaller $c = 0.2$} 
\vspace{-6 pt} 
One advantage of the gradient flow running coupling is that it is straightforward to re-run analyses for an entire family of renormalization schemes parameterized by $c = \sqrt{8t} / L$.
In general the renormalized coupling has smaller statistical uncertainties for smaller $c$, while larger $c$ can help to reduce systematic effects~\cite{Fritzsch:2013je}.
We have already seen in figures~\ref{fig:extrap} and \ref{fig:extrap_compare} that $c = 0.3$ reduces cutoff effects and improves the quality of $(a / L)^2 \to 0$ extrapolations compared to $c = 0.25$.
In \appref{app:s2} we discussed how analyses including $L = 12$ require $c \geq 0.3$ in order to obtain good agreement between results employing the clover vs.\ plaquette discretizations of the energy density $E(t)$ in \eq{eq:t-shift}.
This agreement persists even with $c = 0.25$ when $L \geq 16$ as in the body of the paper, motivating our choice to focus on $c = 0.25$ and 0.3 for our main analyses.

\begin{figure}[btp]
  \includegraphics[width=0.45\textwidth]{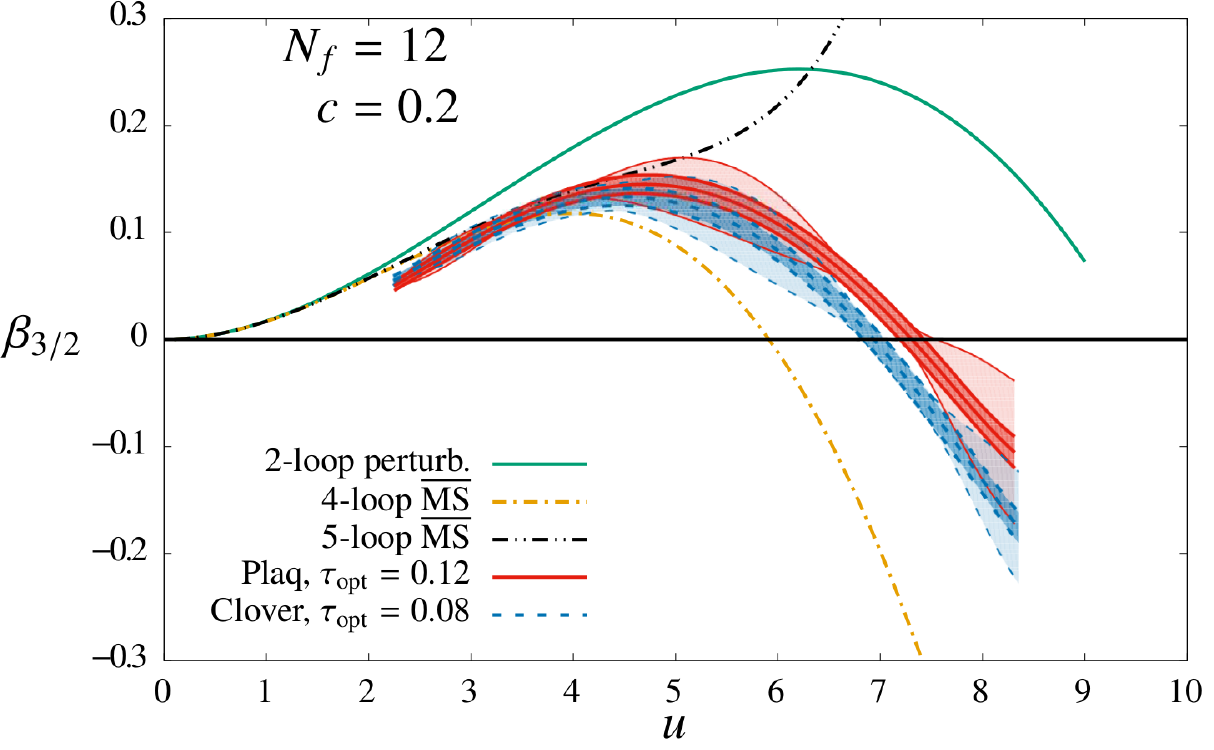}\hfill \includegraphics[width=0.45\textwidth]{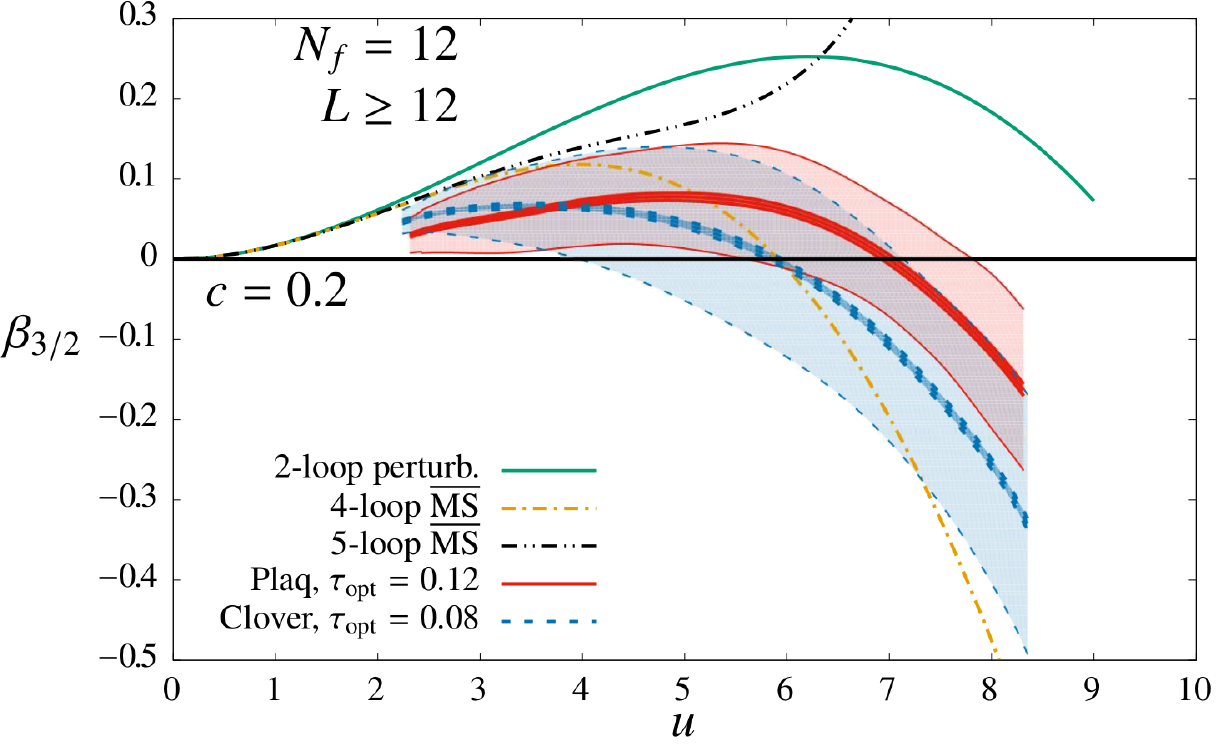} \\[12 pt]
  \includegraphics[width=0.45\textwidth]{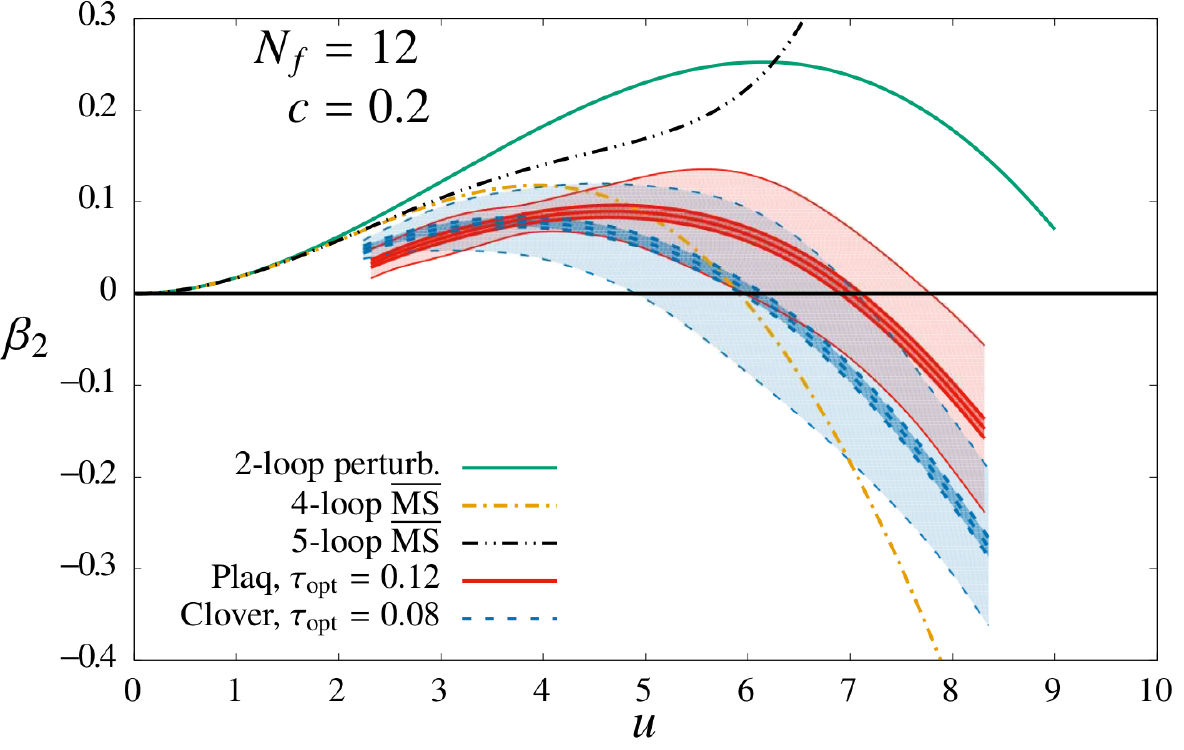}\hfill \includegraphics[width=0.45\textwidth]{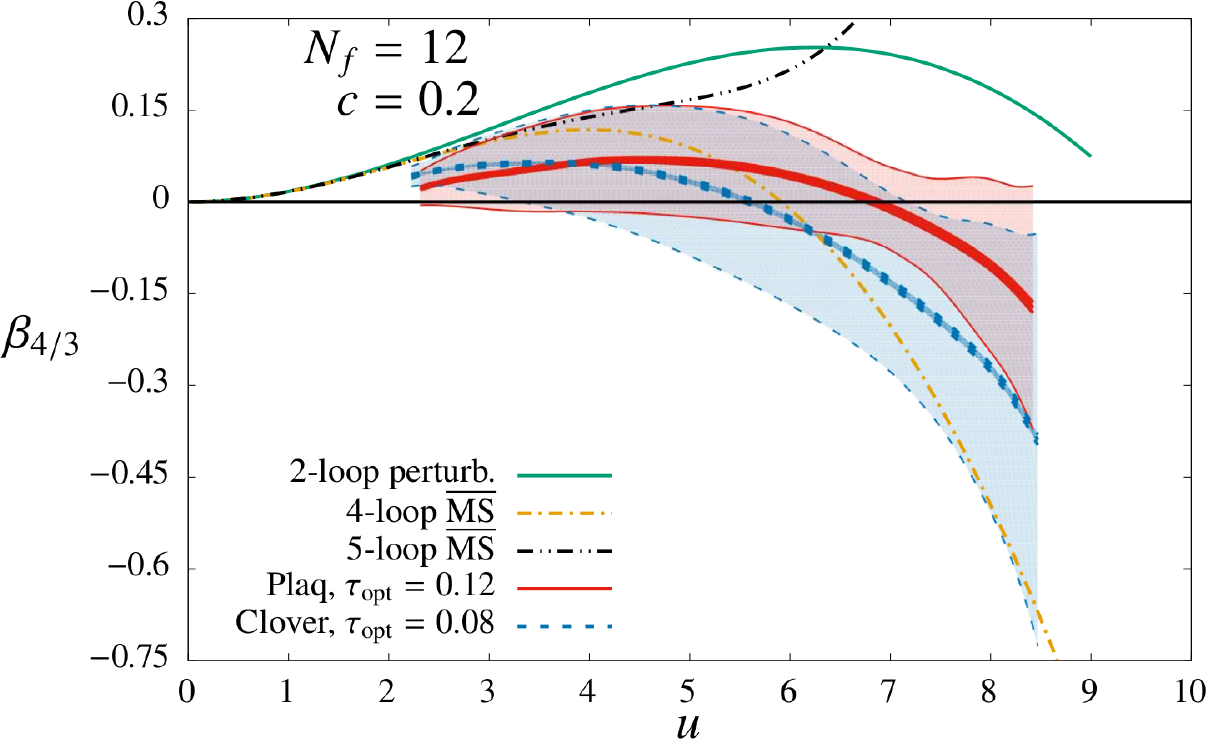}
  \caption{\label{fig:beta_c02}Continuum-extrapolated discrete \be function for $c = 0.2$, plotted in the same style as \protect\fig{fig:beta}.  In the top row the scale change is $s = 3 / 2$, with $L \geq 16$ for the top-left plot (as in the body of the paper).  In the top-right plot, including $L \geq 12$ dramatically increases the systematic uncertainties, and leads to questionable plaquette-based results $\be_s < 0$ for all couplings we can access.  Similar behavior persists in the bottom row of plots considering $s = 2$ (left) and $s = 4 / 3$ (right), both of which necessarily include $L \geq 12$.  As always, all errors are computed exactly as in figures~\protect\ref{fig:beta} and \protect\ref{fig:beta_s}. \vspace{-6 pt}} 
\end{figure}

\begin{table}[htbp]
  \centering
  \renewcommand\arraystretch{1.2}  
  \addtolength{\tabcolsep}{3 pt}   
  \begin{tabular}{c|ccc}
    \hline
                                                & $c = 0.2$                                                  & $c = 0.25$                                             & $c = 0.3$                                              \\
    \hline
                                                & \multicolumn{3}{c}{Main analyses}                                                                                                                                            \\
    \multirow{3}{*}{$s = 3 / 2$, \ $L \geq 16$} & $\gstar = 6.93\left(_{-10}^{+11}\right)_{\text{stat}}$     & $\gstar = 7.26\left(_{-17}^{+18}\right)_{\text{stat}}$ & $\gstar = 7.26\left(_{-25}^{+25}\right)_{\text{stat}}$ \\
                                                & $\gstar = 6.93\left(_{-11}^{+12}\right)_{\text{clov}}$     & $\gstar = 7.26\left(_{-17}^{+33}\right)_{\text{clov}}$ & $\gstar = 7.26\left(_{-25}^{+41}\right)_{\text{clov}}$ \\
                                                & $\gstar = 6.93\left(_{-11}^{+61}\right)_{\text{tot}}$      & $\gstar = 7.26\left(_{-17}^{+80}\right)_{\text{tot}}$  & $\gstar = 7.26\left(_{-25}^{+64}\right)_{\text{tot}}$  \\
    \hline
    \hline
                                                & \multicolumn{3}{c}{Supplemental checks}                                                                                                                                      \\
    \multirow{3}{*}{$s = 2$, \ $L \geq 12$}     & $\gstar = 6.04\left(_{-10}^{+11}\right)_{\text{stat}}$     & $\gstar = 7.15\left(_{-16}^{+16}\right)_{\text{stat}}$ & $\gstar = 7.47\left(_{-22}^{+21}\right)_{\text{stat}}$ \\
                                                & $\gstar = 6.04\left(_{-1.14}^{+1.02}\right)_{\text{clov}}$ & $\gstar = 7.15\left(_{-17}^{+16}\right)_{\text{clov}}$ & $\gstar \geq 7.25_{\text{clov}}$                       \\
                                                & $\gstar = 6.04\left(_{-1.14}^{+1.74}\right)_{\text{tot}}$  & $\gstar \geq 6.98_{\text{tot}}$                        & $\gstar \geq 7.25_{\text{tot}}$                        \\
    \hline
    \multirow{3}{*}{$s = 3 / 2$, \ $L \geq 12$} & $\gstar = 5.88\left(_{-8}^{+7}\right)_{\text{stat}}$       & $\gstar = 7.20\left(_{-10}^{+10}\right)_{\text{stat}}$ & $\gstar = 7.55\left(_{-13}^{+12}\right)_{\text{stat}}$ \\
                                                & $\gstar = 5.88\left(_{-1.91}^{+1.25}\right)_{\text{clov}}$ & $\gstar = 7.20\left(_{-25}^{+19}\right)_{\text{clov}}$ & $\gstar = 7.55\left(_{-13}^{+12}\right)_{\text{clov}}$ \\
                                                & $\gstar = 5.88\left(_{-1.91}^{+1.93}\right)_{\text{tot}}$  & $\gstar = 7.20\left(_{-25}^{+56}\right)_{\text{tot}}$  & $\gstar = 7.55\left(_{-13}^{+25}\right)_{\text{tot}}$  \\
    \hline
    \multirow{3}{*}{$s = 4 / 3$, \ $L \geq 12$} & $\gstar = 5.60\left(_{-7}^{+7}\right)_{\text{stat}}$       & $\gstar = 7.19\left(_{-13}^{+14}\right)_{\text{stat}}$ & $\gstar = 7.81\left(_{-18}^{+17}\right)_{\text{stat}}$ \\
                                                & $\gstar = 5.60\left(_{-2.27}^{+1.56}\right)_{\text{clov}}$ & $\gstar \geq 6.34_{\text{clov}}$                       & $\gstar \geq 7.60_{\text{clov}}$                       \\
                                                & Unconstrained$_{\text{tot}}$                               & $\gstar \geq 6.34_{\text{tot}}$                        & $\gstar \geq 7.60_{\text{tot}}$                        \\
    \hline
  \end{tabular}
  \caption{\label{tab:gstar}Results for \gstar from various combinations of scale change $s$, gradient flow renormalization scheme parameter $c$, and (in the case of $s = 3 / 2$) restriction on the lattice volume.  The central values and statistical uncertainties in the top row of each entry come from the clover discretization of $E(t)$.  The middle row of each entry continues to consider the clover discretization, also accounting for the three sources of systematic uncertainties summarized in \protect\secref{sec:results}.  The third row presents the total uncertainties that include all systematics for both the clover and plaquette discretizations.}
\end{table}

However, since some previous works~\cite{Cheng:2014jba, Fodor:2016zil} used $c = 0.2$, here we consider what results our current data and analyses would produce in this scheme.
Following the same procedures described in \secref{sec:results} leads to the continuum-extrapolated discrete \be function results shown in \fig{fig:beta_c02} for scale changes $s = 3 / 2$ (top), 2 (bottom left) and $4 / 3$ (bottom right).
In the top row of plots we contrast $s = 3 / 2$ analyses with $L \geq 16$ as in the body of the paper (left), or $L \geq 12$ as is required for the other scale changes (right).
While the $L \geq 16$ plot is well behaved and predicts an IR fixed point at $\gstar = 6.93\left(_{-11}^{+61}\right)$, adding statistical and systematic uncertainties in quadrature, the combination of $L = 12$ and $c = 0.2$ dramatically increases the systematic uncertainties.
Even though the other three analyses still produce an IR fixed point with the clover discretization, they prefer significantly smaller $\gstar = 6.04$, 5.88 and 5.60 for $s = 2$, $3 / 2$ and $4 / 3$, respectively, with significantly larger systematic uncertainties.
This is relevant since the result $\gstar = 6.2(2)$ from \refcite{Cheng:2014jba} came from using $c = 0.2$ and $L \geq 12$, without comprehensively considering the systematic uncertainties that we investigate in this work.

For ease of reference, in \tab{tab:gstar} we summarize predictions for \gstar from all the different scale changes $s$ and values of $c$ we have analyzed.
In each case we take the central value for \gstar from the clover discretization, and present three different estimates for the uncertainties.
First, in the top row of each entry, we consider only the statistical uncertainties on the clover-discretization results, corresponding to the dark blue error bands in Figs.~\ref{fig:beta}, \ref{fig:beta_s} and \ref{fig:beta_c02}.
In the middle row we include as well the three sources of systematic error discussed in the body of the paper (and summarized in \secref{sec:results}), again considering only the clover discretization.
These uncertainty estimates correspond to the light blue error bands in Figs.~\ref{fig:beta}, \ref{fig:beta_s} and \ref{fig:beta_c02}.
Finally, in the bottom row of each entry we combine all sources of uncertainties for both the clover and plaquette discretizations, including both the blue and red error bands in Figs.~\ref{fig:beta}, \ref{fig:beta_s} and \ref{fig:beta_c02}.\footnote{\textbf{Note added:} While this paper was under review we corrected a minor numerical bug in the analysis of the plaquette-discretization results, which affected the combined uncertainty estimates in the bottom row of each entry in \tab{tab:gstar}.}

\begin{figure}[btp]
  \includegraphics[width=0.45\textwidth]{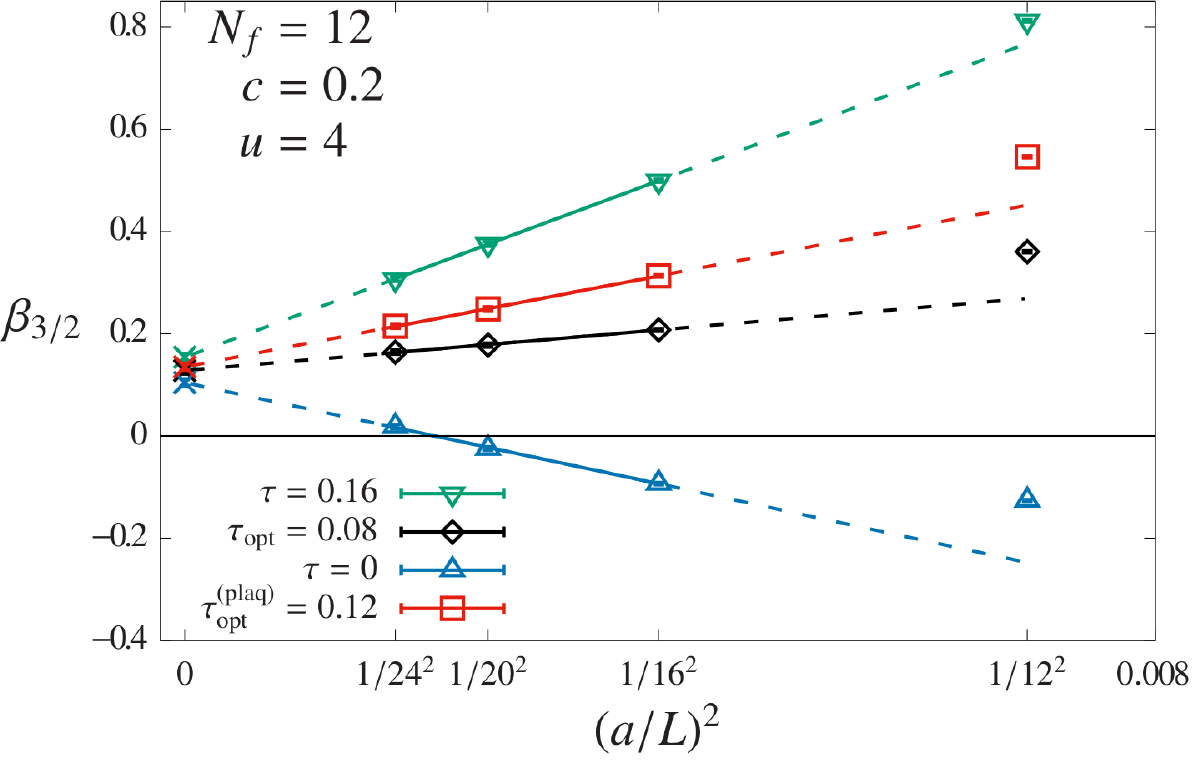}\hfill \includegraphics[width=0.45\textwidth]{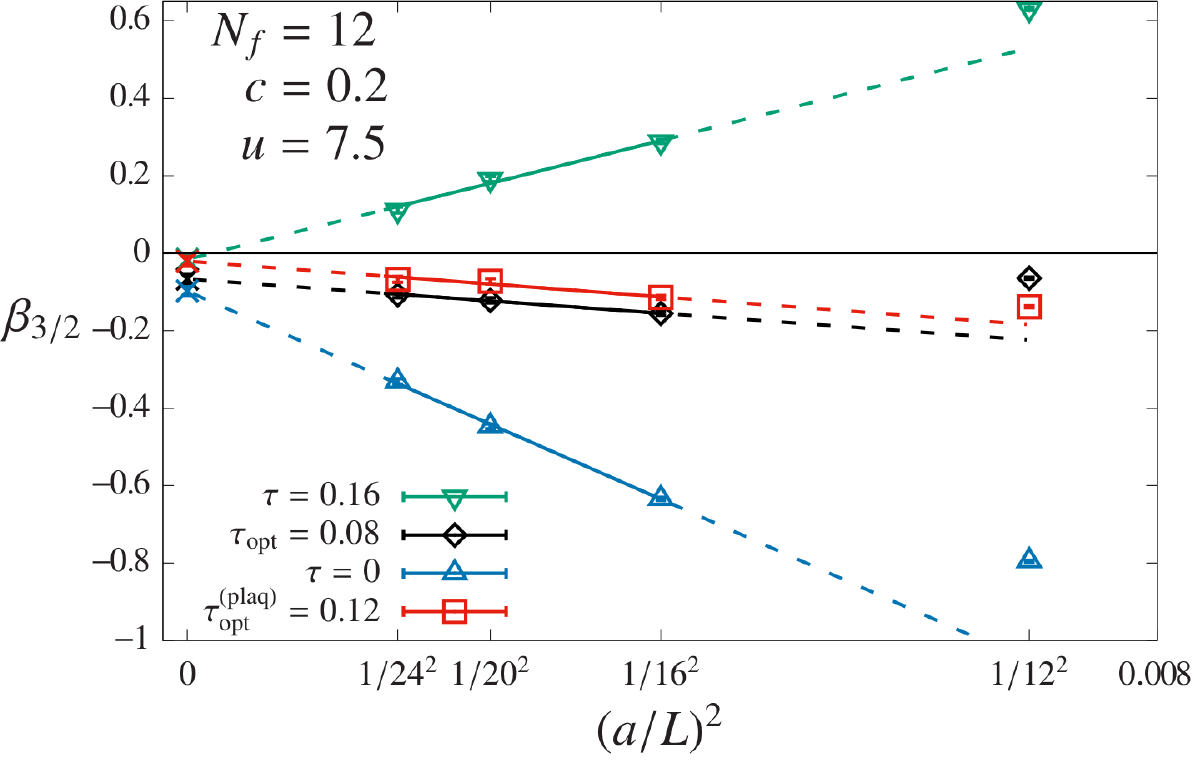}
  \caption{\label{fig:extrap02}Linear $(a / L)^2 \to 0$ extrapolations of the $s = 3 / 2$ discrete \be function for $c = 0.2$, at two values of $u = 4$ (left) and 7.5 (right) on either side of the IR fixed point.  In each plot we compare $\tau_0 = 0$ and 0.16 to the optimal $\topt = 0.08$, and also include results from the plaquette discretization of the energy density $E(t)$ in \protect\eq{eq:t-shift} at the corresponding optimal $\topt^{\text{(plaq)}} = 0.12$.  Unlike \protect\fig{fig:extrap}, the different $\tau_0$ do not extrapolate to consistent values in the $(a / L)^2 \to 0$ limit.}
\end{figure}

It is worthwhile to try to understand the origin of the large systematic uncertainties that arise when $c = 0.2$ and $L \geq 12$.
One issue when $c = 0.2$ is that different values of the $t$-shift improvement parameter $\tau_0$ no longer produce consistent results for $\be_s(\gc)$ upon extrapolating $(a / L)^2 \to 0$.
This is shown in \fig{fig:extrap02}, for renormalized couplings $u = 4$ and 7.5 similar to those considered in \fig{fig:extrap}.
(With $c = 0.2$ and $\tau_0 = 0.16$ we access only $u \leq 7.97$, and can't consider the $u = 8$ shown in \fig{fig:extrap}.)
Although the uncertainties on the points are rather small, it is possible to see statistically significant discrepancies between the extrapolated values.

\begin{figure}[btp]
  \includegraphics[width=0.45\textwidth]{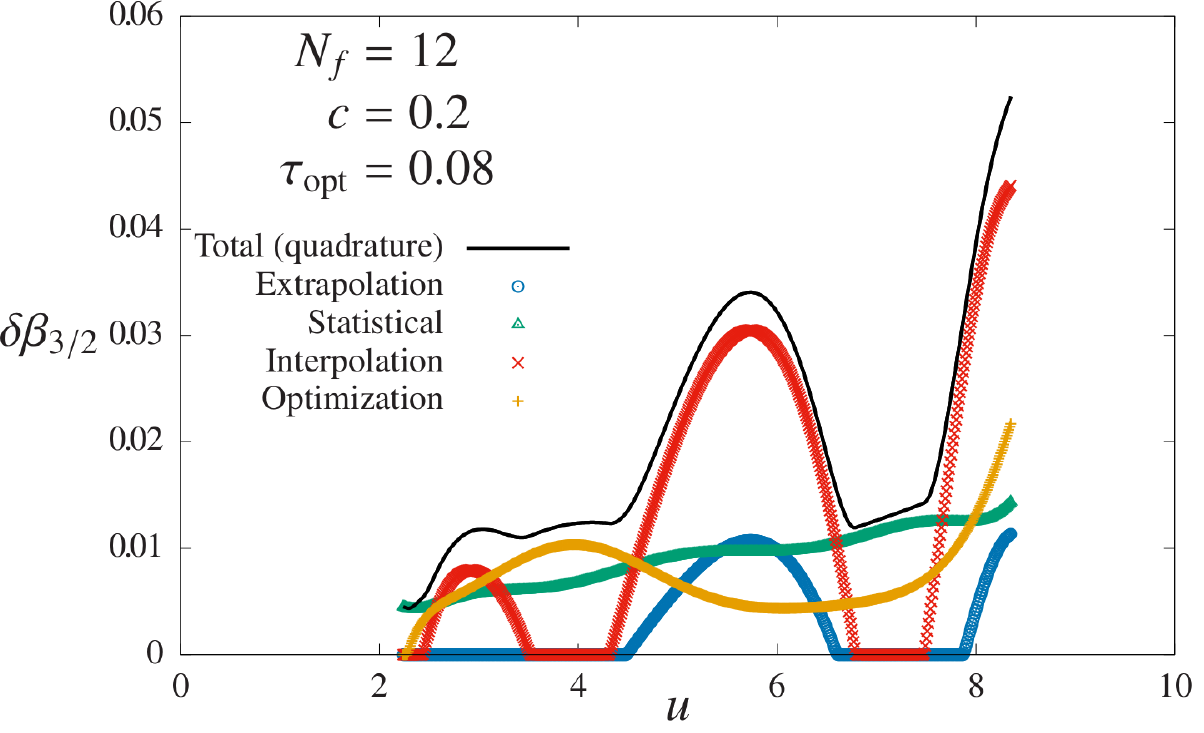}\hfill \includegraphics[width=0.45\textwidth]{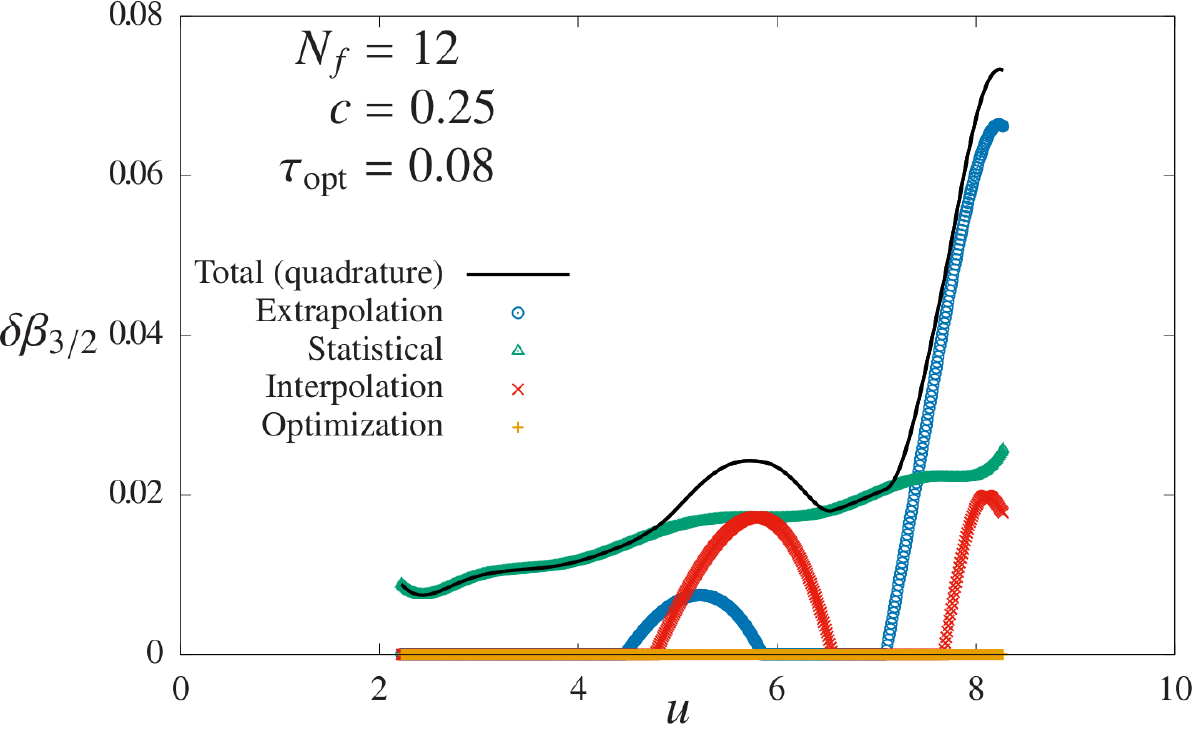} \\[12 pt]
  \includegraphics[width=0.45\textwidth]{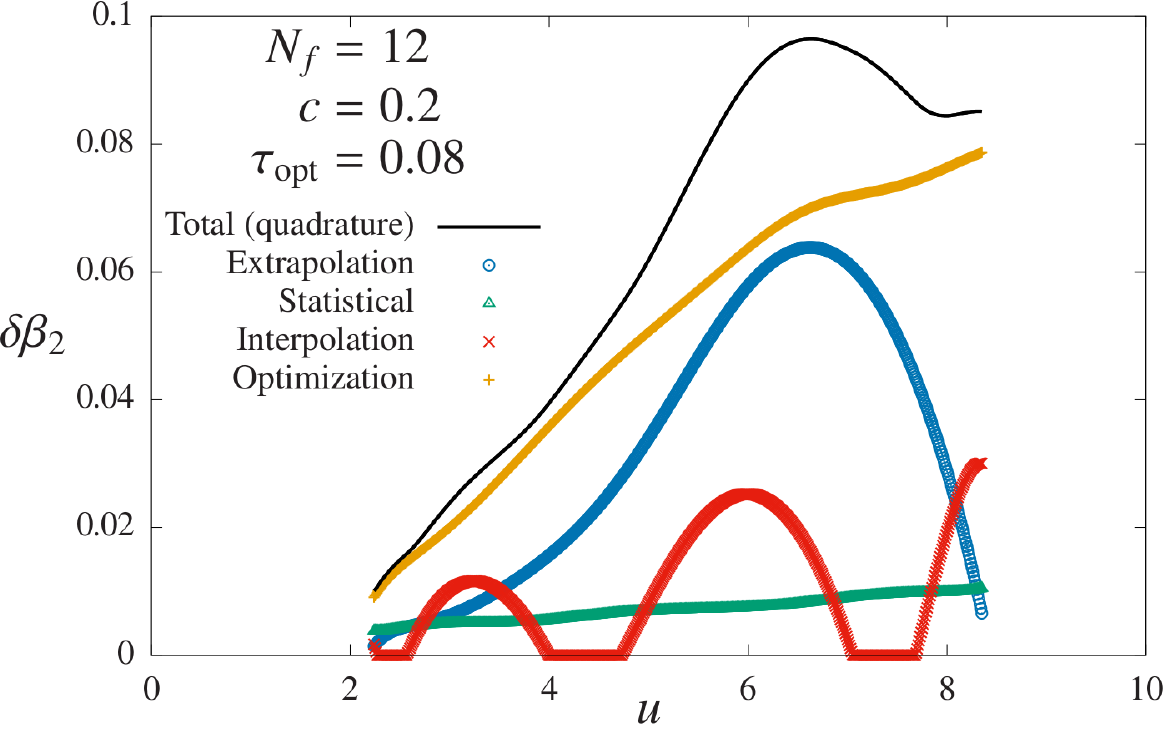}\hfill \includegraphics[width=0.45\textwidth]{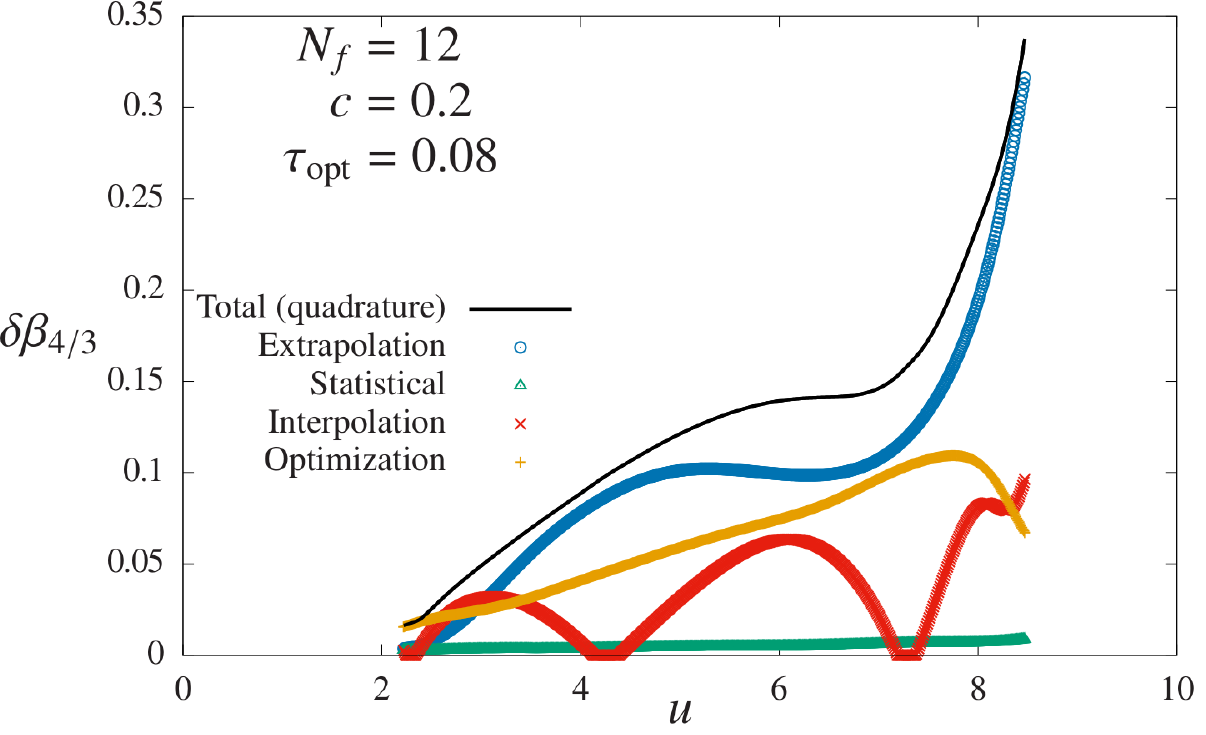}
  \caption{\label{fig:errorBudgets}Statistical and systematic `error budgets' for four of our analyses, as functions of the renormalized coupling $u$.  Each plot shows the statistical uncertainty and the three systematic uncertainties summarized in \protect\secref{sec:results}, as well as their combination in quadrature.  In the top row we compare $s = 3 / 2$ analyses with $L \geq 16$ and $c = 0.2$ (left) vs.\ 0.25 (right).  Even with $L \geq 16$ the small $c = 0.2$ introduces non-zero optimization uncertainties as in accordance with \protect\fig{fig:extrap02}, though these remain comparable to the statistical uncertainties.  When $L = 12$ is included the optimization uncertainties become much larger than the statistical uncertainties, as shown in the bottom row of plots for $s = 2$ (left) and $s = 4 / 3$ (right).  Note the different vertical scale in each plot.}
\end{figure}

Since we account for such discrepancies as a source of systematic error, an easier way to assess them is to inspect the `error budgets' shown in \fig{fig:errorBudgets}.
For each renormalized coupling $u$ these plots show the statistical uncertainties and the three systematic uncertainties summarized in \secref{sec:results}, along with their combination in quadrature.
(Recall from \secref{sec:results} that we take systematic errors to vanish when their effects are indistinguishable from statistical fluctuations, to avoid double-counting the latter.)
The top-right plot corresponds to one of the main analyses discussed in the body of the paper, with $s = 3 / 2$, $c = 0.25$ and $L \geq 16$.
As described in \secref{sec:results}, the optimization uncertainties vanish for all $u$, the interpolation uncertainties are comparable to the statistical uncertainties for intermediate $u \approx 5$--6, and the extrapolation uncertainties dominate for stronger couplings $u \gsim 7$ (where the larger volumes $L \geq 20$ would produce $\be_s$ farther below zero).
When we move to $c = 0.2$ in the top-left plot we see that the optimization uncertainties are now non-zero, in accordance with \fig{fig:extrap02}.

Thanks to $L \geq 16$, in the top-left plot of \fig{fig:errorBudgets} the optimization uncertainties remain comparable to the statistical uncertainties.
This changes when $L = 12$ is included in the bottom row of plots.
Now the optimization uncertainties are much larger than the statistical uncertainties, and for $s = 2$ (bottom left) they dominate the total error budget.
The even larger extrapolation uncertainties for $s = 4 / 3$ (bottom right) are likely related to the difficulty resolving the slow flow of the coupling in such a small change of scale.

\newpage 

\begin{figure}[btp]
  \includegraphics[width=0.45\textwidth]{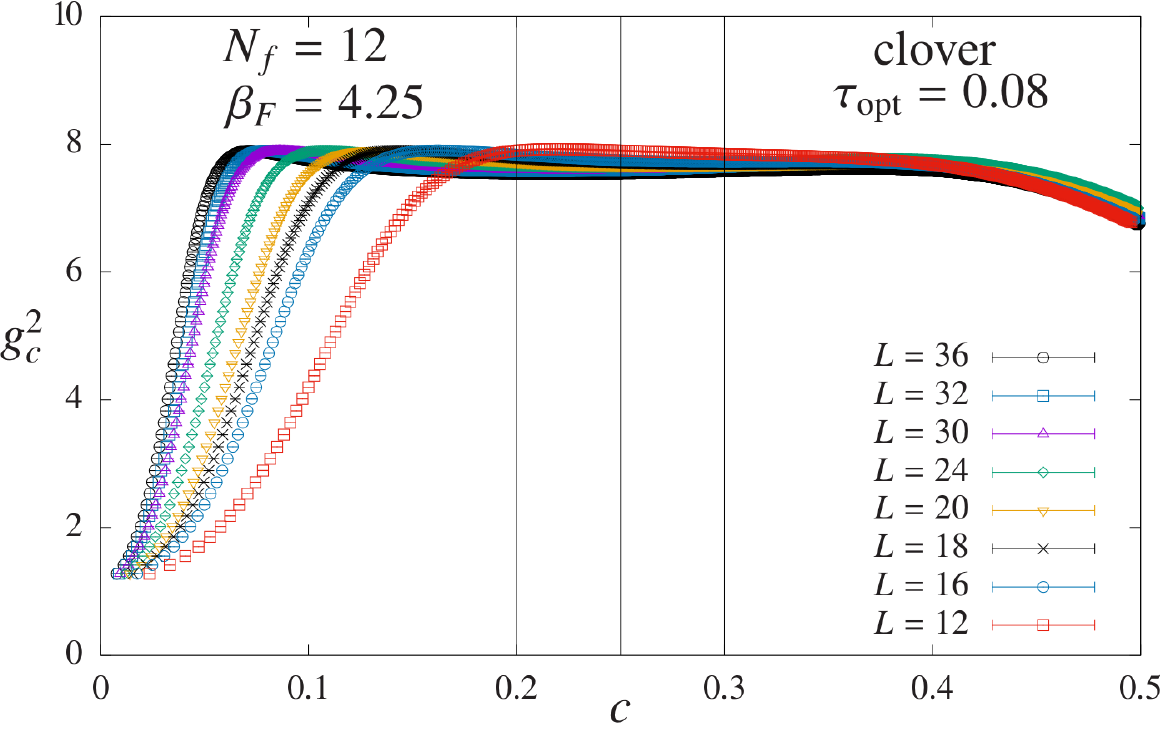}\hfill \includegraphics[width=0.45\textwidth]{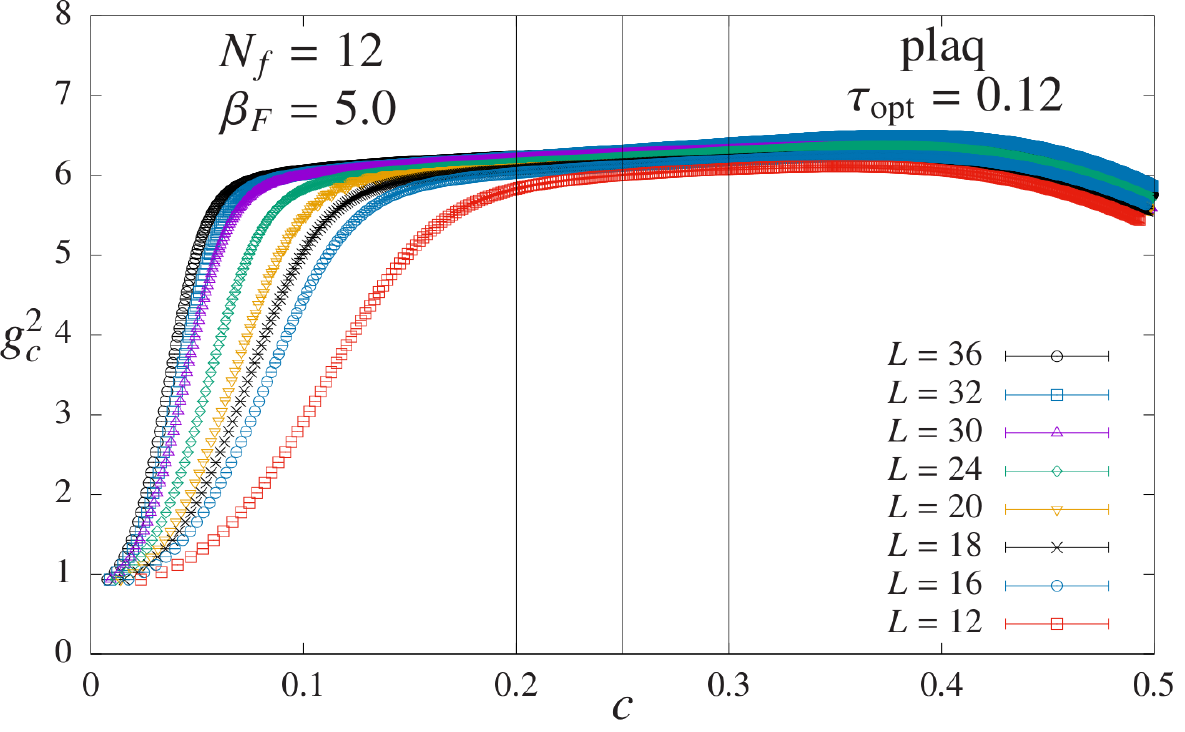}
  \caption{\label{fig:gcSq_vs_c}Renormalized couplings $\gtc(L)$ vs.\ the finite-volume gradient flow scale $c = \sqrt{8t} / L$ for all eight $L^4$ lattice volumes we study.  Two representative values of $\be_F$ are considered, $\be_F = 4.25$ for the clover discretization of the energy density (left) and $\be_F = 5$ for the plaquette discretization (right).  Lattice artifacts are non-negligible in the initial rise of the coupling from $c = 0$, and for $L = 12$ it is not clear whether this initial rise is complete for $c = 0.2$.  Larger values of $L$ are not affected by this issue at $c = 0.2$, while larger $c \geq 0.25$ are needed for $L = 12$, in agreement with the other results discussed in the text.  Vertical lines mark the $c = 0.2$, 0.25 and 0.3 that we analyze.}
\end{figure}

Finally, we can also go back to the basics and investigate the `raw data' going into our step-scaling analyses, namely the renormalized couplings $\gtc(L)$ as functions of the finite-volume gradient flow scale $c = \sqrt{8t} / L$.
Representative samples of these data are shown in \fig{fig:gcSq_vs_c}, for the clover discretization of the energy density $E(t)$ at $\be_F = 4.25$ and the plaquette discretization at $\be_F = 5$.
As $c \to 0$ for fixed $L$, the renormalized couplings are dominated by lattice artifacts and fall to unphysically small values.
The initial rise from $c = 0$ occurs as the gradient flow removes those short-distance cutoff effects, and we must ensure that these artifacts are sufficiently well removed for the values of $c$ at which we carry out our analyses.
Although \fig{fig:gcSq_vs_c} shows that $c = 0.2$ is acceptable for $L \geq 16$, for $L = 12$ it is not clear whether this initial rise is complete before $c = 0.2$.
Larger values of $c \geq 0.25$ appear to be needed for $L = 12$, in agreement with the other results discussed in the text.
The key conclusion is that $c = 0.2$ was a poor choice in our previous $L \geq 12$ study~\cite{Cheng:2014jba}, which we have now corrected in this work by using both larger $c$ and larger $L$.

\section{\label{app:data}Data sets and interpolations} 
Tables~\ref{tab:ensembles12}--\ref{tab:ensembles36} summarize the lattice ensembles considered in this work, with a separate table for each $L = 12$, 16, 18, 20, 24, 30, 32 and 36.
In all cases we use exactly massless fermions with anti-periodic BCs in all four directions, while the gauge fields are periodic.
For each ensemble specified by $L$ and the bare coupling $\be_F$, the tables report results for the renormalized gradient flow couplings $\gc(L; a)$ with $\tau_0 = 0$ and $\gtc(L; a)$ with the optimal $\topt = 0.08$, in both cases considering the clover discretization of $E(t)$ for two values of $c = 0.25$ and 0.3.
Tables~\ref{tab:ensembles02_12}--\ref{tab:ensembles02_32} contain the corresponding data for $c = 0.2$, while \mbox{tables~\ref{tab:plaq12}--\ref{tab:plaq36}} provide the corresponding data for the plaquette discretization of $E(t)$ for all three $c = 0.2$, 0.25 and 0.3.
All these results are obtained from the number of thermalized measurements listed in tables~\ref{tab:ensembles12}--\ref{tab:ensembles36}.
Each measurement is separated by ten molecular dynamics time units (MDTU) generated with the HMC algorithm, and combined into ten-measurement (100-MDTU) jackknife blocks to reduce autocorrelations.
The data in these tables provide all the necessary information for interested readers to reproduce our results or experiment with alternate systematic error analyses.
We also list the average plaquette (normalized to 3) in tables~\ref{tab:ensembles12}--\ref{tab:ensembles36}, to illustrate the roughness of the gauge fields.

Tables~\ref{tab:interp02}--\ref{tab:interp03} summarize the quality of the $\gtc(L; a)$ interpolations vs.\ $\be_F$ on \mbox{each lattice} volume, with a separate table for each $c = 0.2$, 0.25 and 0.3.
Considering both \mbox{the rational} function interpolations that produce our final results (\eq{eq:pade}) \mbox{as well as the polynomial} \mbox{interpolations that we use} to check for potential systematic effects from our choice of interpolating function (\eq{eq:poly}), each table records the $\chi^2$, the number of degrees of \mbox{freedom (d.o.f.)} and the corresponding confidence level
\begin{equation}
  \label{eq:CL}
  \mbox{CL} = 1 - P(a, x) = \frac{1}{\Ga(a)} \int_x^{\infty} dt\ e^{-t}\ t^{a - 1},
\end{equation}
where $a = \mbox{d.o.f.} / 2$ and $x = \chi^2 / 2$.

\begin{table}[htbp]
  \centering
  \begin{tabular}{cc|cc|cc|c}
    \hline
    $\be_F$ & Meas. & \gceq{0.25}    & \gtceq{0.25}     & \gceq{0.3}     & \gtceq{0.3}      & Plaq.         \\
            &       & ($\tau_0 = 0$) & ($\topt = 0.08$) & ($\tau_0 = 0$) & ($\topt = 0.08$) &               \\
    \hline
     3.40   &  910  & 11.688(12)     & 10.052(12)       & 10.919(18)     & 9.826(17)        & 0.868310(54)  \\
     3.60   &  950  & 11.041(12)     &  9.502(11)       & 10.349(18)     & 9.317(17)        & 0.915143(41)  \\
     3.75   & 1020  & 10.603(9)      &  9.131(9)        &  9.962(13)     & 8.971(12)        & 0.950552(39)  \\
     3.80   &  970  & 10.449(9)      &  9.000(9)        &  9.827(14)     & 8.850(13)        & 0.962506(41)  \\
     4.00   &  910  &  9.856(9)      &  8.497(8)        &  9.314(13)     & 8.393(12)        & 1.010129(45)  \\
     4.20   &  940  &  9.315(10)     &  8.041(9)        &  8.853(14)     & 7.983(14)        & 1.057876(59)  \\
     4.25   & 1060  &  9.153(9)      &  7.901(8)        &  8.704(12)     & 7.848(12)        & 1.070025(40)  \\
     4.40   &  950  &  8.731(10)     &  7.542(9)        &  8.326(15)     & 7.509(14)        & 1.106201(65)  \\
     4.50   &  960  &  8.466(8)      &  7.317(8)        &  8.100(12)     & 7.309(12)        & 1.130441(43)  \\
     4.60   &  940  &  8.187(9)      &  7.078(8)        &  7.839(13)     & 7.073(12)        & 1.154627(57)  \\
     4.75   &  920  &  7.787(8)      &  6.738(8)        &  7.485(13)     & 6.757(12)        & 1.190920(44)  \\
     4.80   &  890  &  7.647(7)      &  6.617(7)        &  7.349(11)     & 6.634(10)        & 1.202951(54)  \\
     5.00   &  970  &  7.159(8)      &  6.202(7)        &  6.919(12)     & 6.250(12)        & 1.251122(65)  \\
     5.20   &  970  &  6.667(8)      &  5.781(8)        &  6.471(13)     & 5.847(12)        & 1.298844(68)  \\
     5.40   &  950  &  6.204(7)      &  5.384(6)        &  6.043(10)     & 5.463(9)         & 1.346092(69)  \\
     5.50   &  980  &  5.975(6)      &  5.185(6)        &  5.821(10)     & 5.262(9)         & 1.369364(46)  \\
     5.60   &  970  &  5.772(7)      &  5.013(6)        &  5.641(11)     & 5.102(10)        & 1.392192(64)  \\
     5.80   &  960  &  5.355(5)      &  4.653(5)        &  5.247(8)      & 4.746(7)         & 1.437232(63)  \\
     6.00   &  910  &  4.977(6)      &  4.326(6)        &  4.882(9)      & 4.416(9)         & 1.480518(56)  \\
     6.20   &  960  &  4.658(5)      &  4.052(5)        &  4.593(8)      & 4.158(8)         & 1.522388(53)  \\
     6.40   &  960  &  4.343(5)      &  3.780(4)        &  4.287(7)      & 3.881(7)         & 1.562354(48)  \\
     6.50   & 1050  &  4.185(4)      &  3.642(4)        &  4.130(6)      & 3.738(5)         & 1.581715(40)  \\
     6.60   &  970  &  4.060(4)      &  3.535(4)        &  4.016(7)      & 3.636(7)         & 1.600593(50)  \\
     6.80   &  950  &  3.796(4)      &  3.305(4)        &  3.754(6)      & 3.399(6)         & 1.636906(69)  \\
     7.00   &  970  &  3.570(4)      &  3.109(4)        &  3.534(6)      & 3.200(6)         & 1.671457(67)  \\
     7.20   &  960  &  3.374(3)      &  2.940(3)        &  3.349(5)      & 3.033(5)         & 1.704372(71)  \\
     7.40   &  970  &  3.185(4)      &  2.775(4)        &  3.158(6)      & 2.860(6)         & 1.735602(103) \\
     7.50   & 1070  &  3.098(3)      &  2.700(3)        &  3.077(5)      & 2.787(5)         & 1.750695(45)  \\
     7.60   &  970  &  3.012(3)      &  2.624(3)        &  2.989(5)      & 2.707(5)         & 1.765354(97)  \\
     7.80   &  940  &  2.868(3)      &  2.500(3)        &  2.850(5)      & 2.581(5)         & 1.793602(71)  \\
     8.00   &  970  &  2.733(2)      &  2.382(2)        &  2.719(4)      & 2.463(4)         & 1.820579(49)  \\
     8.50   &  940  &  2.435(2)      &  2.123(2)        &  2.425(3)      & 2.197(3)         & 1.882565(52)  \\
     9.00   &  970  &  2.198(2)      &  1.917(2)        &  2.189(3)      & 1.984(3)         & 1.937947(46)  \\
    \hline
  \end{tabular}
  \caption{\label{tab:ensembles12}$12^4$ lattice ensembles used in this work.  For each bare coupling $\be_F$ we report the renormalized gradient flow couplings for the clover discretization of $E(t)$ at two values of $c = 0.25$ and 0.3 with both $\tau_0 = 0$ and the optimal $\topt = 0.08$, all obtained from the given number of thermalized measurements.  The thermalized measurements are separated by ten molecular dynamics time units (MDTU) generated with the HMC algorithm, and combined into ten-measurement (100-MDTU) jackknife blocks to reduce autocorrelations.  We also list the average plaquette (normalized to 3), to illustrate the roughness of the gauge fields.} 
\end{table}

\begin{table}[htbp]
  \centering
\begin{tabular}{cc|cc|cc|c}
    \hline
    $\be_F$ & Meas. & \gceq{0.25}    & \gtceq{0.25}     & \gceq{0.3}     & \gtceq{0.3}      & Plaq.        \\
            &       & ($\tau_0 = 0$) & ($\topt = 0.08$) & ($\tau_0 = 0$) & ($\topt = 0.08$) &              \\
    \hline
     3.00   & 970   & 11.503(12)     & 10.568(12)       & 10.989(20)     & 10.364(19)       & 0.775557(25) \\
     3.20   & 960   & 10.969(13)     & 10.082(13)       & 10.505(21)     &  9.909(21)       & 0.821721(29) \\
     3.40   & 950   & 10.418(14)     &  9.577(13)       &  9.989(21)     &  9.423(20)       & 0.868279(33) \\
     3.60   & 920   &  9.957(11)     &  9.158(11)       &  9.593(18)     &  9.052(17)       & 0.915119(34) \\
     3.75   & 950   &  9.585(12)     &  8.818(12)       &  9.252(19)     &  8.731(19)       & 0.950548(29) \\
     3.80   & 920   &  9.470(12)     &  8.712(11)       &  9.141(18)     &  8.626(18)       & 0.962439(26) \\
\hline
  \end{tabular}
  \caption{\label{tab:ensembles16}$16^4$ lattice ensembles used in this work, with columns as in \protect\tab{tab:ensembles12}.} 
\end{table}

\begin{table}[htbp]
  \centering
  \begin{tabular}{cc|cc|cc|c}
    \hline
    $\be_F$ & Meas. & \gceq{0.25}    & \gtceq{0.25}     & \gceq{0.3}     & \gtceq{0.3}      & Plaq.        \\
            &       & ($\tau_0 = 0$) & ($\topt = 0.08$) & ($\tau_0 = 0$) & ($\topt = 0.08$) &              \\
    \hline
     3.00   &  940  & 11.085(20)     & 10.375(20)       & 10.703(30)     & 10.223(29)       & 0.775545(36) \\
     3.50   &  960  &  9.841(12)     &  9.215(12)       &  9.543(19)     &  9.118(19)       & 0.891614(20) \\
     3.75   &  970  &  9.287(11)     &  8.699(11)       &  9.027(18)     &  8.625(18)       & 0.950557(20) \\
     4.00   & 1960  &  8.755(9)      &  8.204(9)        &  8.546(15)     &  8.167(14)       & 1.010030(20) \\
     4.25   & 1970  &  8.224(8)      &  7.708(7)        &  8.048(12)     &  7.692(12)       & 1.070044(17) \\
     4.50   & 1970  &  7.694(7)      &  7.214(7)        &  7.554(11)     &  7.221(11)       & 1.130368(16) \\
\hline
  \end{tabular}
  \caption{\label{tab:ensembles18}$18^4$ lattice ensembles used in this work, with columns as in \protect\tab{tab:ensembles12}.} 
\end{table}

\begin{table}[htbp]
  \centering
  \begin{tabular}{cc|cc|cc|c}
    \hline
    $\be_F$ & Meas. & \gceq{0.25}    & \gtceq{0.25}      & \gceq{0.3}     & \gtceq{0.3}      & Plaq.        \\
            &       & ($\tau_0 = 0$) & ($\topt = 0.08$)  & ($\tau_0 = 0$) & ($\topt = 0.08$) &              \\
    \hline
     3.75   & 600   & 9.108(13)      & 8.641(13)         & 8.934(20)      & 8.612(20)        & 0.950548(19) \\
     4.00   & 610   & 8.595(14)      & 8.156(14)         & 8.445(22)      & 8.142(22)        & 1.010043(21) \\
     4.25   & 500   & 8.080(13)      & 7.668(13)         & 7.958(21)      & 7.673(20)        & 1.069997(23) \\
     4.50   & 620   & 7.595(11)      & 7.210(11)         & 7.507(18)      & 7.238(18)        & 1.130303(33) \\
     4.75   & 580   & 7.064(11)      & 6.707(10)         & 6.982(17)      & 6.732(17)        & 1.190868(27) \\
     5.00   & 680   & 6.578(13)      & 6.246(13)         & 6.517(21)      & 6.285(20)        & 1.251121(33) \\
     5.50   & 540   & 5.647(12)      & 5.364(11)         & 5.632(18)      & 5.433(17)        & 1.369242(48) \\
     6.00   & 530   & 4.792(9)       & 4.553(9)          & 4.785(14)      & 4.616(14)        & 1.480568(29) \\
     6.50   & 510   & 4.072(7)       & 3.869(7)          & 4.064(12)      & 3.920(12)        & 1.581665(30) \\
     7.00   & 550   & 3.531(7)       & 3.356(7)          & 3.544(11)      & 3.419(10)        & 1.671436(40) \\
     7.50   & 550   & 3.064(6)       & 2.913(6)          & 3.070(10)      & 2.962(10)        & 1.750695(30) \\
     8.00   & 540   & 2.718(5)       & 2.584(5)          & 2.726(8)       & 2.630(8)         & 1.820554(38) \\
     8.50   & 550   & 2.430(5)       & 2.310(5)          & 2.436(8)       & 2.350(7)         & 1.882595(35) \\
     9.00   & 540   & 2.193(3)       & 2.084(3)          & 2.195(6)       & 2.117(6)         & 1.937907(32) \\
    \hline
  \end{tabular}
  \caption{\label{tab:ensembles20}$20^4$ lattice ensembles used in this work, with columns as in \protect\tab{tab:ensembles12}.} 
\end{table}

\begin{table}[htbp]
  \centering
  \begin{tabular}{cc|cc|cc|c}
    \hline
    $\be_F$ & Meas. & \gceq{0.25}    & \gtceq{0.25}     & \gceq{0.3}     & \gtceq{0.3}      & Plaq.        \\
            &       & ($\tau_0 = 0$) & ($\topt = 0.08$) & ($\tau_0 = 0$) & ($\topt = 0.08$) &              \\
    \hline
     3.75   & 540   & 8.816(15)      & 8.500(15)        & 8.688(23)      & 8.470(23)        & 0.950567(15) \\
     4.00   & 580   & 8.365(15)      & 8.066(15)        & 8.279(24)      & 8.072(24)        & 1.010030(16) \\
     4.25   & 590   & 7.929(19)      & 7.648(19)        & 7.884(30)      & 7.688(30)        & 1.070018(20) \\
     4.50   & 550   & 7.439(14)      & 7.175(14)        & 7.381(23)      & 7.197(22)        & 1.130384(15) \\
     4.75   & 520   & 6.992(15)      & 6.746(15)        & 6.976(25)      & 6.803(25)        & 1.190821(18) \\
     5.00   & 570   & 6.506(13)      & 6.278(13)        & 6.497(21)      & 6.336(21)        & 1.251152(21) \\
\hline
  \end{tabular}
  \caption{\label{tab:ensembles24}$24^4$ lattice ensembles used in this work, with columns as in \protect\tab{tab:ensembles12}.} 
\end{table}

\begin{table}[htbp]
  \centering
  \begin{tabular}{cc|cc|cc|c}
    \hline
    $\be_F$ & Meas. & \gceq{0.25}    & \gtceq{0.25}     & \gceq{0.3}     & \gtceq{0.3}      & Plaq.        \\
            &       & ($\tau_0 = 0$) & ($\topt = 0.08$) & ($\tau_0 = 0$) & ($\topt = 0.08$) &              \\
    \hline
     3.75   & 470   & 8.621(18)      & 8.423(18)        & 8.562(30)      & 8.424(30)        & 0.950577(11) \\
     4.00   & 510   & 8.211(22)      & 8.024(22)        & 8.187(33)      & 8.056(33)        & 1.010042(11) \\
     4.25   & 560   & 7.778(18)      & 7.601(18)        & 7.761(29)      & 7.637(29)        & 1.070013(9)  \\
     4.50   & 550   & 7.323(21)      & 7.157(21)        & 7.323(34)      & 7.207(34)        & 1.130373(10) \\
     4.75   & 560   & 6.898(16)      & 6.742(15)        & 6.902(25)      & 6.792(25)        & 1.190859(9)  \\
     5.00   & 550   & 6.446(14)      & 6.300(14)        & 6.444(22)      & 6.341(22)        & 1.251157(10) \\
     5.50   & 550   & 5.593(13)      & 5.467(13)        & 5.629(21)      & 5.540(21)        & 1.369320(9)  \\
     6.00   & 520   & 4.765(12)      & 4.658(12)        & 4.784(20)      & 4.708(19)        & 1.480543(11) \\
     6.50   & 500   & 4.107(11)      & 4.015(11)        & 4.138(18)      & 4.072(18)        & 1.581666(12) \\
     7.00   & 530   & 3.542(9)       & 3.462(9)         & 3.557(13)      & 3.501(13)        & 1.671488(12) \\
     7.50   & 540   & 3.103(6)       & 3.034(6)         & 3.128(10)      & 3.079(10)        & 1.750674(10) \\
     8.00   & 510   & 2.749(7)       & 2.688(7)         & 2.768(10)      & 2.725(10)        & 1.820560(10) \\
     8.50   & 520   & 2.459(6)       & 2.404(6)         & 2.478(9)       & 2.439(9)         & 1.882567(9)  \\
     9.00   & 520   & 2.209(5)       & 2.159(5)         & 2.215(8)       & 2.179(7)         & 1.937922(10) \\
    \hline
  \end{tabular}
  \caption{\label{tab:ensembles30}$30^4$ lattice ensembles used in this work, with columns as in \protect\tab{tab:ensembles12}.} 
\end{table}

\begin{table}[htbp]
  \centering
  \begin{tabular}{cc|cc|cc|c}
    \hline
    $\be_F$ & Meas. & \gceq{0.25}    & \gtceq{0.25}     & \gceq{0.3}     & \gtceq{0.3}      & Plaq.        \\
            &       & ($\tau_0 = 0$) & ($\topt = 0.08$) & ($\tau_0 = 0$) & ($\topt = 0.08$) &              \\
    \hline
     3.75   & 440   & 8.610(24)      & 8.437(23)        & 8.562(37)      & 8.441(37)        & 0.950563(12) \\
     4.00   & 950   & 8.166(16)      & 8.002(16)        & 8.134(26)      & 8.019(25)        & 1.010050(7)  \\
     4.25   & 420   & 7.755(19)      & 7.600(19)        & 7.758(32)      & 7.649(32)        & 1.070015(9)  \\
     4.50   & 960   & 7.320(15)      & 7.174(15)        & 7.336(25)      & 7.234(25)        & 1.130357(7)  \\
     4.75   & 950   & 6.873(13)      & 6.736(13)        & 6.890(20)      & 6.794(20)        & 1.190867(9)  \\
     5.00   & 960   & 6.469(14)      & 6.341(14)        & 6.512(22)      & 6.421(22)        & 1.251145(7)  \\
     5.25   & 930   & 6.007(13)      & 5.888(13)        & 6.043(22)      & 5.959(21)        & 1.310802(9)  \\
     5.50   & 960   & 5.576(12)      & 5.466(12)        & 5.608(19)      & 5.530(19)        & 1.369291(8)  \\
     5.75   & 950   & 5.176(11)      & 5.073(11)        & 5.211(17)      & 5.138(17)        & 1.426046(9)  \\
     6.00   & 940   & 4.800(7)       & 4.705(7)         & 4.847(12)      & 4.780(12)        & 1.480556(7)  \\
     6.50   & 950   & 4.095(8)       & 4.014(8)         & 4.117(13)      & 4.059(13)        & 1.581660(9)  \\
     7.00   & 950   & 3.573(7)       & 3.503(7)         & 3.615(12)      & 3.565(12)        & 1.671509(8)  \\
     8.00   & 920   & 2.743(5)       & 2.688(5)         & 2.756(9)       & 2.718(9)         & 1.820556(9)  \\
     9.00   & 540   & 2.225(6)       & 2.181(6)         & 2.240(9)       & 2.209(9)         & 1.937924(9)  \\
    \hline
  \end{tabular}
  \caption{\label{tab:ensembles32}$32^4$ lattice ensembles used in this work, with columns as in \protect\tab{tab:ensembles12}.} 
\end{table}

\begin{table}[htbp]
  \centering
  \begin{tabular}{cc|cc|cc|c}
    \hline
    $\be_F$ & Meas. & \gceq{0.25}    & \gtceq{0.25}     & \gceq{0.3}     & \gtceq{0.3}      & Plaq.        \\
            &       & ($\tau_0 = 0$) & ($\topt = 0.08$) & ($\tau_0 = 0$) & ($\topt = 0.08$) &              \\
    \hline
     3.75   & 350   & 8.413(22)      & 8.278(21)        & 8.333(36)      & 8.239(36)        & 0.950560(7)  \\
     4.00   & 390   & 8.097(24)      & 7.969(23)        & 8.086(34)      & 7.996(33)        & 1.010043(10) \\
     4.25   & 460   & 7.668(20)      & 7.547(19)        & 7.663(31)      & 7.578(31)        & 1.070014(7)  \\
     4.50   & 550   & 7.297(20)      & 7.182(19)        & 7.328(30)      & 7.247(30)        & 1.130365(7)  \\
     4.75   & 510   & 6.855(17)      & 6.747(17)        & 6.865(26)      & 6.789(26)        & 1.190839(7)  \\
     5.00   & 550   & 6.423(18)      & 6.322(18)        & 6.454(29)      & 6.383(29)        & 1.251143(9)  \\
     5.50   & 550   & 5.612(14)      & 5.525(14)        & 5.657(21)      & 5.594(21)        & 1.369282(9)  \\
     6.00   & 560   & 4.817(12)      & 4.742(12)        & 4.865(19)      & 4.811(19)        & 1.480557(11) \\
     6.50   & 550   & 4.114(14)      & 4.050(13)        & 4.145(20)      & 4.099(20)        & 1.581667(10) \\
     7.00   & 390   & 3.577(10)      & 3.521(10)        & 3.606(16)      & 3.567(16)        & 1.671496(10) \\
     7.50   & 390   & 3.111(10)      & 3.062(10)        & 3.138(16)      & 3.103(15)        & 1.750672(9)  \\
     8.00   & 560   & 2.770(8)       & 2.726(8)         & 2.794(12)      & 2.764(12)        & 1.820547(9)  \\
     8.50   & 380   & 2.470(7)       & 2.431(7)         & 2.489(11)      & 2.462(11)        & 1.882548(9)  \\
     9.00   & 390   & 2.239(7)       & 2.204(7)         & 2.255(10)      & 2.230(10)        & 1.937906(8)  \\
    \hline
  \end{tabular}
  \caption{\label{tab:ensembles36}$36^4$ lattice ensembles used in this work, with columns as in \protect\tab{tab:ensembles12}.} 
\end{table}

\begin{table}[htbp]
  \centering
  \begin{tabular}{ccc|ccc}
    \hline
    \multicolumn{3}{c|}{$L = 12$}               & \multicolumn{3}{|c}{$L = 16$}               \\
    $\be_F$ & \gceq{0.2}     & \gtceq{0.2}      & $\be_F$ & \gceq{0.2}     & \gtceq{0.2}      \\
            & ($\tau_0 = 0$) & ($\topt = 0.08$) &         & ($\tau_0 = 0$) & ($\topt = 0.08$) \\
    \hline
            &                &                  &  3.00   & 12.551(7)      & 10.980(7)        \\
            &                &                  &  3.20   & 11.915(8)      & 10.430(7)        \\
     3.40   & 12.838(7)      & 10.268(6)        &  3.40   & 11.286(8)      &  9.885(8)        \\
     3.60   & 12.092(7)      &  9.673(6)        &  3.60   & 10.718(6)      &  9.396(6)        \\
     3.75   & 11.566(6)      &  9.256(5)        &  3.75   & 10.282(6)      &  9.018(6)        \\
\hline
  \end{tabular}
  \caption{\label{tab:ensembles02_12}Renormalized gradient flow couplings in the $c = 0.2$ scheme for the $12^4$ and $16^4$ lattice ensembles used in this work, for the clover discretization of $E(t)$ with both $\tau_0 = 0$ and the optimal $\topt = 0.08$.}
\end{table}

\begin{table}[htbp]
  \centering
  \begin{tabular}{ccc|ccc}
    \hline
    \multicolumn{3}{c|}{$L = 18$}               & \multicolumn{3}{|c}{$L = 20$}               \\
    $\be_F$ & \gceq{0.2}     & \gtceq{0.2}      & $\be_F$ & \gceq{0.2}     & \gtceq{0.2}      \\
            & ($\tau_0 = 0$) & ($\topt = 0.08$) &         & ($\tau_0 = 0$) & ($\topt = 0.08$) \\
    \hline
     3.00   & 11.912(12)     & 10.725(11)       &         &                &                  \\
     3.50   & 10.487(7)      &  9.452(6)        &         &                &                  \\
     3.75   &  9.841(6)      &  8.877(6)        &  3.75   & 9.535(7)       & 8.774(7)         \\
     4.00   &  9.223(5)      &  8.325(4)        &  4.00   & 8.961(7)       & 8.250(7)         \\
     4.25   &  8.615(4)      &  7.782(4)        &  4.25   & 8.386(7)       & 7.724(7)         \\
     4.50   &  8.018(4)      &  7.247(3)        &  4.50   & 7.839(6)       & 7.224(6)         \\
     4.75   &  7.431(3)      &  6.721(3)        &  4.75   & 7.273(6)       & 6.705(6)         \\
     5.00   &  6.864(3)      &  6.212(3)        &  5.00   & 6.740(7)       & 6.217(7)         \\
     5.50   &  5.812(3)      &  5.267(3)        &  5.50   & 5.731(6)       & 5.291(6)         \\
     6.00   &  4.882(4)      &  4.428(4)        &  6.00   & 4.840(4)       & 4.471(4)         \\
     6.50   &  4.128(3)      &  3.746(3)        &  6.50   & 4.103(3)       & 3.791(3)         \\
     7.00   &  3.539(3)      &  3.213(3)        &  7.00   & 3.536(4)       & 3.269(4)         \\
     7.50   &  3.083(2)      &  2.800(2)        &  7.50   & 3.070(3)       & 2.839(3)         \\
     8.00   &  2.717(1)      &  2.468(1)        &  8.00   & 2.717(2)       & 2.513(2)         \\
     8.50   &  2.429(2)      &  2.206(2)        &  8.50   & 2.429(2)       & 2.246(2)         \\
     9.00   &  2.197(2)      &  1.996(2)        &  9.00   & 2.193(1)       & 2.028(1)         \\
    \hline
  \end{tabular}
  \caption{\label{tab:ensembles02_18}Renormalized gradient flow couplings in the $c = 0.2$ scheme for the $18^4$ and $20^4$ lattice ensembles used in this work, with columns as in \protect\tab{tab:ensembles02_12}.}
\end{table}

\begin{table}[htbp]
  \centering
  \begin{tabular}{ccc|ccc}
    \hline
    \multicolumn{3}{c|}{$L = 24$}               & \multicolumn{3}{|c}{$L = 30$}               \\
    $\be_F$ & \gceq{0.2}     & \gtceq{0.2}      & $\be_F$ & \gceq{0.2}     & \gtceq{0.2}      \\
            & ($\tau_0 = 0$) & ($\topt = 0.08$) &         & ($\tau_0 = 0$) & ($\topt = 0.08$) \\
    \hline
     3.75   & 9.117(8)       & 8.609(8)         &  3.75   & 8.796(10)      & 8.482(10)        \\
     4.00   & 8.606(9)       & 8.129(9)         &  4.00   & 8.342(12)      & 8.045(12)        \\
     4.25   & 8.113(10)      & 7.666(10)        &  4.25   & 7.883(10)      & 7.604(10)        \\
     4.50   & 7.595(8)       & 7.179(7)         &  4.50   & 7.400(11)      & 7.139(11)        \\
     4.75   & 7.101(7)       & 6.714(7)         &  4.75   & 6.948(8)       & 6.704(8)         \\
\hline
  \end{tabular}
  \caption{\label{tab:ensembles02_24}Renormalized gradient flow couplings in the $c = 0.2$ scheme for the $24^4$ and $30^4$ lattice ensembles used in this work, with columns as in \protect\tab{tab:ensembles02_12}.}
\end{table}

\begin{table}[htbp]
  \centering
  \begin{tabular}{ccc|ccc}
    \hline
    \multicolumn{3}{c|}{$L = 32$}               & \multicolumn{3}{|c}{$L = 36$}               \\
    $\be_F$ & \gceq{0.2}     & \gtceq{0.2}      & $\be_F$ & \gceq{0.2}     & \gtceq{0.2}      \\
            & ($\tau_0 = 0$) & ($\topt = 0.08$) &         & ($\tau_0 = 0$) & ($\topt = 0.08$) \\
    \hline
     3.75   & 8.749(13)      &  8.475(13)       &  3.75   &  8.565(11)     & 8.352(11)        \\
     4.00   & 8.281(9)       &  8.022(9)        &  4.00   &  8.183(13)     & 7.981(13)        \\
     4.25   & 7.830(11)      &  7.587(10)       &  4.25   &  7.734(11)     & 7.544(11)        \\
     4.50   & 7.375(8)       &  7.146(8)        &  4.50   &  7.323(11)     & 7.144(11)        \\
     4.75   & 6.911(7)       &  6.697(7)        &  4.75   &  6.880(10)     & 6.711(10)        \\
     5.00   & 6.473(7)       &  6.274(7)        &  5.00   &  6.430(10)     & 6.273(10)        \\
     5.25   & 6.010(7)       &  5.825(7)        &         &                &                  \\
     5.50   & 5.572(6)       &  5.401(6)        &  5.50   &  5.582(8)      & 5.447(8)         \\
     5.75   & 5.160(6)       &  5.002(6)        &         &                &                  \\
     6.00   & 4.772(4)       &  4.626(4)        &  6.00   &  4.781(6)      & 4.665(6)         \\
     6.50   & 4.077(4)       &  3.953(4)        &  6.50   &  4.087(7)      & 3.988(7)         \\
     7.00   & 3.537(3)       &  3.430(3)        &  7.00   &  3.545(5)      & 3.460(5)         \\
            &                &                  &  7.50   &  3.085(5)      & 3.011(5)         \\
     8.00   & 2.726(3)       &  2.643(3)        &  8.00   &  2.743(4)      & 2.677(4)         \\
            &                &                  &  8.50   &  2.450(4)      & 2.391(4)         \\
     9.00   & 2.210(3)       &  2.143(3)        &  9.00   &  2.220(3)      & 2.167(3)         \\
    \hline
  \end{tabular}
  \caption{\label{tab:ensembles02_32}Renormalized gradient flow couplings in the $c = 0.2$ scheme for the $32^4$ and $36^4$ lattice ensembles used in this work, with columns as in \protect\tab{tab:ensembles02_12}.}
\end{table}

\clearpage 

\begin{table}[htbp]
  \centering
  \begin{tabular}{c|cc|cc|cc}
    \hline
    $\be_F$ & \gceq{0.2}     & \gtceq{0.2}      & \gceq{0.25}    & \gtceq{0.25}     & \gceq{0.3}     & \gtceq{0.3}      \\
            & ($\tau_0 = 0$) & ($\topt = 0.12$) & ($\tau_0 = 0$) & ($\topt = 0.12$) & ($\tau_0 = 0$) & ($\topt = 0.12$) \\
    \hline
     3.40   & 16.544(9)      & 10.662(7)        & 12.999(13)       & 10.023(11)     & 11.507(18)     & 9.701(16)        \\
     3.60   & 15.366(8)      &  9.938(6)        & 12.210(12)       &  9.435(11)     & 10.878(18)     & 9.180(16)        \\
     3.75   & 14.541(7)      &  9.439(5)        & 11.678(9)        &  9.040(8)      & 10.455(13)     & 8.827(12)        \\
     3.80   & 14.267(6)      &  9.270(5)        & 11.492(9)        &  8.902(8)      & 10.307(13)     & 8.704(13)        \\
     4.00   & 13.204(6)      &  8.619(5)        & 10.781(9)        &  8.372(8)      &  9.744(12)     & 8.238(11)        \\
     4.20   & 12.219(7)      &  8.019(5)        & 10.133(10)       &  7.891(9)      &  9.239(14)     & 7.821(13)        \\
\hline
  \end{tabular}
  \caption{\label{tab:plaq12}Renormalized gradient flow couplings based on the plaquette discretization of $E(t)$ for the $12^4$ lattice ensembles used in this work at three values of $c = 0.2$, 0.25 and 0.3, with both $\tau_0 = 0$ and the optimal $\topt = 0.12$ as in \protect\tab{tab:ensembles02_12}.}
\end{table}

\begin{table}[htbp]
  \centering
  \begin{tabular}{c|cc|cc|cc}
    \hline
    $\be_F$ & \gceq{0.2}     & \gtceq{0.2}      & \gceq{0.25}    & \gtceq{0.25}     & \gceq{0.3}     & \gtceq{0.3}      \\
            & ($\tau_0 = 0$) & ($\topt = 0.12$) & ($\tau_0 = 0$) & ($\topt = 0.12$) & ($\tau_0 = 0$) & ($\topt = 0.12$) \\
    \hline
     3.00   & 13.811(8)      & 11.015(7)        & 12.016(12)     & 10.500(11)       & 11.258(19)     & 10.278(19)       \\
     3.20   & 13.046(8)      & 10.424(7)        & 11.434(13)     &  9.998(12)       & 10.749(21)     &  9.816(20)       \\
     3.40   & 12.303(9)      &  9.846(8)        & 10.842(14)     &  9.484(13)       & 10.211(21)     &  9.326(20)       \\
     3.60   & 11.632(7)      &  9.328(6)        & 10.343(11)     &  9.056(10)       &  9.797(17)     &  8.952(16)       \\
     3.75   & 11.122(6)      &  8.931(6)        &  9.943(11)     &  8.710(11)       &  9.441(18)     &  8.629(18)       \\
     3.80   & 10.964(7)      &  8.809(6)        &  9.820(11)     &  8.604(11)       &  9.327(18)     &  8.525(17)       \\
\hline
  \end{tabular}
  \caption{\label{tab:plaq16}Renormalized gradient flow couplings based on the plaquette discretization of $E(t)$ for the $16^4$ lattice ensembles used in this work, with columns as in \protect\tab{tab:plaq12}.}
\end{table}

\begin{table}[htbp]
  \centering
  \addtolength{\tabcolsep}{-0.25 pt}    
  \begin{tabular}{c|cc|cc|cc}
    \hline
    $\be_F$ & \gceq{0.2}     & \gtceq{0.2}      & \gceq{0.25}    & \gtceq{0.25}     & \gceq{0.3}     & \gtceq{0.3}      \\
            & ($\tau_0 = 0$) & ($\topt = 0.12$) & ($\tau_0 = 0$) & ($\topt = 0.12$) & ($\tau_0 = 0$) & ($\topt = 0.12$) \\
    \hline
     3.00   & 12.683(12)     & 10.691(11)       & 11.428(20)     & 10.303(19)       & 10.892(30)     & 10.148(29)       \\
     3.50   & 11.087(7)      &  9.369(6)        & 10.113(12)     &  9.125(11)       &  9.694(19)     &  9.036(18)       \\
     3.75   & 10.371(6)      &  8.777(6)        &  9.530(11)     &  8.604(10)       &  9.163(17)     &  8.541(17)       \\
     4.00   &  9.688(5)      &  8.211(4)        &  8.972(9)      &  8.105(8)        &  8.667(14)     &  8.082(14)       \\
     4.25   &  9.022(4)      &  7.657(4)        &  8.415(7)      &  7.606(7)        &  8.155(12)     &  7.607(11)       \\
     4.50   &  8.370(4)      &  7.114(3)        &  7.862(7)      &  7.110(6)        &  7.648(11)     &  7.136(10)       \\
\hline
  \end{tabular}
  \caption{\label{tab:plaq18}Renormalized gradient flow couplings based on the plaquette discretization of $E(t)$ for the $18^4$ lattice ensembles used in this work, with columns as in \protect\tab{tab:plaq12}.}
\end{table}

\begin{table}[htbp]
  \centering
  \begin{tabular}{c|cc|cc|cc}
    \hline
    $\be_F$ & \gceq{0.2}     & \gtceq{0.2}      & \gceq{0.25}    & \gtceq{0.25}     & \gceq{0.3}     & \gtceq{0.3}      \\
            & ($\tau_0 = 0$) & ($\topt = 0.12$) & ($\tau_0 = 0$) & ($\topt = 0.12$) & ($\tau_0 = 0$) & ($\topt = 0.12$) \\
    \hline
     3.75   & 9.901(7)       & 8.678(7)         & 9.287(12)      & 8.560(12)        & 9.036(20)      & 8.543(19)        \\
     4.00   & 9.286(7)       & 8.146(7)         & 8.756(13)      & 8.072(13)        & 8.537(21)      & 8.072(21)        \\
     4.25   & 8.672(7)       & 7.614(7)         & 8.222(12)      & 7.583(12)        & 8.039(20)      & 7.603(20)        \\
     4.50   & 8.089(6)       & 7.109(6)         & 7.721(11)      & 7.124(11)        & 7.579(18)      & 7.169(17)        \\
     4.75   & 7.490(6)       & 6.587(5)         & 7.175(10)      & 6.621(10)        & 7.045(17)      & 6.664(16)        \\
     5.00   & 6.926(7)       & 6.097(7)         & 6.674(13)      & 6.161(12)        & 6.573(20)      & 6.218(20)        \\
\hline
  \end{tabular}
  \caption{\label{tab:plaq20}Renormalized gradient flow couplings based on the plaquette discretization of $E(t)$ for the $20^4$ lattice ensembles used in this work, with columns as in \protect\tab{tab:plaq12}.}
\end{table}

\begin{table}[htbp]
  \centering
    \begin{tabular}{c|cc|cc|cc}
      \hline
      $\be_F$ & \gceq{0.2}     & \gtceq{0.2}      & \gceq{0.25}    & \gtceq{0.25}     & \gceq{0.3}     & \gtceq{0.3}      \\
              & ($\tau_0 = 0$) & ($\topt = 0.12$) & ($\tau_0 = 0$) & ($\topt = 0.12$) & ($\tau_0 = 0$) & ($\topt = 0.12$) \\
      \hline
       3.75   & 9.327(9)       & 8.532(8)         & 8.928(15)      & 8.442(14)        & 8.753(23)      & 8.421(22)        \\
       4.00   & 8.793(9)       & 8.048(8)         & 8.464(15)      & 8.006(15)        & 8.336(23)      & 8.022(23)        \\
       4.25   & 8.279(10)      & 7.582(10)        & 8.018(18)      & 7.587(18)        & 7.936(30)      & 7.638(29)        \\
       4.50   & 7.742(8)       & 7.094(7)         & 7.518(14)      & 7.114(13)        & 7.428(22)      & 7.148(22)        \\
       4.75   & 7.229(7)       & 6.627(7)         & 7.062(15)      & 6.685(14)        & 7.017(25)      & 6.755(25)        \\
       5.00   & 6.704(7)       & 6.147(7)         & 6.567(13)      & 6.217(13)        & 6.533(21)      & 6.289(21)        \\
\hline
    \end{tabular}
  \caption{\label{tab:plaq24}Renormalized gradient flow couplings based on the plaquette discretization of $E(t)$ for the $24^4$ lattice ensembles used in this work, with columns as in \protect\tab{tab:plaq12}.}
\end{table}

\begin{table}[htbp]
  \centering
    \begin{tabular}{c|cc|cc|cc}
      \hline
      $\be_F$ & \gceq{0.2}     & \gtceq{0.2}      & \gceq{0.25}    & \gtceq{0.25}     & \gceq{0.3}     & \gtceq{0.3}      \\
              & ($\tau_0 = 0$) & ($\topt = 0.12$) & ($\tau_0 = 0$) & ($\topt = 0.12$) & ($\tau_0 = 0$) & ($\topt = 0.12$) \\
      \hline
       3.75   & 8.911(10)      & 8.428(9)         & 8.685(18)      & 8.384(18)        & 8.601(30)      & 8.392(30)        \\
       4.00   & 8.445(12)      & 7.990(12)        & 8.268(21)      & 7.984(21)        & 8.222(33)      & 8.024(32)        \\
       4.25   & 7.975(10)      & 7.547(10)        & 7.830(18)      & 7.561(18)        & 7.792(28)      & 7.605(28)        \\
       4.50   & 7.483(11)      & 7.082(11)        & 7.369(21)      & 7.116(20)        & 7.351(33)      & 7.175(33)        \\
       4.75   & 7.021(8)       & 6.647(8)         & 6.939(15)      & 6.702(15)        & 6.927(25)      & 6.761(25)        \\
       5.00   & 6.547(7)       & 6.200(7)         & 6.482(13)      & 6.260(13)        & 6.466(22)      & 6.311(22)        \\
\hline
    \end{tabular}
  \caption{\label{tab:plaq30}Renormalized gradient flow couplings based on the plaquette discretization of $E(t)$ for the $30^4$ lattice ensembles used in this work, with columns as in \protect\tab{tab:plaq12}.}
\end{table}

\begin{table}[htbp]
  \centering
  \begin{tabular}{c|cc|cc|cc}
    \hline
    $\be_F$ & \gceq{0.2}     & \gtceq{0.2}      & \gceq{0.25}    & \gtceq{0.25}     & \gceq{0.3}     & \gtceq{0.3}      \\
            & ($\tau_0 = 0$) & ($\topt = 0.12$) & ($\tau_0 = 0$) & ($\topt = 0.12$) & ($\tau_0 = 0$) & ($\topt = 0.12$) \\
    \hline
     3.75   & 8.846(13)      & 8.427(12)        & 8.666(23)      & 8.402(23)        & 8.596(37)      & 8.413(36)        \\
     4.00   & 8.369(9)       & 7.973(9)         & 8.216(16)      & 7.966(16)        & 8.165(25)      & 7.991(25)        \\
     4.25   & 7.909(11)      & 7.537(10)        & 7.799(19)      & 7.564(19)        & 7.785(31)      & 7.621(31)        \\
     4.50   & 7.445(8)       & 7.095(8)         & 7.360(15)      & 7.139(15)        & 7.360(25)      & 7.205(25)        \\
     4.75   & 6.973(7)       & 6.647(7)         & 6.908(13)      & 6.701(13)        & 6.911(20)      & 6.766(20)        \\
     5.00   & 6.527(7)       & 6.224(7)         & 6.501(13)      & 6.306(13)        & 6.530(22)      & 6.394(21)        \\
\hline
  \end{tabular}
  \caption{\label{tab:plaq32}Renormalized gradient flow couplings based on the plaquette discretization of $E(t)$ for the $32^4$ lattice ensembles used in this work, with columns as in \protect\tab{tab:plaq12}.}
\end{table}

\begin{table}[htbp]
  \centering
  \begin{tabular}{c|cc|cc|cc}
    \hline
    $\be_F$ & \gceq{0.2}     & \gtceq{0.2}      & \gceq{0.25}    & \gtceq{0.25}     & \gceq{0.3}     & \gtceq{0.3}      \\
            & ($\tau_0 = 0$) & ($\topt = 0.12$) & ($\tau_0 = 0$) & ($\topt = 0.12$) & ($\tau_0 = 0$) & ($\topt = 0.12$) \\
    \hline
     3.75   & 8.638(11)      & 8.313(11)        & 8.455(21)      & 8.251(21)        & 8.359(35)      & 8.217(35)        \\
     4.00   & 8.249(13)      & 7.941(12)        & 8.135(23)      & 7.941(23)        & 8.110(33)      & 7.974(33)        \\
     4.25   & 7.794(11)      & 7.503(11)        & 7.703(19)      & 7.519(19)        & 7.684(31)      & 7.555(30)        \\
     4.50   & 7.377(11)      & 7.103(11)        & 7.328(19)      & 7.154(19)        & 7.347(30)      & 7.225(30)        \\
     4.75   & 6.927(10)      & 6.671(10)        & 6.883(17)      & 6.719(17)        & 6.882(25)      & 6.767(25)        \\
     5.00   & 6.472(10)      & 6.233(10)        & 6.447(18)      & 6.294(18)        & 6.469(28)      & 6.362(28)        \\
\hline
  \end{tabular}
  \caption{\label{tab:plaq36}Renormalized gradient flow couplings based on the plaquette discretization of $E(t)$ for the $36^4$ lattice ensembles used in this work, with columns as in \protect\tab{tab:plaq12}.}
\end{table}

\begin{table}[htbp]
  \centering
  \begin{tabular}{cc|cc|cc}
    \hline
     $L$  & ~d.o.f.~ &  \multicolumn{2}{c|}{Rational function} &  \multicolumn{2}{c}{Polynomial}     \\
          &          &  $\chi^2$ &  CL                         &  $\chi^2$ &  CL                     \\
    \hline
    ~12~  &  28      & ~114.9~   &  ~1.8$\times$10$^{-12}$~    & 1294      &  $<$10$^{-16}$          \\
     16   &  30      &   96.0    &   7.8$\times$10$^{-9}$      & 2190      &  $<$10$^{-16}$          \\
     18   &  11      &   25.6    &   0.01                      & ~789.4~   &  $<$10$^{-16}$          \\
     20   &   9      &   20.0    &   0.02                      &   68.2    & ~3.5$\times$10$^{-11}$~ \\
     24   &  13      &    5.1    &   0.97                      &   42.6    &  5.3$\times$10$^{-5}$   \\
     30   &   9      &   14.6    &   0.10                      &   33.9    &  9.2$\times$10$^{-5}$   \\
     32   &   9      &   14.4    &   0.11                      &   46.3    &  5.2$\times$10$^{-7}$   \\
     36   &   9      &   13.0    &   0.16                      &   50.0    &  1.1$\times$10$^{-7}$   \\
    \hline
  \end{tabular}
  \caption{\label{tab:interp02}Quality of $c = 0.2$ renormalized coupling interpolations as functions of the bare coupling, $\gtceq{0.2}(\be_F)$ on each $L^4$ lattice volume.  Both the rational function interpolations using \protect\eq{eq:pade} and the polynomial interpolations using \protect\eq{eq:poly} use the clover discretization of $E(t)$ with optimal $\topt = 0.08$ and involve the same number of fit parameters producing the same number of degrees of freedom.  The confidence level (CL) is computed from the $\chi^2$ and d.o.f.\ through \protect\eq{eq:CL}.}
\end{table}

\begin{table}[htbp]
  \centering
  \begin{tabular}{cc|cc|cc}
    \hline
     $L$  & ~d.o.f.~ &  \multicolumn{2}{c|}{Rational function} &  \multicolumn{2}{c}{Polynomial}    \\
          &          &  $\chi^2$ &  CL                         &  $\chi^2$ &  CL                    \\
    \hline
    ~12~  &  28      & ~64.7~    & ~1.0$\times$10$^{-4}$~      & ~463.6~   &  $<$10$^{-16}$         \\
     16   &  30      &  44.3     &  0.04                       &  705.1    &  $<$10$^{-16}$         \\
     18   &  11      &  10.9     &  0.45                       &  236.8    &  $<$10$^{-16}$         \\
     20   &   9      &  16.4     &  0.06                       &   29.7    & ~5.0$\times$10$^{-4}$~ \\
     24   &  13      &   4.6     &  0.98                       &   17.7    &  0.17                  \\
     30   &   9      &  10.5     &  0.32                       &   13.4    &  0.14                  \\
     32   &   9      &  17.3     &  0.04                       &   27.4    &  1.2$\times$10$^{-3}$  \\
     36   &   9      &  11.6     &  0.24                       &   27.0    &  1.4$\times$10$^{-3}$  \\
    \hline
  \end{tabular}
  \caption{\label{tab:interp025}Quality of $c = 0.25$ renormalized coupling interpolations $\gtceq{0.25}(\be_F)$, with $\topt = 0.08$ and columns as in \protect\tab{tab:interp02}.}
\end{table}

\begin{table}[htbp]
  \centering
  \begin{tabular}{cc|cc|cc}
    \hline
     $L$  & ~d.o.f.~ &  \multicolumn{2}{c|}{Rational function} &  \multicolumn{2}{c}{Polynomial}     \\
          &          &  $\chi^2$ &  CL                         &  $\chi^2$ &  CL                     \\
    \hline
    ~12~  &  28      & ~50.0~    & ~0.01~                      & ~212.2~   &  $<$10$^{-16}$          \\
     16   &  30      &  32.5     &  0.34                       &  281.7    &  $<$10$^{-16}$          \\
     18   &  11      &   7.0     &  0.80                       &   85.8    & ~3.5$\times$10$^{-11}$~ \\
     20   &   9      &  15.3     &  0.08                       &   19.2    &  0.02                   \\
     24   &  13      &   6.2     &  0.94                       &   12.1    &  0.52                   \\
     30   &   9      &  11.2     &  0.26                       &   12.0    &  0.22                   \\
     32   &   9      &  19.9     &  0.02                       &   24.5    &  3.6$\times$10$^{-3}$   \\
     36   &   9      &  11.1     &  0.27                       &   18.9    &  0.03                   \\
    \hline
  \end{tabular}
  \caption{\label{tab:interp03}Quality of $c = 0.3$ renormalized coupling interpolations $\gtceq{0.3}(\be_F)$, with $\topt = 0.08$ and columns as in \protect\tab{tab:interp02}.}
\end{table}

\clearpage 

\bibliographystyle{utphys}
\bibliography{12f_beta}

\providecommand{\href}[2]{#2}\begingroup\raggedright\begin{thebibliography}{100}

\bibitem{Caswell:1974gg}
W.~E. Caswell, ``{Asymptotic Behavior of Nonabelian Gauge Theories to Two Loop
  Order}'', \href{http://dx.doi.org/10.1103/PhysRevLett.33.244}{{\em Phys. Rev.
  Lett.} {\bf 33} (1974) 244}.

\bibitem{Banks:1981nn}
T.~Banks and A.~Zaks, ``{On the Phase Structure of Vector-Like Gauge Theories
  with Massless Fermions}'',
  \href{http://dx.doi.org/10.1016/0550-3213(82)90035-9}{{\em Nucl. Phys.} {\bf
  B196} (1982) 189}.

\bibitem{Aoki:2013zsa}
LatKMI Collaboration: Y.~Aoki, T.~Aoyama, M.~Kurachi, T.~Maskawa, K.-i. Nagai,
  H.~Ohki, E.~Rinaldi, A.~Shibata, K.~Yamawaki and T.~Yamazaki, ``{Light
  composite scalar in twelve-flavor QCD on the lattice}'',
  \href{http://dx.doi.org/10.1103/PhysRevLett.111.162001}{{\em Phys. Rev.
  Lett.} {\bf 111} (2013) 162001} [\href{http://arxiv.org/abs/1305.6006}{{\tt
  arXiv:1305.6006}}].

\bibitem{Fodor:2014pqa}
Z.~Fodor, K.~Holland, J.~Kuti, D.~Nogradi and C.~H. Wong, ``{Can a light Higgs
  impostor hide in composite gauge models?}'',
  \href{http://pos.sissa.it/archive/conferences/187/062/LATTICE
  2013_062.pdf}{{\em PoS} {\bf LATTICE 2013} (2014) 062}
  [\href{http://arxiv.org/abs/1401.2176}{{\tt arXiv:1401.2176}}].

\bibitem{Chatrchyan:2013lba}
CMS Collaboration, ``{Observation of a new boson
  with mass near 125 GeV in pp collisions at $\sqrt{s}$ = 7 and 8 TeV}'',
  \href{http://dx.doi.org/10.1007/JHEP06(2013)081}{{\em JHEP} {\bf 1306} (2013)
  081} [\href{http://arxiv.org/abs/1303.4571}{{\tt arXiv:1303.4571}}].

\bibitem{Aad:2013wqa}
ATLAS Collaboration, ``{Measurements of Higgs boson
  production and couplings in diboson final states with the ATLAS detector at
  the LHC}'', \href{http://dx.doi.org/10.1016/j.physletb.2013.08.010}{{\em
  Phys. Lett.} {\bf B726} (2013) 88--119} [\href{http://arxiv.org/abs/1307.1427}{{\tt
  arXiv:1307.1427}}].

\bibitem{Brower:2015owo}
R.~C. Brower, A.~Hasenfratz, C.~Rebbi, E.~Weinberg and O.~Witzel, ``{Composite
  Higgs model at a conformal fixed point}'',
  \href{http://dx.doi.org/10.1103/PhysRevD.93.075028}{{\em Phys. Rev.} {\bf
  D93} (2016) 075028} [\href{http://arxiv.org/abs/1512.02576}{{\tt
  arXiv:1512.02576}}].

\bibitem{Hasenfratz:2016gut}
A.~Hasenfratz, C.~Rebbi and O.~Witzel, ``{Large scale separation and
  resonances within LHC range from a prototype BSM model}'',
  \href{http://arxiv.org/abs/1609.01401}{{\tt arXiv:1609.01401}}.

\bibitem{Ryttov:2010iz}
T.~A. Ryttov and R.~Shrock, ``Higher-loop corrections to the infrared evolution
  of a gauge theory with fermions'',
  \href{http://dx.doi.org/10.1103/PhysRevD.83.056011}{{\em Phys. Rev.} {\bf
  D83} (2011) 056011} [\href{http://arxiv.org/abs/1011.4542}{{\tt
  arXiv:1011.4542}}].

\bibitem{Pica:2010xq}
C.~Pica and F.~Sannino, ``{UV and IR Zeros of Gauge Theories at The Four Loop
  Order and Beyond}'', \href{http://dx.doi.org/10.1103/PhysRevD.83.035013}{{\em
  Phys. Rev.} {\bf D83} (2011) 035013}
  [\href{http://arxiv.org/abs/1011.5917}{{\tt arXiv:1011.5917}}].

\bibitem{Baikov:2016tgj}
P.~A. Baikov, K.~G. Chetyrkin and J.~H. K{\"u}hn, ``{Five-Loop Running of the
  QCD coupling constant}'',
  \href{http://dx.doi.org/10.1103/PhysRevLett.118.082002}{{\em Phys. Rev.
  Lett.} {\bf 118} (2017) 082002} [\href{http://arxiv.org/abs/1606.08659}{{\tt
  arXiv:1606.08659}}].

\bibitem{Herzog:2017ohr}
F.~Herzog, B.~Ruijl, T.~Ueda, J.~A.~M. Vermaseren and A.~Vogt, ``{The
  five-loop beta function of Yang--Mills theory with fermions}'',
  \href{http://dx.doi.org/10.1007/JHEP02(2017)090}{{\em JHEP} {\bf 1702} (2017)
  090} [\href{http://arxiv.org/abs/1701.01404}{{\tt arXiv:1701.01404}}].

\bibitem{Stevenson:2016mnv}
P.~M. Stevenson, ``{The Banks--Zaks expansion in perturbative QCD: An
  update}'', \href{http://dx.doi.org/10.1142/S0217732316502266}{{\em Mod. Phys.
  Lett.} {\bf A31} (2016) 1650226}
  [\href{http://arxiv.org/abs/1607.01670}{{\tt arXiv:1607.01670}}].

\bibitem{Ryttov:2016ner}
T.~A. Ryttov and R.~Shrock, ``{Infrared Zero of $\beta$ and Value of $\gamma_m$
  for an SU(3) Gauge Theory at the Five-Loop Level}'',
  \href{http://dx.doi.org/10.1103/PhysRevD.94.105015}{{\em Phys. Rev.} {\bf
  D94} (2016) 105015} [\href{http://arxiv.org/abs/1607.06866}{{\tt
  arXiv:1607.06866}}].

\bibitem{Ryttov:2016asb}
T.~A. Ryttov and R.~Shrock, ``{Scheme-independent calculation of
  $\gamma_{\bar\psi\psi,IR}$ for an SU(3) gauge theory}'',
  \href{http://dx.doi.org/10.1103/PhysRevD.94.105014}{{\em Phys. Rev.} {\bf
  D94} (2016) 105014} [\href{http://arxiv.org/abs/1608.00068}{{\tt
  arXiv:1608.00068}}].

\bibitem{Ryttov:2016hal}
T.~A. Ryttov and R.~Shrock, ``{Scheme-Independent Series Expansions at an
  Infrared Zero of the Beta Function in Asymptotically Free Gauge Theories}'',
  \href{http://dx.doi.org/10.1103/PhysRevD.94.125005}{{\em Phys. Rev.} {\bf
  D94} (2016) 125005} [\href{http://arxiv.org/abs/1610.00387}{{\tt
  arXiv:1610.00387}}].

\bibitem{Appelquist:1996dq}
T.~Appelquist, J.~Terning and L.~C.~R. Wijewardhana, ``{The Zero temperature
  chiral phase transition in SU(N) gauge theories}'',
  \href{http://dx.doi.org/10.1103/PhysRevLett.77.1214}{{\em Phys. Rev. Lett.}
  {\bf 77} (1996) 1214--1217} [\href{http://arxiv.org/abs/hep-ph/9602385}{{\tt
  hep-ph/9602385}}].

\bibitem{Appelquist:1998rb}
T.~Appelquist, A.~Ratnaweera, J.~Terning and L.~C.~R. Wijewardhana, ``{The
  Phase structure of an SU($N$) gauge theory with $N_f$ flavors}'',
  \href{http://dx.doi.org/10.1103/PhysRevD.58.105017}{{\em Phys. Rev.} {\bf
  D58} (1998) 105017} [\href{http://arxiv.org/abs/hep-ph/9806472}{{\tt
  hep-ph/9806472}}].

\bibitem{Bashir:2013zha}
A.~Bashir, A.~Raya and J.~Rodriguez-Quintero, ``{QCD: Restoration of Chiral
  Symmetry and Deconfinement for Large $N_f$}'',
  \href{http://dx.doi.org/10.1103/PhysRevD.88.054003}{{\em Phys. Rev.} {\bf
  D88} (2013) 054003} [\href{http://arxiv.org/abs/1302.5829}{{\tt
  arXiv:1302.5829}}].

\bibitem{Dietrich:2006cm}
D.~D. Dietrich and F.~Sannino, ``{Conformal window of SU(N) gauge theories with
  fermions in higher dimensional representations}'',
  \href{http://dx.doi.org/10.1103/PhysRevD.75.085018}{{\em Phys. Rev.} {\bf
  D75} (2007) 085018} [\href{http://arxiv.org/abs/hep-ph/0611341}{{\tt
  hep-ph/0611341}}].

\bibitem{Braun:2006jd}
J.~Braun and H.~Gies, ``{Chiral phase boundary of QCD at finite temperature}'',
  \href{http://dx.doi.org/10.1088/1126-6708/2006/06/024}{{\em JHEP} {\bf 0606}
  (2006) 024} [\href{http://arxiv.org/abs/hep-ph/0602226}{{\tt
  hep-ph/0602226}}].

\bibitem{Braun:2010qs}
J.~Braun, C.~S. Fischer and H.~Gies, ``{Beyond Miransky Scaling}'',
  \href{http://dx.doi.org/10.1103/PhysRevD.84.034045}{{\em Phys. Rev.} {\bf
  D84} (2011) 034045} [\href{http://arxiv.org/abs/1012.4279}{{\tt
  arXiv:1012.4279}}].

\bibitem{Appelquist:1999hr}
T.~Appelquist, A.~G. Cohen and M.~Schmaltz, ``{A new constraint on strongly
  coupled gauge theories}'',
  \href{http://dx.doi.org/10.1103/PhysRevD.60.045003}{{\em Phys. Rev.} {\bf
  D60} (1999) 045003} [\href{http://arxiv.org/abs/hep-th/9901109}{{\tt
  hep-th/9901109}}].

\bibitem{Appelquist:2007hu}
T.~Appelquist, G.~T. Fleming and E.~T. Neil, ``{Lattice study of the conformal
  window in QCD-like theories}'',
  \href{http://dx.doi.org/10.1103/PhysRevLett.100.171607}{{\em Phys. Rev.
  Lett.} {\bf 100} (2008) 171607} [\href{http://arxiv.org/abs/0712.0609}{{\tt
  arXiv:0712.0609}}].

\bibitem{Appelquist:2009ty}
T.~Appelquist, G.~T. Fleming and E.~T. Neil, ``{Lattice Study of Conformal
  Behavior in SU(3) Yang-Mills Theories}'',
  \href{http://dx.doi.org/10.1103/PhysRevD.79.076010}{{\em Phys. Rev.} {\bf
  D79} (2009) 076010} [\href{http://arxiv.org/abs/0901.3766}{{\tt
  arXiv:0901.3766}}].

\bibitem{Bilgici:2009nm}
E.~Bilgici, A.~Flachi, E.~Itou, M.~Kurachi, C.~J.~D. Lin, H.~Matsufuru,
  H.~Ohki, T.~Onogi, E.~Shintani and T.~Yamazaki, ``{Search for the IR fixed
  point in the twisted Polyakov loop scheme}'',
  \href{http://pos.sissa.it/archive/conferences/091/063/LAT2009_063.pdf}{{\em
  PoS} {\bf LAT2009} (2009) 063} [\href{http://arxiv.org/abs/0910.4196}{{\tt
  arXiv:0910.4196}}].

\bibitem{Itou:2010we}
E.~Itou, T.~Aoyama, M.~Kurachi, C.~J.~D. Lin, H.~Matsufuru, H.~Ohki, T.~Onogi,
  E.~Shintani and T.~Yamazaki, ``{Search for the IR fixed point in the Twisted
  Polyakov Loop scheme (II)}'',
  \href{http://pos.sissa.it/archive/conferences/105/054/Lattice
  2010_054.pdf}{{\em PoS} {\bf Lattice 2010} (2010) 054}
  [\href{http://arxiv.org/abs/1011.0516}{{\tt arXiv:1011.0516}}].

\bibitem{Ogawa:2011ki}
K.~Ogawa, T.~Aoyama, H.~Ikeda, E.~Itou, M.~Kurachi, C.-J.~D. Lin, H.~Matsufuru,
  H.~Ohki, T.~Onogi, E.~Shintani and T.~Yamazaki, ``{The Infrared behavior of
  SU(3) Nf=12 gauge theory -- about the existence of conformal fixed point}'',
  \href{http://pos.sissa.it/archive/conferences/139/081/Lattice
  2011_081.pdf}{{\em PoS} {\bf Lattice 2011} (2011) 081}
  [\href{http://arxiv.org/abs/1111.1575}{{\tt arXiv:1111.1575}}].

\bibitem{Lin:2012iw}
C.~J.~D. Lin, K.~Ogawa, H.~Ohki and E.~Shintani, ``{Lattice study of infrared
  behaviour in SU(3) gauge theory with twelve massless flavours}'',
  \href{http://dx.doi.org/10.1007/JHEP08(2012)096}{{\em JHEP} {\bf 1208} (2012)
  096} [\href{http://arxiv.org/abs/1205.6076}{{\tt arXiv:1205.6076}}].

\bibitem{Itou:2012qn}
E.~Itou, ``{Properties of the twisted Polyakov loop coupling and the infrared
  fixed point in the SU(3) gauge theories}'',
  \href{http://dx.doi.org/10.1093/ptep/ptt053}{{\em PTEP} {\bf 2013} (2013)
  083B01} [\href{http://arxiv.org/abs/1212.1353}{{\tt arXiv:1212.1353}}].

\bibitem{Itou:2013faa}
E.~Itou, ``{The twisted Polyakov loop coupling and the search for an IR fixed
  point}'', \href{http://pos.sissa.it/archive/conferences/187/005/LATTICE
  2013_005.pdf}{{\em PoS} {\bf LATTICE 2013} (2014) 005}
  [\href{http://arxiv.org/abs/1311.2676}{{\tt arXiv:1311.2676}}].

\bibitem{Cheng:2014jba}
A.~Cheng, A.~Hasenfratz, Y.~Liu, G.~Petropoulos and D.~Schaich, ``{Improving
  the continuum limit of gradient flow step scaling}'',
  \href{http://dx.doi.org/10.1007/JHEP05(2014)137}{{\em JHEP} {\bf 1405} (2014)
  137} [\href{http://arxiv.org/abs/1404.0984}{{\tt arXiv:1404.0984}}].

\bibitem{Hasenfratz:2015xpa}
A.~Hasenfratz, ``{Improved gradient flow for step scaling function and scale
  setting}'',
  \href{http://pos.sissa.it/archive/conferences/214/257/LATTICE2014_257.pdf}{{\em PoS}
  {\bf LATTICE2014} (2015) 257}
  [\href{http://arxiv.org/abs/1501.07848}{{\tt arXiv:1501.07848}}].

\bibitem{Lin:2015zpa}
C.~J.~D. Lin, K.~Ogawa and A.~Ramos, ``{The Yang-Mills gradient flow and SU(3)
  gauge theory with 12 massless fundamental fermions in a colour-twisted
  box}'', \href{http://dx.doi.org/10.1007/JHEP12(2015)103}{{\em JHEP} {\bf
  1512} (2015) 103} [\href{http://arxiv.org/abs/1510.05755}{{\tt
  arXiv:1510.05755}}].

\bibitem{Fodor:2016zil}
Z.~Fodor, K.~Holland, J.~Kuti, S.~Mondal, D.~Nogradi and C.~H. Wong, ``{Fate
  of the conformal fixed point with twelve massless fermions and SU(3) gauge
  group}'', \href{http://dx.doi.org/10.1103/PhysRevD.94.091501}{{\em Phys.
  Rev.} {\bf D94} (2016) 091501} [\href{http://arxiv.org/abs/1607.06121}{{\tt
  arXiv:1607.06121}}].

\bibitem{Deuzeman:2009mh}
A.~Deuzeman, M.~P. Lombardo and E.~Pallante, ``{Evidence for a conformal phase
  in SU(N) gauge theories}'',
  \href{http://dx.doi.org/10.1103/PhysRevD.82.074503}{{\em Phys. Rev.} {\bf
  D82} (2010) 074503} [\href{http://arxiv.org/abs/0904.4662}{{\tt
  arXiv:0904.4662}}].

\bibitem{Jin:2009mc}
X.-Y. Jin and R.~D. Mawhinney, ``{Lattice QCD with 8 and 12 degenerate quark
  flavors}'',
  \href{http://pos.sissa.it/archive/conferences/091/049/LAT2009_049.pdf}{{\em
  PoS} {\bf LAT2009} (2009) 049} [\href{http://arxiv.org/abs/0910.3216}{{\tt
  arXiv:0910.3216}}].

\bibitem{Deuzeman:2010fn}
A.~Deuzeman, E.~Pallante and M.~P. Lombardo, ``{The Bulk transition of
  many-flavour QCD and the search for a UVFP at strong coupling}'',
  \href{http://pos.sissa.it/archive/conferences/105/067/Lattice
  2010_067.pdf}{{\em PoS} {\bf Lattice 2010} (2010) 067}
  [\href{http://arxiv.org/abs/1012.5971}{{\tt arXiv:1012.5971}}].

\bibitem{Cheng:2011ic}
A.~Cheng, A.~Hasenfratz and D.~Schaich, ``{Novel phase in SU(3) lattice gauge
  theory with 12 light fermions}'',
  \href{http://dx.doi.org/10.1103/PhysRevD.85.094509}{{\em Phys. Rev.} {\bf
  D85} (2012) 094509} [\href{http://arxiv.org/abs/1111.2317}{{\tt
  arXiv:1111.2317}}].

\bibitem{Deuzeman:2011pa}
A.~Deuzeman, M.~P. Lombardo, T.~Nunes~da Silva and E.~Pallante, ``{Bulk
  transitions of twelve flavor QCD and $U_A(1)$ symmetry}'',
  \href{http://pos.sissa.it/archive/conferences/139/321/Lattice
  2011_321.pdf}{{\em PoS} {\bf Lattice 2011} (2011) 321}
  [\href{http://arxiv.org/abs/1111.2590}{{\tt arXiv:1111.2590}}].

\bibitem{Jin:2012dw}
X.-Y. Jin and R.~D. Mawhinney, ``{Lattice QCD with 12 Degenerate Quark
  Flavors}'', \href{http://pos.sissa.it/archive/conferences/139/066/Lattice
  2011_066.pdf}{{\em PoS} {\bf Lattice 2011} (2011) 066}
  [\href{http://arxiv.org/abs/1203.5855}{{\tt arXiv:1203.5855}}].

\bibitem{Fodor:2012uu}
Z.~Fodor, K.~Holland, J.~Kuti, D.~Nogradi, C.~Schroeder and C.~H. Wong,
  ``{Twelve fundamental and two sextet fermion flavors}'',
  \href{http://pos.sissa.it/archive/conferences/139/073/Lattice
  2011_073.pdf}{{\em PoS} {\bf Lattice 2011} (2011) 073}
  [\href{http://arxiv.org/abs/1205.1878}{{\tt arXiv:1205.1878}}].

\bibitem{Schaich:2012fr}
D.~Schaich, A.~Cheng, A.~Hasenfratz and G.~Petropoulos, ``{Bulk and
  finite-temperature transitions in SU(3) gauge theories with many light
  fermions}'', \href{http://pos.sissa.it/archive/conferences/164/028/Lattice
  2012_028.pdf}{{\em PoS} {\bf Lattice 2012} (2012) 028}
  [\href{http://arxiv.org/abs/1207.7164}{{\tt arXiv:1207.7164}}].

\bibitem{Deuzeman:2012ee}
A.~Deuzeman, M.~P. Lombardo, T.~Nunes Da~Silva and E.~Pallante, ``{The bulk
  transition of QCD with twelve flavors and the role of improvement}'',
  \href{http://dx.doi.org/10.1016/j.physletb.2013.02.030}{{\em Phys. Lett.}
  {\bf B720} (2013) 358--365} [\href{http://arxiv.org/abs/1209.5720}{{\tt
  arXiv:1209.5720}}].

\bibitem{daSilva:2012wg}
T.~Nunes~da Silva and E.~Pallante, ``{The strong coupling regime of twelve
  flavors QCD}'', \href{http://pos.sissa.it/archive/conferences/164/052/Lattice
  2012_052.pdf}{{\em PoS} {\bf Lattice 2012} (2012) 052}
  [\href{http://arxiv.org/abs/1211.3656}{{\tt arXiv:1211.3656}}].

\bibitem{Hasenfratz:2013uha}
A.~Hasenfratz, A.~Cheng, G.~Petropoulos and D.~Schaich,
  ``\href{http://dx.doi.org/10.1142/9789814566254_0004}{Reaching the chiral
  limit in many flavor systems}'', in {\em Strong Coupling Gauge Theories in
  the LHC Perspective (SCGT12)} (2014) 44--50
  [\href{http://arxiv.org/abs/1303.7129}{{\tt arXiv:1303.7129}}].

\bibitem{Ishikawa:2013tua}
K.~I. Ishikawa, Y.~Iwasaki, Y.~Nakayama and T.~Yoshie, ``{Global Structure of
  Conformal Theories in the SU(3) Gauge Theory}'',
  \href{http://dx.doi.org/10.1103/PhysRevD.89.114503}{{\em Phys. Rev.} {\bf
  D89} (2014) 114503} [\href{http://arxiv.org/abs/1310.5049}{{\tt
  arXiv:1310.5049}}].

\bibitem{Ishikawa:2015iwa}
K.~I. Ishikawa, Y.~Iwasaki, Y.~Nakayama and T.~Yoshie, ``{IR fixed points in
  $SU(3)$ gauge Theories}'',
  \href{http://dx.doi.org/10.1016/j.physletb.2015.07.019}{{\em Phys. Lett.}
  {\bf B748} (2015) 289--294} [\href{http://arxiv.org/abs/1503.02359}{{\tt
  arXiv:1503.02359}}].

\bibitem{Fodor:2009wk}
Z.~Fodor, K.~Holland, J.~Kuti, D.~Nogradi and C.~Schroeder, ``{Nearly
  conformal gauge theories in finite volume}'',
  \href{http://dx.doi.org/10.1016/j.physletb.2009.10.040}{{\em Phys. Lett.}
  {\bf B681} (2009) 353--361} [\href{http://arxiv.org/abs/0907.4562}{{\tt
  arXiv:0907.4562}}].

\bibitem{Fodor:2011tu}
Z.~Fodor, K.~Holland, J.~Kuti, D.~Nogradi and C.~Schroeder, ``{Twelve massless
  flavors and three colors below the conformal window}'',
  \href{http://dx.doi.org/10.1016/j.physletb.2011.07.037}{{\em Phys. Lett.}
  {\bf B703} (2011) 348--358} [\href{http://arxiv.org/abs/1104.3124}{{\tt
  arXiv:1104.3124}}].

\bibitem{Appelquist:2011dp}
T.~Appelquist, G.~T. Fleming, M.~F. Lin, E.~T. Neil and D.~Schaich, ``{Lattice
  Simulations and Infrared Conformality}'',
  \href{http://dx.doi.org/10.1103/PhysRevD.84.054501}{{\em Phys. Rev.} {\bf
  D84} (2011) 054501} [\href{http://arxiv.org/abs/1106.2148}{{\tt
  arXiv:1106.2148}}].

\bibitem{DeGrand:2011cu}
T.~DeGrand, ``{Finite-size scaling tests for spectra in SU(3) lattice gauge
  theory coupled to 12 fundamental flavor fermions}'',
  \href{http://dx.doi.org/10.1103/PhysRevD.84.116901}{{\em Phys. Rev.} {\bf
  D84} (2011) 116901} [\href{http://arxiv.org/abs/1109.1237}{{\tt
  arXiv:1109.1237}}].

\bibitem{Deuzeman:2012pv}
A.~Deuzeman, M.~P. Lombardo and E.~Pallante, ``{On the spectrum of QCD-like
  theories and the conformal window}'',
  \href{http://pos.sissa.it/archive/conferences/139/083/Lattice
  2011_083.pdf}{{\em PoS} {\bf Lattice 2011} (2012) 083}
  [\href{http://arxiv.org/abs/1201.1863}{{\tt arXiv:1201.1863}}].

\bibitem{Aoki:2012eq}
LatKMI Collaboration: Y.~Aoki, T.~Aoyama, M.~Kurachi, T.~Maskawa, K.-i. Nagai,
  H.~Ohki, A.~Shibata, K.~Yamawaki and T.~Yamazaki, ``{Lattice study of
  conformality in twelve-flavor QCD}'',
  \href{http://dx.doi.org/10.1103/PhysRevD.86.059903,
  10.1103/PhysRevD.86.054506}{{\em Phys. Rev.} {\bf D86} (2012) 054506}
  [\href{http://arxiv.org/abs/1207.3060}{{\tt arXiv:1207.3060}}].

\bibitem{Fodor:2012et}
Z.~Fodor, K.~Holland, J.~Kuti, D.~Nogradi, C.~Schroeder and C.~H. Wong,
  ``{Conformal finite size scaling of twelve fermion flavors}'',
  \href{http://pos.sissa.it/archive/conferences/164/279/Lattice
  2012_279.pdf}{{\em PoS} {\bf Lattice 2012} (2012) 279}
  [\href{http://arxiv.org/abs/1211.4238}{{\tt arXiv:1211.4238}}].

\bibitem{Deuzeman:2013kma}
A.~Deuzeman, M.~P. Lombardo, K.~Miura, T.~Nunes~da Silva and E.~Pallante,
  ``{Phases of many flavors QCD: lattice results}'',
  \href{http://pos.sissa.it/archive/conferences/171/274/Confinement
  X_274.pdf}{{\em PoS} {\bf Confinement X} (2012) 274}
  [\href{http://arxiv.org/abs/1304.3245}{{\tt arXiv:1304.3245}}].

\bibitem{Cheng:2013xha}
A.~Cheng, A.~Hasenfratz, Y.~Liu, G.~Petropoulos and D.~Schaich, ``{Finite size
  scaling of conformal theories in the presence of a near-marginal operator}'',
  \href{http://dx.doi.org/10.1103/PhysRevD.90.014509}{{\em Phys. Rev.} {\bf
  D90} (2014) 014509} [\href{http://arxiv.org/abs/1401.0195}{{\tt
  arXiv:1401.0195}}].

\bibitem{Lombardo:2014cqa}
M.~P. Lombardo, K.~Miura, T.~Nunes~da Silva, E.~Pallante and N.~S. Montoro,
  ``{More results on theories inside the conformal window}'',
  \href{http://pos.sissa.it/archive/conferences/187/082/LATTICE
  2013_082.pdf}{{\em PoS} {\bf LATTICE 2013} (2014)}.

\bibitem{Lombardo:2014pda}
M.~P. Lombardo, K.~Miura, T.~J. Nunes~da Silva and E.~Pallante, ``{On the
  particle spectrum and the conformal window}'',
  \href{http://dx.doi.org/10.1007/JHEP12(2014)183}{{\em JHEP} {\bf 1412} (2014)
  183} [\href{http://arxiv.org/abs/1410.0298}{{\tt arXiv:1410.0298}}].

\bibitem{Lombardo:2014mda}
M.~P. Lombardo, K.~Miura, T.~J. Nunes~da Silva and E.~Pallante, ``{One, two,
  zero: Scales of strong interactions}'',
  \href{http://dx.doi.org/10.1142/S0217751X14450079}{{\em Int. J. Mod. Phys.}
  {\bf A29} (2014) 1445007} [\href{http://arxiv.org/abs/1410.2036}{{\tt
  arXiv:1410.2036}}].

\bibitem{Hasenfratz:2012fp}
A.~Hasenfratz, A.~Cheng, G.~Petropoulos and D.~Schaich, ``{Mass anomalous
  dimension from Dirac eigenmode scaling in conformal and confining systems}'',
  \href{http://pos.sissa.it/archive/conferences/164/034/Lattice
  2012_034.pdf}{{\em PoS} {\bf Lattice 2012} (2012) 034}
  [\href{http://arxiv.org/abs/1207.7162}{{\tt arXiv:1207.7162}}].

\bibitem{Cheng:2013eu}
A.~Cheng, A.~Hasenfratz, G.~Petropoulos and D.~Schaich, ``{Scale-dependent
  mass anomalous dimension from Dirac eigenmodes}'',
  \href{http://dx.doi.org/10.1007/JHEP07(2013)061}{{\em JHEP} {\bf 1307} (2013)
  061} [\href{http://arxiv.org/abs/1301.1355}{{\tt arXiv:1301.1355}}].

\bibitem{Cheng:2013bca}
A.~Cheng, A.~Hasenfratz, G.~Petropoulos and D.~Schaich, ``{Determining the
  mass anomalous dimension through the eigenmodes of Dirac operator}'',
  \href{http://pos.sissa.it/archive/conferences/187/088/LATTICE
  2013_088.pdf}{{\em PoS} {\bf LATTICE 2013} (2013) 088}
  [\href{http://arxiv.org/abs/1311.1287}{{\tt arXiv:1311.1287}}].

\bibitem{Itou:2014ota}
E.~Itou and A.~Tomiya, ``{Determination of the mass anomalous dimension for
  $N_f = 12$ and $N_f = 9$ SU(3) gauge theories}'',
  \href{http://pos.sissa.it/archive/conferences/214/252/LATTICE2014_252.pdf}{{\em PoS}
  {\bf LATTICE2014} (2014) 252}
  [\href{http://arxiv.org/abs/1411.1155}{{\tt arXiv:1411.1155}}].

\bibitem{Hasenfratz:2010fi}
A.~Hasenfratz, ``{Conformal or Walking? Monte Carlo renormalization group
  studies of SU(3) gauge models with fundamental fermions}'',
  \href{http://dx.doi.org/10.1103/PhysRevD.82.014506}{{\em Phys. Rev.} {\bf
  D82} (2010) 014506} [\href{http://arxiv.org/abs/1004.1004}{{\tt
  arXiv:1004.1004}}].

\bibitem{Hasenfratz:2011xn}
A.~Hasenfratz, ``{Infrared Fixed Point of the 12-Fermion SU(3) Gauge Model
  Based on 2-Lattice Monte Carlo Renomalization-Group Matching}'',
  \href{http://dx.doi.org/10.1103/PhysRevLett.108.061601}{{\em Phys. Rev.
  Lett.} {\bf 108} (2012) 061601} [\href{http://arxiv.org/abs/1106.5293}{{\tt
  arXiv:1106.5293}}].

\bibitem{Hasenfratz:2011da}
A.~Hasenfratz, ``{MCRG study of 12 fundamental flavors with mixed
  fundamental-adjoint gauge action}'',
  \href{http://pos.sissa.it/archive/conferences/139/065/Lattice
  2011_065.pdf}{{\em PoS} {\bf Lattice 2011} (2011) 065}
  [\href{http://arxiv.org/abs/1112.6146}{{\tt arXiv:1112.6146}}].

\bibitem{Meurice:2012sj}
Y.~Meurice, A.~Bazavov, B.~A. Berg, D.~Du, A.~Denbleyker, Y.~Liu, D.~K.
  Sinclair, J.~Unmuth-Yockey and H.~Zou, ``{Fisher zeros and conformality in
  lattice models}'',
  \href{http://pos.sissa.it/archive/conferences/164/229/Lattice
  2012_229.pdf}{{\em PoS} {\bf Lattice 2012} (2012) 051}
  [\href{http://arxiv.org/abs/1210.6969}{{\tt arXiv:1210.6969}}].

\bibitem{Fodor:2012uw}
Z.~Fodor, K.~Holland, J.~Kuti, D.~Nogradi, C.~Schroeder and C.~H. Wong,
  ``{Confining force and running coupling with twelve fundamental and two
  sextet fermions}'',
  \href{http://pos.sissa.it/archive/conferences/164/025/Lattice
  2012_025.pdf}{{\em PoS} {\bf Lattice 2012} (2012) 025}
  [\href{http://arxiv.org/abs/1211.3548}{{\tt arXiv:1211.3548}}].

\bibitem{Petropoulos:2012mg}
G.~Petropoulos, A.~Cheng, A.~Hasenfratz and D.~Schaich, ``{MCRG study of 8 and
  12 fundamental flavors}'',
  \href{http://pos.sissa.it/archive/conferences/164/034/Lattice
  2012_051.pdf}{{\em PoS} {\bf Lattice 2012} (2012) 051}
  [\href{http://arxiv.org/abs/1212.0053}{{\tt arXiv:1212.0053}}].

\bibitem{Itou:2013kaa}
E.~Itou, ``{A novel scheme for the wave function renormalization of the
  composite operators}'', \href{http://dx.doi.org/10.1093/ptep/ptv045}{{\em
  PTEP} {\bf 2015} (2015) 043B08} [\href{http://arxiv.org/abs/1307.6645}{{\tt
  arXiv:1307.6645}}].

\bibitem{Itou:2013ofa}
E.~Itou, ``{The anomalous dimension at the infrared fixed point of $N_f = 12$
  SU(3) theory}'',
  \href{http://pos.sissa.it/archive/conferences/187/481/LATTICE
  2013_481.pdf}{{\em PoS} {\bf LATTICE 2013} (2013) 481}
  [\href{http://arxiv.org/abs/1311.2998}{{\tt arXiv:1311.2998}}].

\bibitem{Aoki:2016yrm}
Y.~Aoki, T.~Aoyama, E.~Bennett, M.~Kurachi, T.~Maskawa, K.~Miura, K.-i. Nagai,
  H.~Ohki, E.~Rinaldi, A.~Shibata, K.~Yamawaki and T.~Yamazaki, ``{Topological
  observables in many-flavour QCD}'', {\em PoS} {\bf LATTICE 2015} (2016) 214
  [\href{http://arxiv.org/abs/1601.04687}{{\tt arXiv:1601.04687}}].

\bibitem{DeGrand:2015zxa}
T.~DeGrand, ``{Lattice tests of beyond Standard Model dynamics}'',
  \href{http://dx.doi.org/10.1103/RevModPhys.88.015001}{{\em Rev. Mod. Phys.}
  {\bf 88} (2016) 015001} [\href{http://arxiv.org/abs/1510.05018}{{\tt
  arXiv:1510.05018}}].

\bibitem{Giedt:2015alr}
J.~Giedt, ``{Anomalous dimensions on the lattice}'',
  \href{http://dx.doi.org/10.1142/S0217751X16300118}{{\em Int. J. Mod. Phys.}
  {\bf A31} (2016) 1630011} [\href{http://arxiv.org/abs/1512.09330}{{\tt
  arXiv:1512.09330}}].

\bibitem{Nogradi:2016qek}
D.~Nogradi and A.~Patella, ``{Strong dynamics, composite Higgs and the
  conformal window}'', \href{http://dx.doi.org/10.1142/S0217751X1643003X}{{\em
  Int. J. Mod. Phys.} {\bf A31} (2016) 1643003}
  [\href{http://arxiv.org/abs/1607.07638}{{\tt arXiv:1607.07638}}].

\bibitem{Calabrese:2002bm}
P.~Calabrese, A.~Pelissetto and E.~Vicari, ``{Multicritical phenomena in
  $O(n_1) \oplus O(n_2)$-symmetric theories}'',
  \href{http://dx.doi.org/10.1103/PhysRevB.67.054505}{{\em Phys. Rev.} {\bf
  B67} (2003) 054505} [\href{http://arxiv.org/abs/cond-mat/0209580}{{\tt
  cond-mat/0209580}}].

\bibitem{Hasenfratz:2015ssa}
A.~Hasenfratz, Y.~Liu and C.~Y.-H. Huang, ``{The renormalization group step
  scaling function of the 2-flavor SU(3) sextet model}'',
  \href{http://arxiv.org/abs/1507.08260}{{\tt arXiv:1507.08260}}.

\bibitem{Hasenfratz:2014rna}
A.~Hasenfratz, D.~Schaich and A.~Veernala, ``{Nonperturbative beta function of
  eight-flavor SU(3) gauge theory}'',
  \href{http://dx.doi.org/10.1007/JHEP06(2015)143}{{\em JHEP} {\bf 1506} (2015)
  143} [\href{http://arxiv.org/abs/1410.5886}{{\tt arXiv:1410.5886}}].

\bibitem{Fodor:2015baa}
Z.~Fodor, K.~Holland, J.~Kuti, S.~Mondal, D.~Nogradi and C.~H. Wong, ``{The
  running coupling of 8 flavors and 3 colors}'',
  \href{http://dx.doi.org/10.1007/JHEP06(2015)019}{{\em JHEP} {\bf 1506} (2015)
  019} [\href{http://arxiv.org/abs/1503.01132}{{\tt arXiv:1503.01132}}].

\bibitem{Iha:2016ppj}
H.~Iha, H.~Makino and H.~Suzuki, ``{Upper bound on the mass anomalous
  dimension in many-flavor gauge theories: a conformal bootstrap approach}'',
  \href{http://dx.doi.org/10.1093/ptep/ptw046}{{\em PTEP} {\bf 2016} (2016)
  053B03} [\href{http://arxiv.org/abs/1603.01995}{{\tt arXiv:1603.01995}}].

\bibitem{Doff:2016jzk}
A.~Doff and A.~A. Natale, ``{Anomalous mass dimension in multiflavor QCD}'',
  \href{http://dx.doi.org/10.1103/PhysRevD.94.076005}{{\em Phys. Rev.} {\bf
  D94} (2016) 076005} [\href{http://arxiv.org/abs/1610.02564}{{\tt
  arXiv:1610.02564}}].

\bibitem{Ryttov:2016hdp}
T.~A. Ryttov, ``{Consistent Perturbative Fixed Point Calculations in QCD and
  Supersymmetric QCD}'',
  \href{http://dx.doi.org/10.1103/PhysRevLett.117.071601}{{\em Phys. Rev.
  Lett.} {\bf 117} (2016) 071601} [\href{http://arxiv.org/abs/1604.00687}{{\tt
  arXiv:1604.00687}}].

\bibitem{Hasenfratz:2001hp}
A.~Hasenfratz and F.~Knechtli, ``{Flavor symmetry and the static potential with
  hypercubic blocking}'',
  \href{http://dx.doi.org/10.1103/PhysRevD.64.034504}{{\em Phys. Rev.} {\bf
  D64} (2001) 034504} [\href{http://arxiv.org/abs/hep-lat/0103029}{{\tt
  arXiv:hep-lat/0103029}}].

\bibitem{Hasenfratz:2007rf}
A.~Hasenfratz, R.~Hoffmann and S.~Schaefer, ``{Hypercubic smeared links for
  dynamical fermions}'',
  \href{http://dx.doi.org/10.1088/1126-6708/2007/05/029}{{\em JHEP} {\bf 0705}
  (2007) 029} [\href{http://arxiv.org/abs/hep-lat/0702028}{{\tt
  hep-lat/0702028}}].

\bibitem{Narayanan:2006rf}
R.~Narayanan and H.~Neuberger, ``{Infinite N phase transitions in continuum
  Wilson loop operators}'',
  \href{http://dx.doi.org/10.1088/1126-6708/2006/03/064}{{\em JHEP} {\bf 0603}
  (2006) 064} [\href{http://arxiv.org/abs/hep-th/0601210}{{\tt
  hep-th/0601210}}].

\bibitem{Luscher:2009eq}
M.~Luscher, ``{Trivializing maps, the Wilson flow and the HMC algorithm}'',
  \href{http://dx.doi.org/10.1007/s00220-009-0953-7}{{\em Commun. Math. Phys.}
  {\bf 293} (2010) 899--919} [\href{http://arxiv.org/abs/0907.5491}{{\tt
  arXiv:0907.5491}}].

\bibitem{Luscher:2013vga}
M.~Luscher, ``{Future applications of the Yang-Mills gradient flow in lattice
  QCD}'', \href{http://pos.sissa.it/archive/conferences/187/016/LATTICE
  2013_016.pdf}{{\em PoS} {\bf LATTICE 2013} (2013) 016}
  [\href{http://arxiv.org/abs/1308.5598}{{\tt arXiv:1308.5598}}].

\bibitem{Luscher:2010iy}
M.~Luscher, ``{Properties and uses of the Wilson flow in lattice QCD}'',
  \href{http://dx.doi.org/10.1007/JHEP08(2010)071}{{\em JHEP} {\bf 1008} (2010)
  071} [\href{http://arxiv.org/abs/1006.4518}{{\tt arXiv:1006.4518}}].

\bibitem{Fodor:2012td}
Z.~Fodor, K.~Holland, J.~Kuti, D.~Nogradi and C.~H. Wong, ``{The Yang-Mills
  gradient flow in finite volume}'',
  \href{http://dx.doi.org/10.1007/JHEP11(2012)007}{{\em JHEP} {\bf 1211} (2012)
  007} [\href{http://arxiv.org/abs/1208.1051}{{\tt arXiv:1208.1051}}].

\bibitem{Fodor:2012qh}
Z.~Fodor, K.~Holland, J.~Kuti, D.~Nogradi and C.~H. Wong, ``{The gradient flow
  running coupling scheme}'',
  \href{http://pos.sissa.it/archive/conferences/164/050/Lattice
  2012_050.pdf}{{\em PoS} {\bf Lattice 2012} (2012) 050}
  [\href{http://arxiv.org/abs/1211.3247}{{\tt arXiv:1211.3247}}].

\bibitem{Fritzsch:2013je}
P.~Fritzsch and A.~Ramos, ``{The gradient flow coupling in the Schr{\"o}dinger
  Functional}'', \href{http://dx.doi.org/10.1007/JHEP10(2013)008}{{\em JHEP}
  {\bf 1310} (2013) 008} [\href{http://arxiv.org/abs/1301.4388}{{\tt
  arXiv:1301.4388}}].

\bibitem{Ramos:2014kka}
S.~Sint and A.~Ramos, ``{On O($a^2$) effects in gradient flow observables}'',
  \href{http://pos.sissa.it/archive/conferences/214/329/LATTICE2014_329.pdf}{{\em PoS}
  {\bf LATTICE2014} (2015) 329}
  [\href{http://arxiv.org/abs/1411.6706}{{\tt arXiv:1411.6706}}].

\bibitem{Ramos:2015baa}
A.~Ramos and S.~Sint, ``{Symanzik improvement of the gradient flow in lattice
  gauge theories}'',
  \href{http://dx.doi.org/10.1140/epjc/s10052-015-3831-9}{{\em Eur. Phys. J.}
  {\bf C76} (2016) 15} [\href{http://arxiv.org/abs/1508.05552}{{\tt
  arXiv:1508.05552}}].

\bibitem{Fodor:2014cpa}
Z.~Fodor, K.~Holland, J.~Kuti, S.~Mondal, D.~Nogradi and C.~H. Wong, ``{The
  lattice gradient flow at tree-level and its improvement}'',
  \href{http://dx.doi.org/10.1007/JHEP09(2014)018}{{\em JHEP} {\bf 1409} (2014)
  018} [\href{http://arxiv.org/abs/1406.0827}{{\tt arXiv:1406.0827}}].

\bibitem{Hasenfratz:2014SCGT}
A.~Hasenfratz, A.~Cheng, Y.~Liu, G.~Petropoulos and D.~Schaich,
 ``\href{http://www.kmi.nagoya-u.ac.jp/workshop/SCGT14Mini/program/poster-abstracts.html#0044}{Running
  coupling from gradient flow for the $N_f = 12$ SU(3) model}'', in {\em
  {Sakata Memorial KMI Mini-Workshop on Strong Coupling Gauge Theories Beyond
  the Standard Model}}, 5 March 2014.

\bibitem{Karavirta:2011zg}
T.~Karavirta, J.~Rantaharju, K.~Rummukainen and K.~Tuominen, ``{Determining
  the conformal window: SU(2) gauge theory with $N_f = 4$, 6 and 10 fermion
  flavours}'', \href{http://dx.doi.org/10.1007/JHEP05(2012)003}{{\em JHEP} {\bf
  1205} (2012) 003} [\href{http://arxiv.org/abs/1111.4104}{{\tt
  arXiv:1111.4104}}].

\bibitem{Kuti:2017BU}
J.~Kuti, Z.~Fodor, K.~Holland, D.~Nogradi and C.~H. Wong,
  ``\href{http://www-hep.colorado.edu/~eneil/lbsm17/talks/Kuti.pdf}{Dilaton
  signatures with lattice BSM tools}'', in {\em {Lattice for Beyond the
  Standard Model Physics}}, 20 April 2017.

\bibitem{Fodor:2017Lat}
Z.~Fodor, K.~Holland, J.~Kuti, D.~Nogradi and C.~H. Wong,
  ``\href{https://makondo.ugr.es/event/0/session/24/contribution/405}{Extended
  investigation of the 12 flavor beta function}'', in {\em {35th International
  Symposium on Lattice Field Theory}}, 20 June 2017.

\bibitem{Hayakawa:2010yn}
M.~Hayakawa, K.~I. Ishikawa, Y.~Osaki, S.~Takeda, S.~Uno and N.~Yamada,
  ``{Running coupling constant of ten-flavor QCD with the Schr\"odinger
  functional method}'',
  \href{http://dx.doi.org/10.1103/PhysRevD.83.074509}{{\em Phys. Rev.} {\bf
  D83} (2011) 074509} [\href{http://arxiv.org/abs/1011.2577}{{\tt
  arXiv:1011.2577}}].

\bibitem{Appelquist:2012nz}
LSD Collaboration: T.~Appelquist, R.~C. Brower, M.~I. Buchoff, M.~Cheng, S.~D.
  Cohen, G.~T. Fleming, J.~Kiskis, M.~Lin, H.~Na, E.~T. Neil, J.~C. Osborn,
  C.~Rebbi, D.~Schaich, C.~Schroeder, G.~Voronov and P.~Vranas, ``{Approaching
  Conformality with Ten Flavors}'', \href{http://arxiv.org/abs/1204.6000}{{\tt
  arXiv:1204.6000}}.

\bibitem{Chiu:2016uui}
T.-W. Chiu, ``{The $\beta$-function of $SU(3)$ gauge theory with $N_f = 10$
  massless fermions in the fundamental representation}'',
  \href{http://arxiv.org/abs/1603.08854}{{\tt arXiv:1603.08854}}.

\end{thebibliography}\endgroup
\end{document}